\documentclass[aps,twocolumn,showpacs,superscriptaddress,pre,floatfix]{revtex4-1}
\usepackage{amsmath}
\usepackage{amssymb}
\usepackage{epsfig}
\usepackage{multirow}
\usepackage{graphicx}
\usepackage{filemod}
\usepackage{float}
\usepackage{textcomp}
\usepackage{datetime}
\graphicspath{{./figures/}}
\def\d{\mathrm{d}\hspace*{-0.1ex}}

\begin{document}

\title{Understanding population annealing Monte Carlo simulations}
\date{\Filemodtoday{\jobname}}

\author{Martin Weigel}
\affiliation{Applied Mathematics Research Centre, Coventry University, Coventry, CV1 5FB, United Kingdom}
\affiliation{Institut f\"ur Physik, Technische Universit\"at Chemnitz, 09107 Chemnitz, Germany}

\author{Lev Barash}
\affiliation{Landau Institute for Theoretical Physics, 142432 Chernogolovka, Russia}

\author{Lev Shchur}
\affiliation{Landau Institute for Theoretical Physics, 142432 Chernogolovka, Russia}
\affiliation{National Research University Higher School of Economics, 101000 Moscow, Russia}

\author{Wolfhard Janke}
\affiliation{Institut f\"{u}r Theoretische Physik, Leipzig University, IPF 231101, 04081 Leipzig, 
  Germany}

\begin{abstract}
  Population annealing is a recent addition to the arsenal of the practitioner in
  computer simulations in statistical physics and beyond that is found to deal well
  with systems with complex free-energy landscapes. Above all else, it promises to
  deliver unrivaled parallel scaling qualities, being suitable for parallel machines
  of the biggest calibre. Here we study population annealing using as the main
  example the two-dimensional Ising model which allows for particularly clean
  comparisons due to the available exact results and the wealth of published
  simulational studies employing other approaches. We analyze in depth the accuracy
  and precision of the method, highlighting its relation to older techniques such as
  simulated annealing and thermodynamic integration. We introduce intrinsic
  approaches for the analysis of statistical and systematic errors, and provide a
  detailed picture of the dependence of such errors on the simulation parameters. The
  results are benchmarked against canonical and parallel tempering simulations.
\end{abstract}


\maketitle


\section{Introduction}

Over the last 70 years or so, computer simulations have undoubtedly developed into
what can be considered a third pillar in science, complementing the classic duality
of experiment and analytical theory \cite{gramelsberger:11}. While this would not
have been possible without the enormous increase of computational power available
(spanning at least seven orders of magnitude), the invention of new and the
improvement of existing algorithms for many problems has made a contribution of
similar significance. The recent acceleration of the move towards higher and higher
degrees of parallelism in computational architectures has added a new twist to this
challenge: invent powerful algorithms that also have the potential to parallelize
well over thousands or even millions of cores.

The hardest computational problems relate to systems with complex free-energy
landscapes, often featuring phase transitions, entropic barriers and slow relaxation
\cite{janke:07}. To cope with such systems, that cover a broad range of areas from
(spin) glasses through biopolymers to constraint optimization problems, simulators
require methods that ensure a wide exploration of phase space, overcome energetic and
entropic barriers, and artificially speed up slow dynamics. In Monte Carlo
simulations, besides methods relating to the set of moves employed, that are
typically highly tailored to the specific systems at hand such as in cluster updates
for spin systems \cite{swendsen-wang:87a,wolff:89a}, the most important approaches
are meta-algorithms that simulate a problem in a generalized ensemble. The most
well-known schemes of this type are multicanonical simulations \cite{berg:92b},
simulated and parallel tempering \cite{geyer:91,hukushima:96a}, and Wang-Landau
sampling \cite{wang:01a}.

A more recent addition to this simulational toolbox are population annealing Monte
Carlo simulations, where an ensemble of system copies is propagated in parallel while
undergoing a gradual cooling process accompanied by periodic resampling steps
\cite{iba:01,hukushima:03,machta:10a}. While this technique has only received
significant attention in recent years
\cite{wang:15a,wang:15b,barash:16,barash:17,callaham:17,barash:18,amey:18,barzegar:17,christiansen:18,rose:19,perera:20},
related approaches have been known and used as particle filters in statistics
\cite{doucet:11} and as diffusion methods in quantum Monte Carlo
\cite{vonderlinden:92}. Also, there are closely related algorithms such as
replication techniques \cite{garel:90} and the pruned-enriched Rosenbluth method
(PERM) \cite{grassberger:97a} for walks and polymers, or the nested sampling method
for determining the density of states \cite{partay:14}. In contrast to the more
widely used Markov chain Monte Carlo (MCMC) methods, these approaches are based on
the sequential sampling paradigm \cite{doucet:13}. While originating in Monte Carlo,
it was recently demonstrated that it is indeed possible to generalize the scheme to
molecular dynamics simulations \cite{christiansen:18}, and further generalization are
likely to be found in the future.

Due to the employment of a population of system copies, the method is ideally suited
for highly parallelized implementations, and excellent scaling results have been
reported in applications
\cite{barash:16,barzegar:17,weigel:18,christiansen:18,russkov:20}. The method hence
bears great potential for attacking hard simulational problems, and indeed
significant successes have been reported for problems such as spin glasses
\cite{wang:15b,perera:20,barash:18}, hard disks \cite{amey:18} and biopolymers
\cite{christiansen:18}. The theoretical understanding of the approach, however, is
still in its infancy. While population annealing formally is a sequential Monte Carlo
method \cite{doucet:13}, it requires as a crucial ingredient a way of additionally
randomizing configurations through embedded single-replica Monte Carlo steps. This
element is conveniently chosen to be a MCMC method, bearing the additional advantage
of further driving the population towards equilibrium. As a consequence of this
hybrid composition of the approach, its performance hence depends on the subtle
interplay of correlations introduced into the population through resampling and the
decorrelating effect of the MCMC moves. It is the purpose of the present paper to
illustrate the implementation and properties of the method for a simple reference
system, to provide a clear picture of how the statistical and systematic errors
depend on the parameters of the method, and to provide techniques for monitoring the
convergence of the simulation and analyzing the resulting data. The above mentioned
success reports for non-trivial problems notwithstanding, the method has previously
not been used for a non-trivial but well controlled model system (but see
Ref.~\cite{machta:11} for a simple two-well problem). We fill this gap here by
providing an in-depth study of population annealing applied to the fruit fly of
statistical physics, the two-dimensional Ising model.

The rest of the paper is organized as follows. In
Secs.~\ref{sec:algo}--\ref{sec:initial} we formally introduce the population
annealing algorithm, followed by a short illustration of potential issues in applying
it to the Ising model. Section \ref{sec:correlations} is devoted to an exploration of
correlations introduced into the population through resampling. We show how these can
be quantified through a blocking analysis on the tree-ordered population, leading to
the introduction of the effective population size $R_\mathrm{eff}$, and how this
approach can be used to estimate statistical errors from a single run. The dependence
of statistical errors on the simulation parameters is further explored in
Sec.~\ref{sec:errors}. In Sec.~\ref{sec:free} we introduce the free-energy estimator
and illustrate its relation to thermodynamic integration, as well as discussing the
relevance and choice of weights for averaging over independent runs. Section
\ref{sec:bias} is devoted to a detailed analysis of the systematic errors of the
method and their dependence on the simulation parameters. Finally, in
Sec.~\ref{sec:performance} we survey the performance of the approach and compare it
to some more standard methods before presenting our conclusions in
Sec.~\ref{sec:conclusions}.

\section{Algorithm}
\label{sec:algo}

As outlined above, the approach is a hybrid of sequential algorithm and MCMC that
simulates a population of configurations at each time, updating them with an embedded
MCMC step and resampling the population periodically as the temperature is gradually
lowered. Population annealing (PA) can hence be summarized as follows:
\begin{enumerate}
\item Set up an equilibrium ensemble of $R_0 = R$ independent copies (replicas) of
  the system at inverse temperature $\beta_0$. Typically $\beta_0 = 0$, where this
  can be easily achieved.
\item Change the inverse temperature from $\beta_{i-1}$ to $\beta_i >
  \beta_{i-1}$. To maintain uniform weights, resample configurations
  $j = 1,\ldots, R_{i-1}$ with their relative Boltzmann weight
  $\tau_i(E_j) = \exp[-(\beta_i-\beta_{i-1})E_j]/Q_i$, where
  \begin{equation}
    Q_i \equiv Q(\beta_{i-1},\beta_i) = \frac{1}{R_{i-1}}
    \sum_{j=1}^{R_{i-1}} \exp[-(\beta_i-\beta_{i-1})E_j],
    \label{eq:Q}
  \end{equation}
  resulting in a new population of size $R_i$.
\item Update each replica by $\theta$ rounds of an MCMC algorithm at inverse
  temperature $\beta_i$.
\item Calculate estimates for observable quantities ${\cal O}$ as population averages
  $\sum_{j=1}^{R_i} {\cal O}_j/R_i$.
\item Goto step 2 unless the target temperature $\beta_\mathrm{f}$ has been reached.
\end{enumerate}
If we choose $\beta_0 = 0$, equilibrium configurations for the replicas can be
generated by simple sampling, i.e., by assigning independent, purely random
configurations to each copy. For systems without hard constraints such as typical
spin models, this process can be implemented straightforwardly. For constrained
systems such as polymers, one might need to revert to an MCMC simulation at a
sufficiently high temperature instead.

To understand the origin of the resampling factors, it is useful to consider a more
general algorithm, where resampling occurs less frequently or it is completely
omitted. It is then necessary to keep track of the weight of each replica in the
annealing process \cite{iba:01,doucet:11}. On changing the temperature from
$\beta_{i-1}$ to $\beta_i$ the Boltzmann weight $W_{i-1}^j$ of a replica at energy
$E_j$ is multiplied by the incremental importance weight $\gamma_i^j$ to arrive at
the new weight $W_i^j$ at inverse temperature $\beta_i$,
\begin{equation}
  \label{eq:boltzmann-weight}
  W_i^j = W_{i-1}^j \gamma_i^j,\;\;\; \gamma_i^j =
  \frac{Z_{\beta_{i-1}}}{Z_{\beta_i}} e^{-(\beta_i-\beta_{i-1})E_j^{i-1}},
\end{equation}
where $Z_{\beta_i}$ is the canonical partition function. Note that at each
temperature step it is the current energy $E_j^{i-1}$ of replica $j$ before
resampling that enters here, hence the additional superscript $i-1$. The initial
configurations are drawn from the uniform distribution, $W_0 = 1/Z_0$. If no MCMC
steps according to step 3 would be performed such that $E_j^i = E_j^0$ for all $i$,
this would correspond to simple sampling and the total weight would yield the
canonical probability distribution, i.e,
$W_i^j = Z_{\beta_i}^{-1} \exp(-\beta_i E_j)$.  In any case, if resampling occurs
according to the prescription outlined in step 2 of the above algorithm, the average
number of copies created for each replica equals $\tau_i(E_j)$. To keep the overall
distribution invariant, the weight of each surviving copy needs to be reduced by a
factor $1/\tau_i(E_j)$, resulting in a modified weight of
\begin{equation}
  \begin{split}
    \widetilde{W}_i^j &= \widetilde{W}_{i-1}^j \frac{\gamma_i^j}{\tau_i(E_j)} =
    \widetilde{W}_{i-1}^j \frac{Z_{\beta_{i-1}}}{Z_{\beta_i}} Q_i \\
    &= W_0^j \frac{Z_0}{Z_{\beta_1}} \cdots  \frac{Z_{\beta_{i-1}}}{Z_{\beta_i}}
    \prod_{k=1}^i Q_k = \frac{1}{Z_{\beta{i}}}  \prod_{k=1}^i Q_k.    
  \end{split}
  \label{eq:resampling-weights}
\end{equation}
These weights are now independent of the replica number $j$, such that estimates of
observables follow from plain averages over the population as indicated in step
4. The product of factors $Q_k$, however, depends on the particular realization of PA
run. As we discuss in Sec.~\ref{sec:weighted-averages} this has some consequences for
combining results from different runs.

The resampling process in step 2 can be implemented in different ways
\cite{hukushima:03,machta:10a}. As described in Ref.~\cite{hukushima:03}, in the
process of resampling a total of $R_{i-1}$ replicas are chosen according to the
probabilities $\tau_i(E_j)/R_{i-1}$ using a multinomial distribution. As
$R_i = R_{i-1} = R$, this amounts to a simulation at a fixed population size that
might be particularly useful for an implementation of the algorithm on a distributed
machine. Alternatively, one could use other resampling schemes with a fixed
population size with a similar effect \cite{gessert:prep}. On the other hand,
following Ref.~\cite{machta:10a} it is also possible to use a sum of Poisson
distributions, leading to a fluctuating population size. To keep the fluctuations of
the population size small in this case, it is useful to define the rescaled
probabilities
\begin{equation}
  \hat{\tau}_i(E_j) = (R/R_{i-1})\tau_i(E_j)
  \label{eq:taus}
\end{equation}
and draw the number of copies to make of replica $j$ according to a Poisson
distribution of mean $\hat{\tau}_i(E_j)$. Sampling from a multinomial or Poisson
distribution using uniform (pseudo) random numbers efficiently is not completely
straightforward \cite{gentle:03}, so a useful and particularly simple alternative is
to draw a random number $r$ uniformly in $[0,1)$ and take the number of copies of
replica $j$ in the new population to be
\begin{equation}
  r_i^j = \left\{
    \begin{array}{ll}
      \lfloor  \hat{\tau}_i(E_j) \rfloor & \mathrm{if}\;\;r >  \hat{\tau}_i(E_j) - \lfloor
                                           \hat{\tau}_i(E_j) \rfloor \\
      \lfloor  \hat{\tau}_i(E_j) \rfloor + 1 & \mathrm{otherwise}
    \end{array}
  \right..
  \label{eq:resampling-factors}
\end{equation}
Here, $\lfloor x\rfloor$ denotes the largest integer that is less than or equal to
$x$ (i.e., rounding down). The new population size is
$R_i = \sum_{j=1}^{R_{i-1}} r_i^j$. This method requires only a single call to the
random number generator for each replica in the current population and no lookup
tables, and additionally leads to very small fluctuations in the total population
size. In general, it is possible to use an arbitrary distribution
${\cal P}(r_i^1,\ldots,r_i^{R_{i-1}})$ as long as
$\langle r_i^j\rangle = \hat{\tau}_i(E_j)$, but the comparison of resampling methods
is outside of the scope of the present work \cite{li:20,gessert:prep}. Note that it
is common that $r_i^j = 0$ for some replicas, in which case these copies disappear
from the population, while other configurations will be replicated several times.

In the standard setup, steps of equal size in inverse temperature are taken, i.e.,
\begin{equation}
  \beta_i = \beta_{i-1} + \Delta\beta,
\end{equation}
and $\Delta\beta$ is an adjustable parameter. It is also possible to make the
temperature steps self-adaptive \cite{barash:16}, but we do not discuss this
possibility in the present paper that is focused on an analysis of the plain vanilla
algorithm in the (possibly) simplest non-trivial system. Regarding the MCMC updates
in step 3, we focus here on single-spin flip Metropolis and heatbath updates, i.e.,
local moves. The algorithm is completely general, however, and a combination with
other techniques such as non-local cluster updates is straightforward
\cite{barzegar:17}. Some possible effects of the choice of spin-update algorithm are
discussed below in Sec.~\ref{sec:updates}.

To allow for a fair comparison between PA and standard approaches, we usually
consider simulations of the same overall run-time. As will be shown below in
Sec.~\ref{sec:performance}, the time overhead for resampling is quite negligible in
most situations, almost all simulation time is spent flipping spins. For each
population member, $\theta$ calls to the MCMC subroutine are performed after each
resampling step. In the most general case, during these updates $\mu$ measurements
are taken at equidistant time steps, followed by a temperature step
$\beta \to \beta' = \beta + \Delta\beta$. As a result, at each temperature step a
sample of size $N = \mu R$ is available for all quantities considered. For the sake
of simplicity and because we want to analyze the behavior of the algorithm as
initially proposed, in the present work we focus on $\mu=1$, but we point out that
choosing $\mu>1$ will in general improve the results \cite{rose:19}.

\section{Model and observables}
\label{sec:model}

While many of the considerations developed below are fairly general, the numerical
work is focused on the case of the two-dimensional ferromagnetic Ising model in zero
field with Hamiltonian
\begin{equation}
  \label{eq:hamiltonian}
  {\cal H} = -J\sum_{\langle i,j\rangle}s_i s_j,
\end{equation}
where interactions are only considered between nearest neighbors $\langle i,j\rangle$
on an $L\times L$ square lattice, and periodic boundary conditions are applied
throughout. In the following we set $J=1$ to fix units. As is well known, this model
undergoes a continuous phase transition at the inverse temperature
$\beta_c = \frac{1}{2}\ln(1+\sqrt{2})$ \cite{mccoy:book}. In addition to the
closed-form solution of the model first derived by Onsager \cite{onsager:44}, many
results are available also for finite systems
\cite{kaufman:49a,ferdinand:69a,beale:96a}, such that we can compare our simulation
results to exact data.

For the simulations discussed here, at each temperature step we recorded the
configurational energy $E_k$ and the magnetization $M_k$, $k = 1, \ldots, N$
\footnote{Here and in the following sections for notational convenience we assume a
  constant population size even though a resampling scheme with fluctuating
  population size could be used. It should be clear how to generalize the resulting
  expressions to cover this case by replacing $R$ with $R_i$ at each step.}. From
these we calculate the average energy per spin,
\begin{equation}
e = \frac{1}{L^d N} \sum_{k=1}^N E_k,
\end{equation}
where $d=2$ for the present case, the specific heat per spin,
\begin{equation}
C_V = \beta^2 L^d\left(\frac{1}{L^{2d} N} \sum_{k=1}^N E_k^2 - e^2\right),
\end{equation}
as well as the (modulus of the) magnetization per spin,
\begin{equation}
m = \frac{1}{L^d N} \sum_{k=1}^N |M_k|, 
\end{equation}
and the magnetic susceptibility per spin,
\begin{equation}
\chi = \beta L^d\left(\frac{1}{L^{2d} N} \sum_{k=1}^N M_k^2 - m^2\right).
\end{equation}
It is of course easily possible to add other observables such as correlation lengths,
overlaps, or correlation functions, but these will not be discussed explicitly here.

\begin{figure}[tb]
  \centering
  \includegraphics[clip=true,keepaspectratio=true,width=0.95\columnwidth]{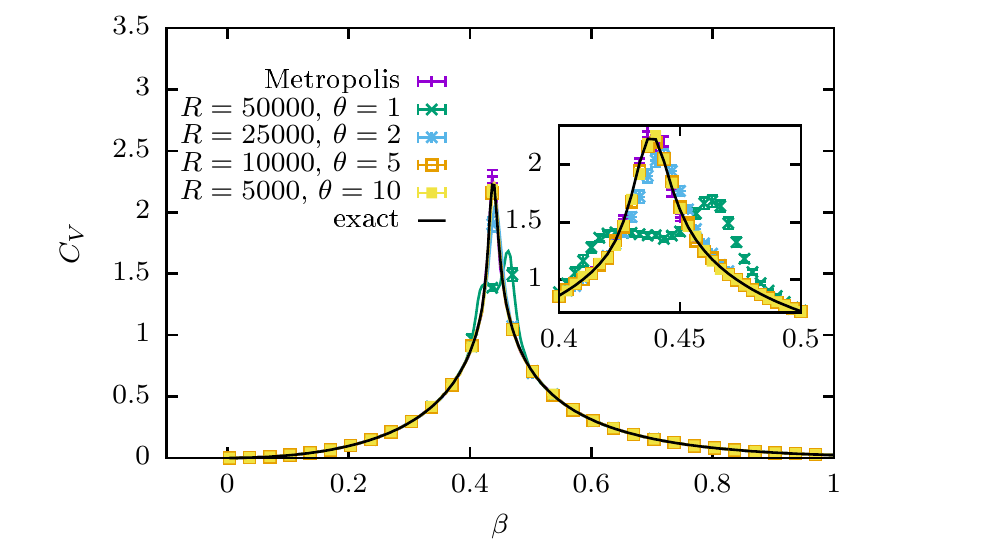}
  \includegraphics[clip=true,keepaspectratio=true,width=0.95\columnwidth]{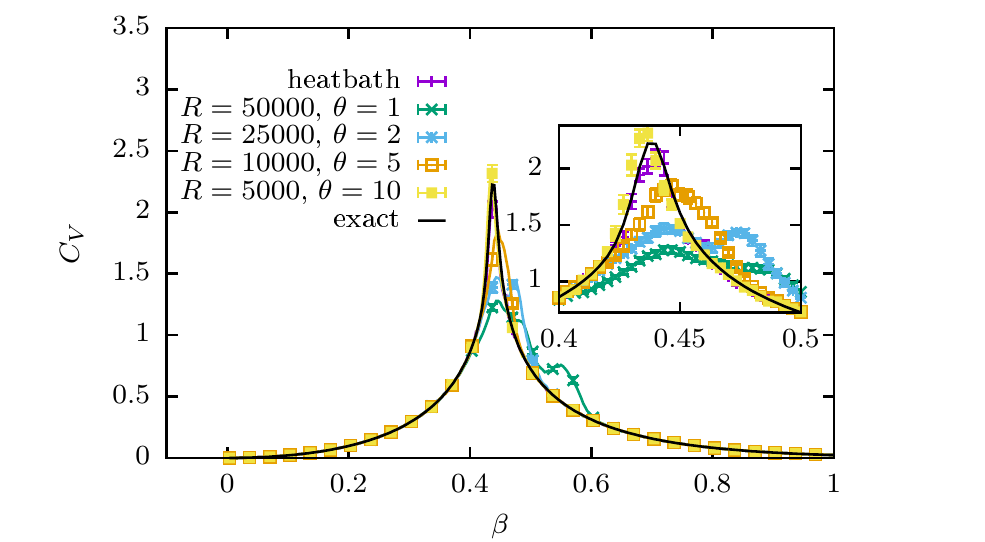}
  \includegraphics[clip=true,keepaspectratio=true,width=0.95\columnwidth]{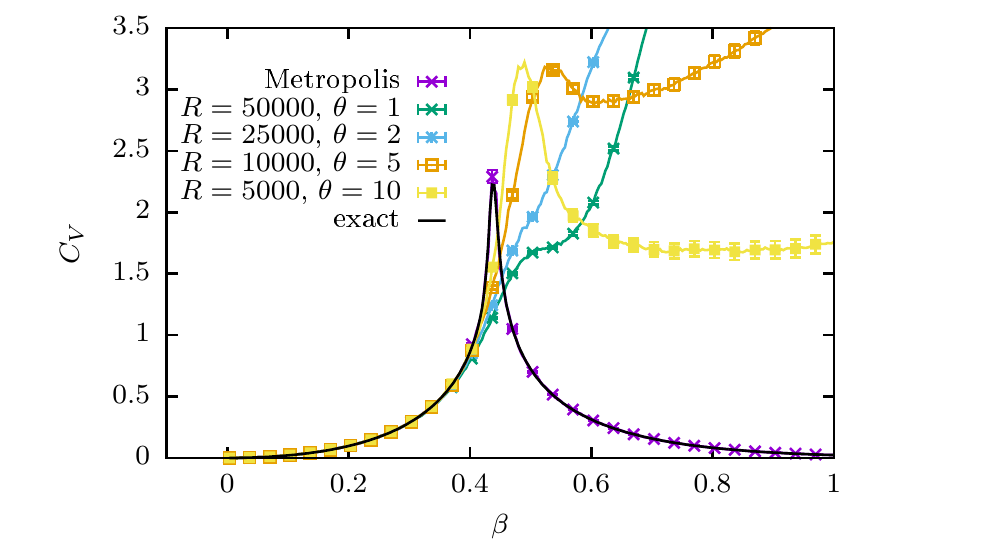}
  \caption
  {Top: Specific heat as estimated from population annealing with Metropolis update
    and fixed temperature steps $\Delta\beta = 1/300$ for $L=64$. The pure Metropolis
    simulation shown for comparison uses $R=1$ with a total of $50\,000$ sweeps and
    measurements. The exact reference data are calculated according to
    Ref.~\cite{kaufman:49a,ferdinand:69a}.  Middle: Same as top panel, but for the
    heatbath update. Bottom: Results of PA simulations with the same parameters as in
    the top panel, but with the resampling step turned off. To make the comparison as
    fair as possible, in contrast to the rest of the paper all runs shown here use
    $\mu = \theta$ measurements per temperature step. The insets in the top and
    middle panels show detail around the peak close to the critical point of the
    system.}
  \label{fig:specific_heat}
\end{figure}

\section{Initial assessment}
\label{sec:initial}

\subsection{Equilibration}
\label{sec:illustration}

All comparisons in this section are for simulations of approximately the same overall
run-time. That is, if $R$ is the population size, $\theta$ is the number of
equilibration steps, and $N_T$ is the number of simulation temperatures,
respectively, we keep the product $R\theta N_T$ constant. Also, the total number of
measurements (samples in averages) is the same for each data set.

As is seen in the top panel of Fig.~\ref{fig:specific_heat}, a small number $\theta$
of equilibration steps is not sufficient to keep the population in equilibrium, at
least in the critical region. On the other hand, it is clear that the resampling
itself also has an equilibrating effect, as the deviations are much larger for
identical runs with the resampling step switched off, cf.\ the bottom panel of
Fig.~\ref{fig:specific_heat}.

From these initial experiments it becomes clear that a more systematic understanding
of the behavior of the algorithm is required. In particular, we want to analyze the
dependence of the quality of estimation on the parameters of the approach. In the
following, we will present a principled study of the statistical and systematic error
(bias) as a function of the main control parameters, the population size $R$, the
size $\Delta\beta$ of the inverse temperature step, and the number $\theta$ of calls
to the MCMC subroutine.

\subsection{Spin updates}
\label{sec:updates}

The efficiency of the MCMC step clearly depends on the spin update employed. This is
also visible comparing the top to the middle panel of Fig.~\ref{fig:specific_heat},
where the latter shows the specific heat for the discussed simulations, but using the
heatbath update. For the Ising model studied here, single-spin flip updates
essentially come in the Metropolis and heatbath variants (the Glauber update is
equivalent to heatbath for Ising spins see, e.g.,
Ref.~\cite{janke:08}). Alternatively, one could also employ cluster updates which
would clearly lead to vastly improved decorrelation in the critical region, but an
investigation of these updates embedded in the population annealing heuristic is
postponed to a later study.

\begin{figure}[tb]
  \centering
  \includegraphics[clip=true,keepaspectratio=true,width=0.95\columnwidth]{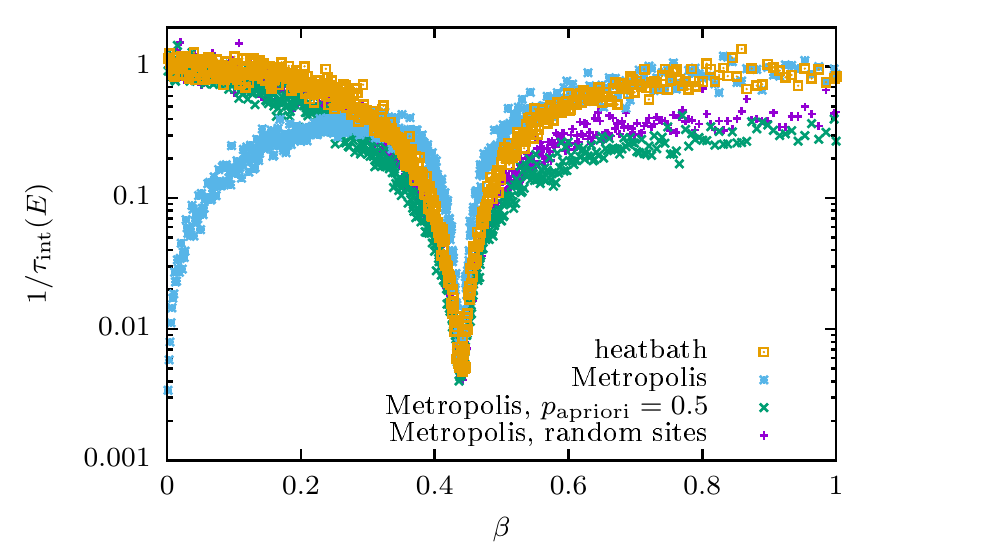}
  \caption
  {Inverse of the integrated autocorrelation time $\tau_\mathrm{int}(E)$ of the
    energy estimated from simulations of a $L=128$ system with a single replica and
    different spin updates.}
  \label{fig:updates}
\end{figure}

Figure \ref{fig:updates} shows the (inverse) integrated autocorrelation times of the
internal energy for different variants of Metropolis updates as compared to the
heatbath algorithm. For most temperatures, the sequential Metropolis update is found
to lead to the smallest values of $\tau_\mathrm{int}(E)$. It is well known that
Metropolis updates are more efficient than heatbath for Ising spins (in contrast to
systems with more micro-states such as the $q$-state Potts model with $q > 2$, see
Ref.~\cite{loison:04}). This effect is well visible in Fig.~\ref{fig:updates},
especially in the critical region. Also, sequential updating, while (or because) it
violates detailed balance and only fulfills the necessary condition of balance, in
general leads to faster decorrelation \cite{berg:04,ren:06}. Initially surprisingly,
however, the sequential Metropolis update does not work well at high temperatures. As
is easily seen, the Metropolis acceptance probability
$\min[1,\exp(-\beta\Delta E)] \to 1$ as $\beta \to 0$, hence virtually all proposed
spin flips are accepted for very small $\beta$. In a sequential scheme, however, this
means that the full spin configuration is being almost perfectly inverted through
each sweep, which clearly does {\em not\/} lead to a proper decorrelation of
configurations. In other words, the sequential Metropolis update is not ergodic in
the limit $\beta \to 0$. A number of modifications can be applied to amend this, for
instance the introduction of an {\em a priori\/} flipping probablity less than one in
the sequential update or a reversion to the random-order update. As can be seen from
Fig.~\ref{fig:updates}, these recitify the problem of non-ergodicity for
$\beta \to 0$, but at the expense of somewhat increased autocorrelation times for all
the other temperatures. In practice, the non-ergodicity of the sequential Metropolis
update does not cause any major problems in PA simulations unless one is interested
in results at very small $\beta$, and we have hence used this update for a range of
the simulations reported here. The effects of deviations for the smallest $\beta$ can
be seen in some of the figures, for instance Figs.~\ref{fig:independent_samples} and
\ref{fig:population_size}.

\section{Correlations}
\label{sec:correlations}

\subsection{Families}
\label{sec:families}

It is clear that the quality of approximation and, in particular, the statistical
errors will crucially depend on the level of correlations built up through
resampling in the population. A conservative way of estimating these effects is based
on the study of {\em families\/} \cite{wang:15,wang:15a}, i.e., the descendants of a
single configuration in the initial population. If the simulation is started with
configurations created by simple sampling at $\beta = 0$, these are rigorously
independent of each other. (This is only approximately the case for systems with
constraints where the initial configurations might be generated by sampling them from
a single simulation at high temperatures \cite{christiansen:18}.) To assess
statistical errors in a PA simulation, we would like to estimate the variance of the
mean
\begin{equation}
  \overline{\cal O} = \frac{1}{R} \sum_{j=1}^R {\cal O}_j  
\end{equation}
of an observable ${\cal O}$. Following arguments proposed in Ref.~\cite{wang:15a}, we
first consider the case that ${\cal O}$ takes the same value for all replicas of each
family, corresponding to the limit $\theta = 0$. Denote the fraction of the present
population that descends from an initial replica $k$ as $\mathfrak{n}_k$. Then,
\begin{equation}
  \overline{\cal O} = \sum_{k=1}^f \mathfrak{n}_k {\cal O}_k,
\end{equation}
where $\mathcal{O}_k$ denotes the value of the observable ${\cal O}$ in the $k$-th
family (at the current temperature), and $f$ is the number of surviving families at
the current step. To estimate the variance $\sigma^2(\overline{\cal O})$, we further
assume that $\sigma^2({\cal O}_k) = \sigma^2({\cal O})$, implying that the variance
is not correlated with the family size. Since the families are uncorrelated, the
individual variances add up and one finds
\begin{equation}
  \sigma^2(\overline{\cal O}) = \sigma^2({\cal O}) \sum_k \mathfrak{n}_k^2.
\end{equation}
In the fully uncorrelated case where each family has only one member, one has
$\mathfrak{n}_k = 1/R$ and hence
\begin{equation}
   \sigma^2(\overline{\cal O}) = \frac{\sigma^2({\cal O})}{R},
\end{equation}
as expected for the variance of the mean of uncorrelated random variables. More
generally, we might consider the quantity $R_t = R/\rho_t$ with
\begin{equation}
  \rho_t = R \sum_k{\frak n}_k^2
  \label{eq:rhot}
\end{equation}
an effective number of uncorrelated replicas in the $\theta = 0$ limit \footnote{But
  note that this incorporates the additional assumption of an absence of correlations
  between the variance and the family size. In Ref.~\cite{wang:15a} $\rho_t$ is
  defined as the limit of Eq.~\eqref{eq:rhot} for $R\to\infty$, but a consideration
  for finite population sizes is more appropriate in our formalism of considering an
  effective number of independent replicas.}. $\rho_t$ represents the mean square
family size \cite{wang:15a}.

\begin{figure}[tb]
  \centering
  \includegraphics[clip=true,keepaspectratio=true,width=0.95\columnwidth]{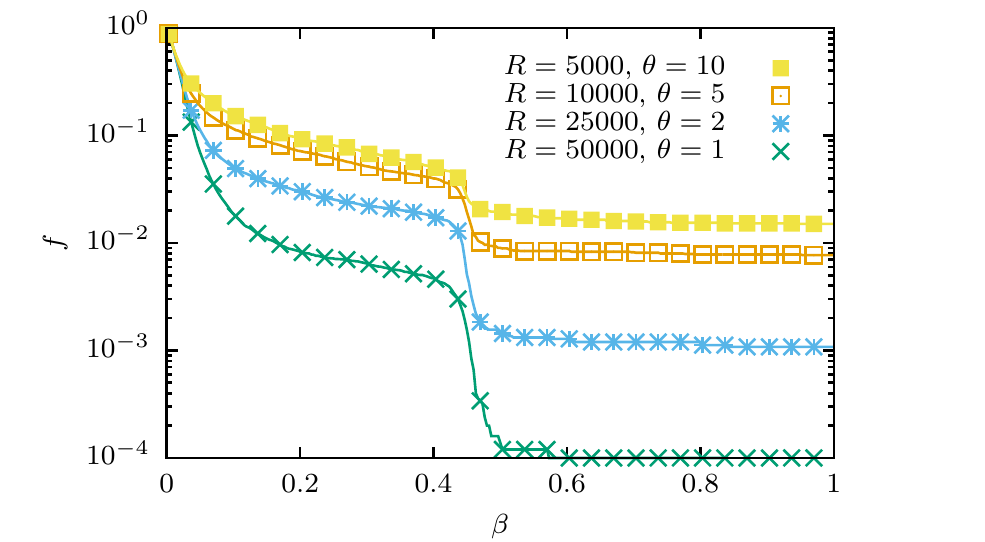}
  \includegraphics[clip=true,keepaspectratio=true,width=0.95\columnwidth]{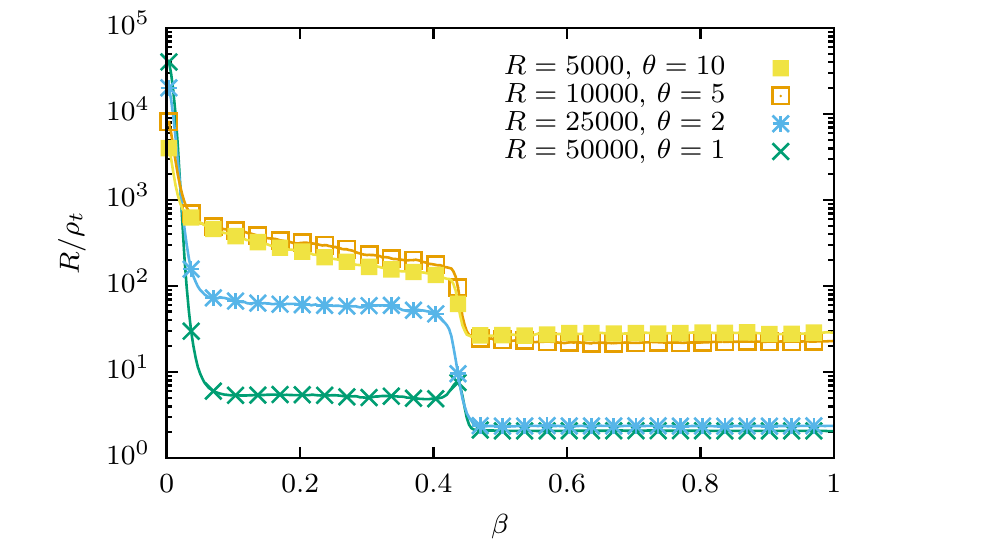}
  \includegraphics[clip=true,keepaspectratio=true,width=0.95\columnwidth]{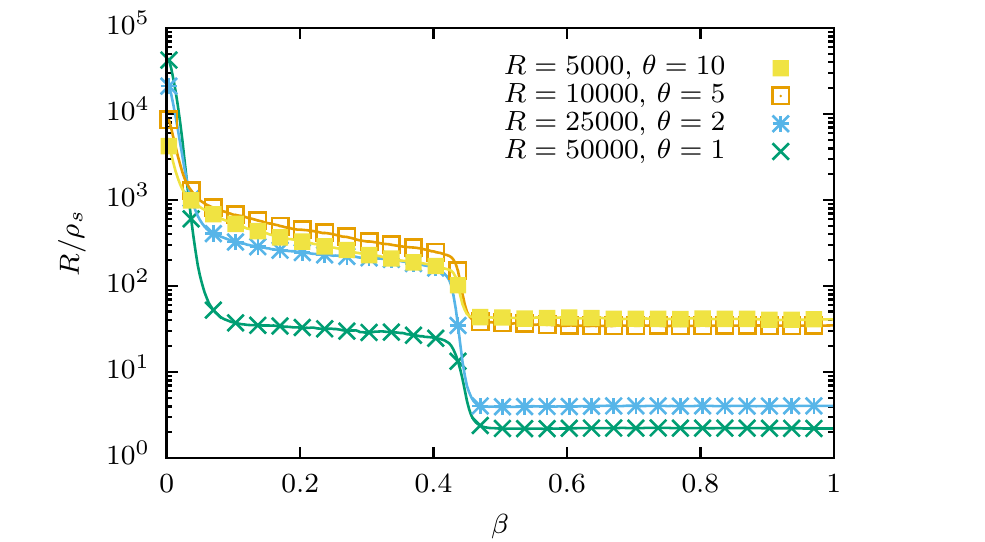}
  \caption
  {Effective number of independent replicas as estimated from the family statistics
    for PA simulations with $L=64$ (Metropolis update, $\Delta\beta = 1/300$). Top:
    number $f$ of surviving families. Middle: $R/\rho_t$ according to
    Eq.~(\ref{eq:rhot}). Bottom: $R/\rho_s$ according to Eq.~(\ref{eq:rhos}).}
  \label{fig:families}
\end{figure}

Alternatively, it was also proposed in Ref.~\cite{wang:15a} to consider the entropy
of the family size distribution,
\begin{equation}
  S_f = -\sum_k \mathfrak{n}_k \ln \mathfrak{n}_k,
\end{equation}
such that $R_s = R/\rho_s$ with
\begin{equation}
  \rho_s = R/\exp(S_f)
  \label{eq:rhos}
\end{equation}
can be considered as an alternative measure of the effective number of independent
measurements. Figure \ref{fig:families} summarizes the behavior of these
family-related correlation measures for runs of the 2D Ising model. It is clear that
they are quite similar to each other, showing a general decline through the loss of
diversity from resampling. It can be shown that at the early stages of the process
the decay in family numbers is exponential, as in each step a fraction of
$1-\kappa_i$ is lost, where $\kappa_i$ is the overlap of the energy histograms (i.e.,
the probability distribution of energies) at temperature steps $i$ and $i+1$. At
intermediate stages, however, the decay levels off as families consist of more than
one member and so the probability of their extinction is decreased. In the vicinity
of the critical point there is a further steep decline in all three quantities. This
effect has two causes, namely (1) due to the overlap of energy histograms at
neighboring temperatures having a minimum close to $\beta_c$, there is a stronger
multiplication of replicas in the low-energy wing of the distribution leading to a
stronger correlation in surviving families, and (2) at least for small $\theta$ the
population members are not fully relaxed at a given temperature, leading to a further
reduction of the overlap of the actual histogram at the higher temperature and the
equilibrium histogram at the lower one.

\subsection{Effective population size}
\label{sec:correlation}

These measures related to the family statistics, however, neglect the effect of the
spin flips which, as is seen in Fig.~\ref{fig:specific_heat} are of crucial
importance for the effectiveness of the full algorithm (in order to get correct
results from resampling alone exponentially large population sizes would be
required). We expect, for instance, that at low temperatures $\beta \gg \beta_c$ the
local dynamics will have no problem at equilibrating the population (although,
formally, the typical single spin-flip dynamics are not rapidly mixing as there
remains a barrier between the pure phases, but this is not seen in the behavior of
energetic quantities). Hence, it is clearly not the case that the ``diversity'' in
the population is minimal at the lowest temperature, but we rather expect it to
become minimal in the vicinity of the critical point and then to come up again due to
the spin-flip updates.

This behavior is captured in correlations between different members of the
population, which are the property we are actually trying to measure when considering
the family statistics. This is analogous to temporal correlations in standard MCMC
simulations which can be analyzed with a well-known toolbox of techniques
\cite{weigel:09,sokal:97}. Consider an observable ${\cal O}$. In an uncorrelated
sample, we know that the variance of the mean $\overline{\cal O}$ is inversely linear
in the sample size $R$ \cite{feller:68},
\begin{equation}
  \sigma^2(\overline{{\cal O}}) = \frac{\sigma^2({\cal O})}{R}.
  \label{eq:variance-of-the-mean}
\end{equation}
In reality, however, there are correlations introduced through the resampling process
(but reduced by the effect of the MCMC subroutine), such that the variance of the
mean decays only with an effective number of samples
$R_\mathrm{eff}({\cal O}) \le R$, i.e.,
\begin{equation}
  \sigma^2(\overline{{\cal O}}) = \frac{\sigma^2({\cal O})}{R_\mathrm{eff}} =
  \frac{\sigma^2({\cal O})}{R/2\tau_R^\mathrm{int}}, 
  \label{eq:variance-of-the-mean-correlated}
\end{equation}
where $\tau_R^\mathrm{int} = \tau_R^\mathrm{int}({\cal O})$ measures the degree of
correlation in replica space.  Assuming that we can estimate
$\sigma^2(\overline{{\cal O}})$ and $\sigma^2({\cal O})$, this provides an estimate
of the effective number of independent measurements \footnote{Note that we use
  somewhat sloppy notation here in not clearly distinguishing between probabilistic
  parameters and their estimates.},
\begin{equation}
  R_\mathrm{eff}({\cal O}) = \frac{\sigma^2({\cal O})}{\sigma^2(\overline{{\cal O}})}.
  \label{eq:neff}
\end{equation}
In general, the variance of the mean is given by
\begin{equation}
  \sigma^2(\overline{{\cal O}}) = \frac{1}{R^2}\sum_{i,j=1}^R \left(\langle {\cal O}_i {\cal
      O}_j \rangle - \langle{\cal O}_i\rangle \langle{\cal O}_j\rangle\right) =
  \frac{1}{R^2} \sum_{i,j=1}^R \Gamma_{ij},
\end{equation}
where $\Gamma_{ij}$ is the covariance matrix of the measurements ${\cal O}_i$,
$i = 1, \ldots, R$. Members of the population are more strongly correlated the more
recently in terms of previous temperature steps they have originated from the same
common ancestor. We can localize these correlations by deliberately always putting
offspring of the same parent configuration next to each other in the resampled
population. We then expect the correlations to be {\em local\/}, i.e.,
$\lim_{|i-j|\to\infty} \Gamma_{ij} = 0$ sufficiently fast, such that the variance of
the mean can be determined by considering the statistics of a {\em blocked\/} series
of $n$ bins with averages \cite{flyvbjerg:89a}
\begin{equation}
  {\cal O}_i^\mathrm{B} = \frac{1}{B}\sum_{j=(i-1)B+1}^{iB}{\cal O}_j,\;\;\;\;i=1,\ldots,n,
  \label{eq:blocks}
\end{equation}
where $B = R/n$ is the number of elements per block (for simplicity, we assume that
$n$ is chosen to divide $R$). If blocks are large enough, they will be effectively
uncorrelated and we can use the naive (uncorrelated) variance estimator to find the
variance of the mean,
\begin{equation}
  \hat{\sigma}^2(\overline{{\cal O}}) = \frac{1}{n(n-1)} \sum_{i=1}^n ({\cal O}^\mathrm{B}_i
  - \overline{{\cal O}^\mathrm{B}})^2.
  \label{eq:blocked_variance}
\end{equation}
Alternatively, in particular to reduce bias for non-linear functions of observables,
one might want to use an analysis based on jackknife blocks, where all data {\em
  apart\/} from the $i$-th block of Eq.~\eqref{eq:blocks} are gathered in block $i$,
i.e.,
\begin{equation}
  {\cal O}_i^\mathrm{J} = \frac{B}{R-B}\sum_{j\ne i}{\cal
    O}_j^\mathrm{B},\;\;\;\;i=1,\ldots,n.
  \label{eq:jackknife-blocks}
\end{equation}
The variance of the mean then follows from the corresponding jackknife estimator
\cite{efron:book},
\begin{equation}
  \label{eq:jackknife-variance}
  \hat{\sigma}^2(\overline{{\cal O}}) = \frac{n-1}{n} \sum_{i=1}^n ({\cal O}^\mathrm{J}_i
  - \overline{{\cal O}^\mathrm{J}})^2.
\end{equation}
The variance $\sigma^2({\cal O})$, on the other hand, can be estimated by the
standard (uncorrelated) variance estimator on the unblocked series, where bias
corrections are proportional to $\tau_R^\mathrm{int}/R$ and are hence not relevant in
the desirable case where $R \gg \tau_R^\mathrm{int}$.

\begin{figure}[tb]
  \centering
  \includegraphics[clip=true,keepaspectratio=true,width=0.95\columnwidth]{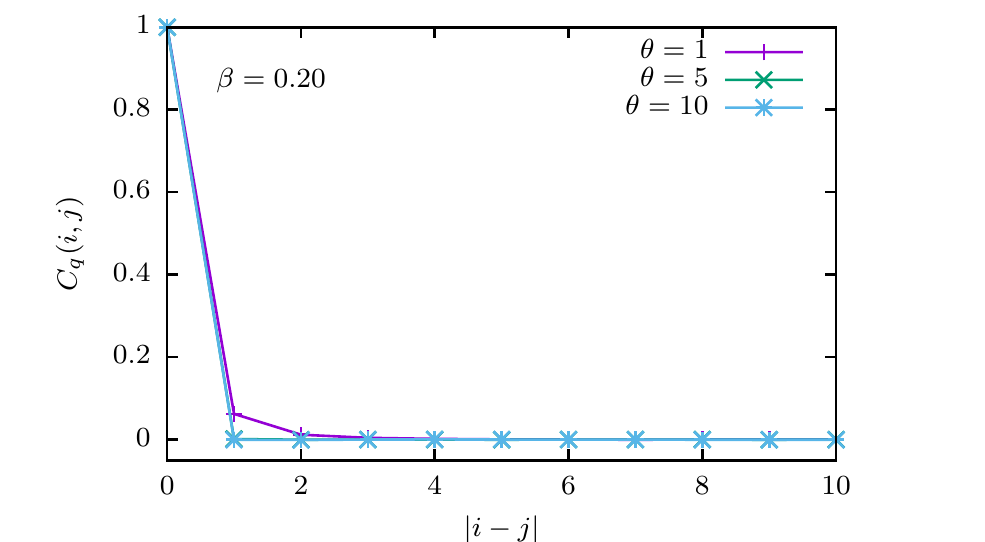}
  \includegraphics[clip=true,keepaspectratio=true,width=0.95\columnwidth]{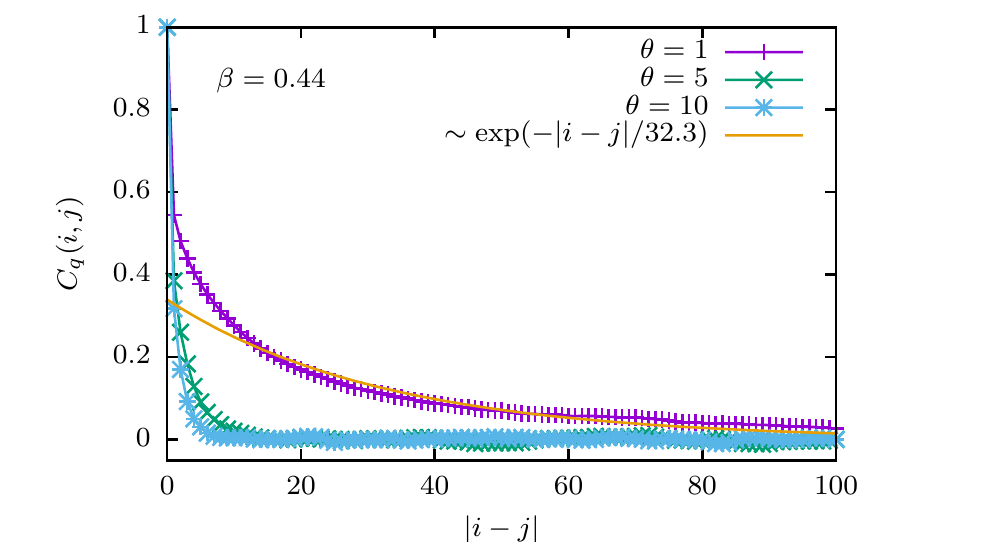}
  \includegraphics[clip=true,keepaspectratio=true,width=0.95\columnwidth]{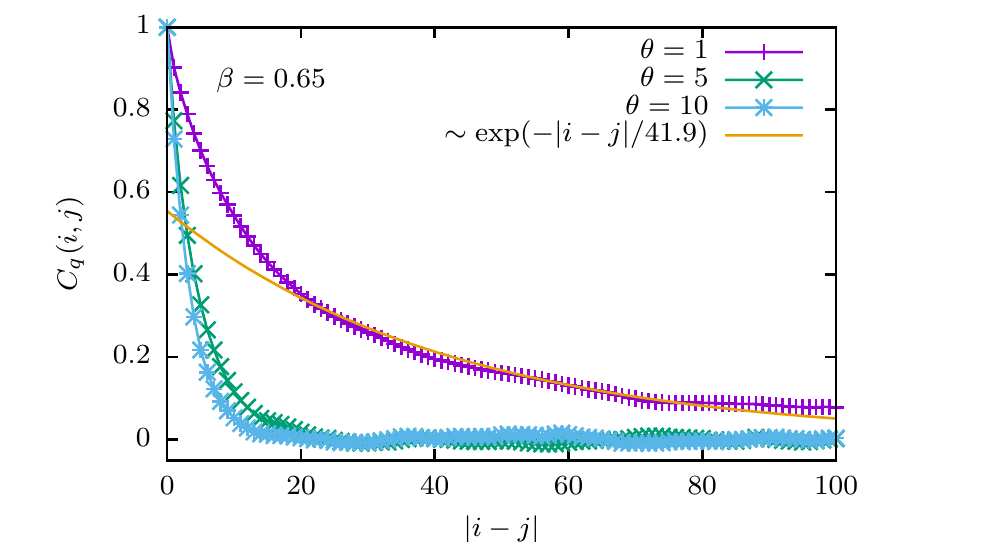}
  \caption
  {Distance dependence of the overlap correlation function $C_q(i,j)$ according to
    Eq.~\eqref{eq:overlap_correlation} for PA runs of a $L=16$ system with
    $R=50\,000$ and $\theta=1$, $5$, and $10$, respectively. Top: at high temperatures
    the population is nearly uncorrelated. Middle: close to criticality, significant
    correlations develop that asymptotically follow the form
    $C_q(i,j) \sim \exp(-|i-j|/\tau_R)$ with $\tau_R = 32.3$ ($\theta=1$),
    $\tau_R = 3.0$ ($\theta=5$), and $\tau_R = 1.8$ ($\theta=10$),
    respectively. Bottom: for the overlap, correlations persist in the ordered phase.
  }
  \label{fig:correlation_function}
\end{figure}

For the present problem, locality of correlations can be ensured through the
resampling process by placing the $r_i^j$ copies of each member of the parent
population according to Eq.~\eqref{eq:resampling-factors} at adjacent indices of the
resampled population. Hence at each stage members of the same families are grouped
together. Correlations between population members then decay with the distance
$|i-j|$ in index space since the larger this separation the further in the past of
the resampling tree do the instances have a common ancestor, with the extreme case
being that of members of different families that are by construction completely
uncorrelated. To illustrate this, we consider the distance dependence of the
configurational overlap between replicas, i.e.,
\begin{equation}
  C_q(i,j) = \frac{1}{L^d}\sum_{k=1}^{L^d} s_k^{(i)} s_k^{(j)},
  \label{eq:overlap_correlation}
\end{equation}
where $s_k^{(i)}$ denotes the $k$-th spin variables in replica $i$. We expect
$C_q(i,j)$ to be translationally invariant in replica-index space, and so
Fig.~\ref{fig:correlation_function} illustrates the behavior of $C_q(|i-j|)$ at
different temperatures for runs of $R=50\,000$ replicas. It is seen that there is a
clear decay of $C_q(i,j)$ with the replica distance $|i-j|$, and it is compatible with
an exponential asymptotic form,
\begin{equation}
  C_q(i,j) \sim \exp(-|i-j|/\tau_R),
\label{eq:exponential_R}
\end{equation}
where $\tau_R$ is negligible for high temperatures. Close to criticality for
$\beta=0.44$, the tail of $C_q(i,j)$ for $\theta=1$ is compatible with the form
\eqref{eq:exponential_R} with $\tau_R \approx 32.3$, while for $\theta=5$ it is
reduced to $\tau_R \approx 3.0$, and for $\theta=10$ we find $\tau_R \approx 1.8$. At
least for $\theta=1$ it is clearly seen that the initial decay does not follow the
same single exponential. This is in line with the behavior of time series in MCMC,
where the decay is in general understood to be a superposition of many exponentials
\cite{sokal:97}. One effect contributing to this behavior for the present case of
correlations in the population of PA is that even for nearby replicas there is a
chance of them belonging to different families, which are by definition completely
uncorrelated. The decay of correlations in the regime of very small $|i-j|$ is
therefore faster than the asymptotic decay, cf.\ the middle and lower panel of
Fig.~\ref{fig:correlation_function}.

\begin{figure}[tb]
  \centering
  \includegraphics[clip=true,keepaspectratio=true,width=0.95\columnwidth]{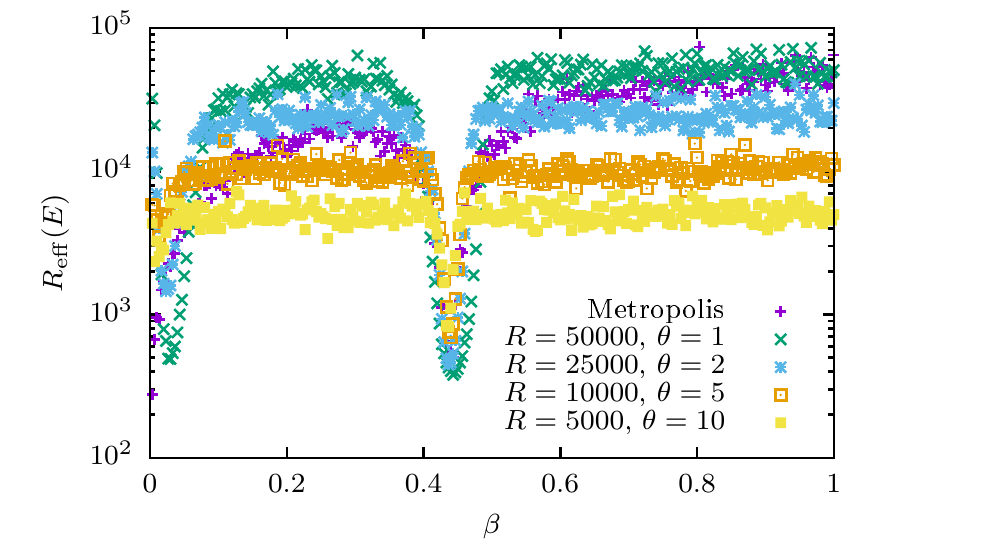}
  \includegraphics[clip=true,keepaspectratio=true,width=0.95\columnwidth]{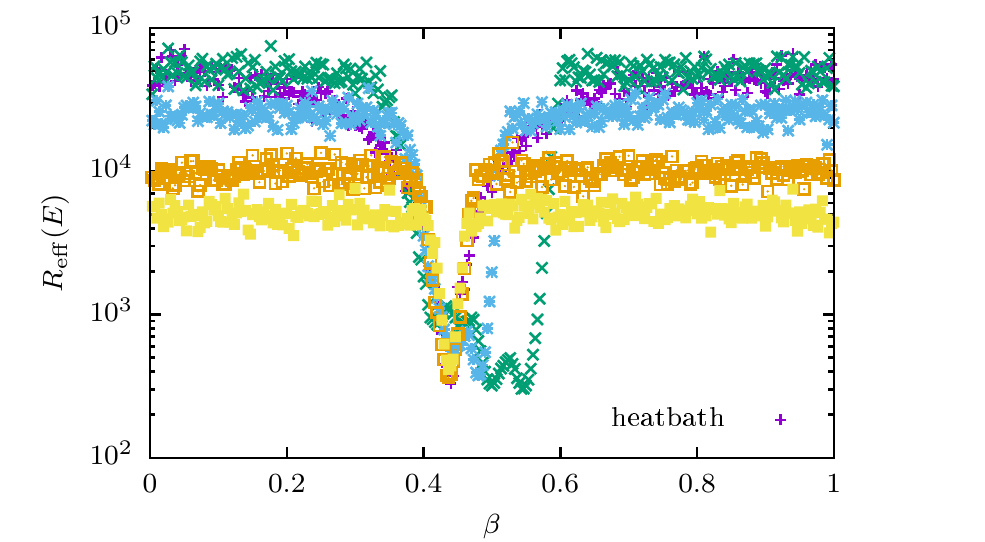}
  \includegraphics[clip=true,keepaspectratio=true,width=0.95\columnwidth]{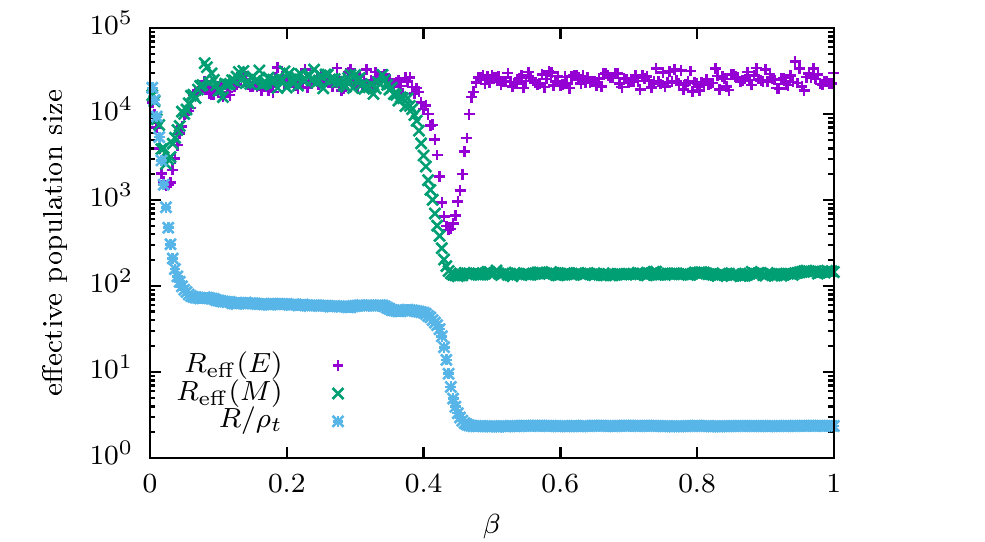}
  \caption
  {Effective number of statistically independent samples (for the energy measurement)
    in PA and standard canonical simulations for an $L=64$ system and different
    population sizes for the Metropolis (top panel) and heatbath updates (middle
    panel; for the legend of data sets see the upper panel). The bottom panel shows
    the effective population sizes $R_\mathrm{eff}(E)$ for the energy and
    $R_\mathrm{eff}(M)$ for the magnetization for $R=25\,000$, $\theta = 2$ in
    comparison to the quantity $R/\rho_t$ based on family statistics alone,
    illustrating that the latter is a lower bound for the former, but it is far from
    being tight. All simulations shown use inverse temperature steps of size
    $\Delta\beta = 1/300$.}
  \label{fig:independent_samples}
\end{figure}

Applying the blocking analysis to the thus ordered population allows one to determine
an effective number of statistically independent replicas according to
Eq.~\eqref{eq:neff}. The resulting values extracted from the variances of the energy
estimates are shown in Fig.~\ref{fig:independent_samples}. Initially, for
$\beta = 0$, the population is uncorrelated and hence $R_\mathrm{eff}=R$. The
resampling generates correlations, leading to a general decay of
$R_\mathrm{eff}$. The spin flips, on the other hand, decorrelate replicas and
therefore work towards increasing $R_\mathrm{eff}$. On approaching the critical
point, spin flips become less and less effective, leading to a decay of
$R_\mathrm{eff}$ there, similar to what is observed for the family-related
observables $R/\rho_t$ and $R/\rho_s$ in Fig.~\ref{fig:families}. In contrast to the
latter quantities that do not feel the effect of spin flips, however,
$R_\mathrm{eff}(E)$ is able to recover to $R_\mathrm{eff}(E)=R$ deep in the ordered
phase. As we shall see below, $R_\mathrm{eff}$ plays a central role in the
characterization of the performance of a PA run. The bottom panel of
Fig.~\ref{fig:independent_samples} illustrates the fact that the family-related
quantity $R/\rho_t$ is a lower bound for $R_\mathrm{eff}$, but it is far from tight
and it can in fact be orders of magnitude below $R_\mathrm{eff}$. Due to the dynamic
ergodicity breaking, $R_\mathrm{eff}(M)$ does not recover in the ordered phase in the
way observed for $R_\mathrm{eff}(E)$. Note, however, that this is dependent on the
update algorithm employed and, for instance, if using a cluster-update method
\cite{swendsen-wang:87a,wolff:88a,wolff:89a} both $R_\mathrm{eff}(E)$ and
$R_\mathrm{eff}(M)$ approach $R$ also in the ordered phase.

\section{Statistical errors}
\label{sec:errors}

Note that the same blocking analysis provides estimates of statistical errors of
quantities sampled in PA from a single simulation run, such that the error estimates
through multiple runs initially proposed in Ref.~\cite{machta:10a} are no longer
necessary. To this end one can use the blocked estimator \eqref{eq:blocked_variance}
or, equivalently, the jackknife estimator \eqref{eq:jackknife-variance} for the
variance of the mean. For non-linear observables such as the specific heat,
correlation length etc.\ one should instead always use the jackknife form
\eqref{eq:jackknife-variance} in order to minimize the statistical bias in error
estimates \cite{efron:book,janke:02,young:12}.

A useful check of self-consistency is to monitor the number of independent samples
estimated through Eq.~(\ref{eq:neff}). The ratio $R/R_\mathrm{eff}$ is like an
integrated autocorrelation time. It should be much smaller than the size $B=R/n$ of a
block for the approach to be self-consistent. This is the case if
\begin{equation}
  R_\mathrm{eff} \gg n.
\end{equation}
Since we typically use $n=100$ to arrive at reliable error estimates,
$R_\mathrm{eff}$ should not fall below a few thousand replicas to avoid bias in the
error estimation. At the same time, however, the PA simulation itself is no longer
reliable if this condition is not met as we do not have sufficient statistically
independent information to sample the energy distribution faithfully. As we shall see
below, $R_\mathrm{eff}$ also affects the simulation bias. Monitoring $R_\mathrm{eff}$
hence serves as an important indicator of the trustability of the simulation results
-- much like the integrated autocorrelation time provides such an indicator for MCMC
simulations \cite{sokal:97}.

\begin{figure}[tb]
  \centering
  \includegraphics[clip=true,keepaspectratio=true,width=0.95\columnwidth]{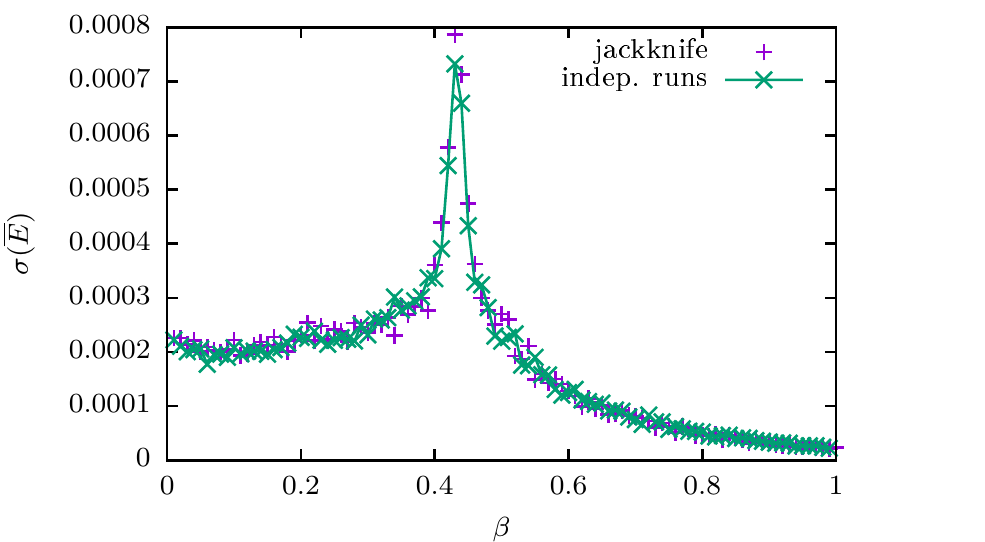}
  \caption
  {Estimated standard deviation of the mean (a.k.a.\ error bar) of the internal
    energy, $\sigma(\overline{E})$, for the $L=32$ Ising model as estimated from PA
    simulations with the blocking technique using Eq.~\eqref{eq:jackknife-variance}
    compared to the unbiased estimate resulting from repeating the full simulation
    200 times with different sequences of random numbers ($R=50\,000$, $\theta=10$,
    $\Delta\beta = 0.01$). All simulations use the Metropolis update.
    \label{fig:statistical_errors}}
\end{figure}

\begin{figure}[tb]
  \centering
  \includegraphics[clip=true,keepaspectratio=true,width=0.95\columnwidth]{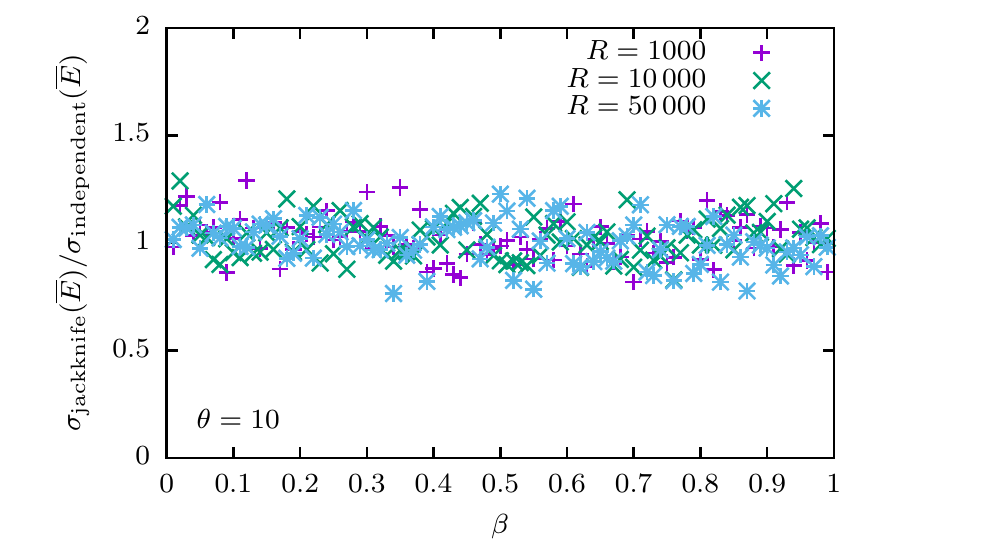}
  \includegraphics[clip=true,keepaspectratio=true,width=0.95\columnwidth]{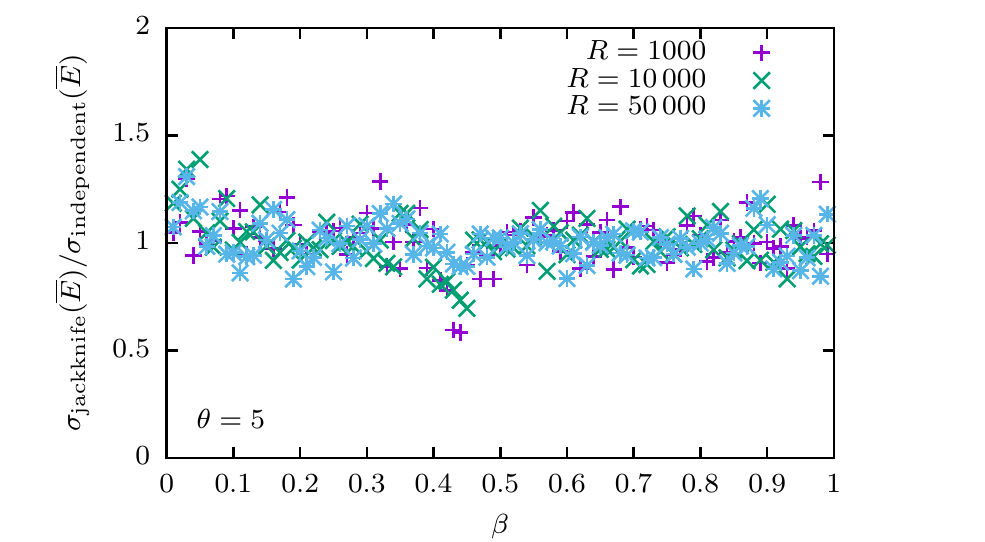}
  \includegraphics[clip=true,keepaspectratio=true,width=0.95\columnwidth]{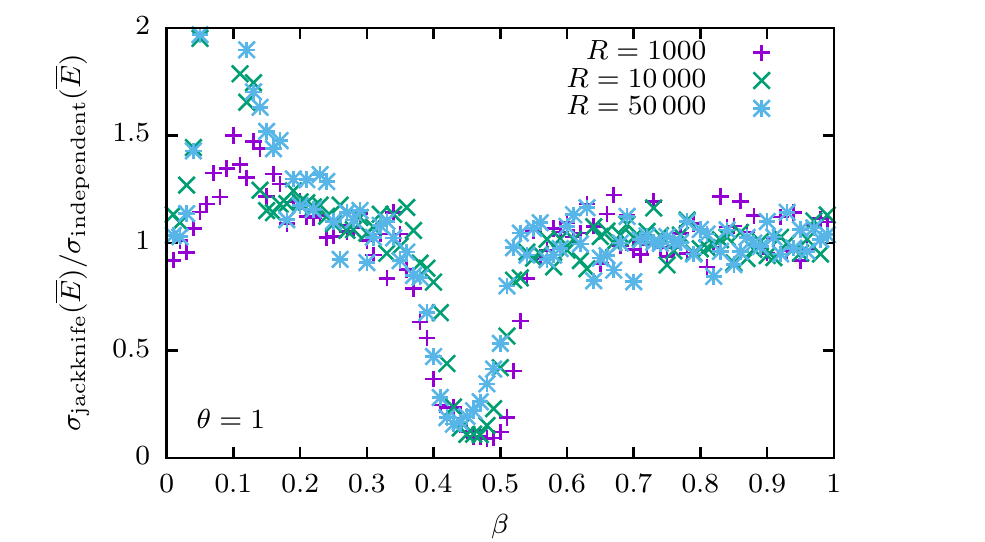}
  \caption
  {Ratio of error bars estimated from the jackknife analysis of a single PA run and
    from the fluctuation between 200 independent runs for PA simulations of the 2D
    Ising model with $L=32$ ($\Delta\beta = 0.01$). Deviations from the target value
    1.0 occur if the number $R_\mathrm{eff}$ of independent measurements becomes too
    small and hence the blocks in the analysis are no longer independent. In this
    case the blocking or jackknifing approach underestimates the errors (bottom
    panel).
    \label{fig:jackknife_problem}}
\end{figure}

To confirm the reliability of this way of estimating statistical errors, we show in
Fig.~\ref{fig:statistical_errors} a comparison of the error bars thus computed to the
errors estimated independently from repeating the PA simulation 200 times with
independent seeds of the random-number generator. These simulations for $L=32$,
$R=50\,000$ and $\theta=10$ show full compatibility between the two approaches. A
more detailed analysis shown in Fig.~\ref{fig:jackknife_problem} illustrates the
dependence on population size (which shows no differences between the two approaches)
and the number $\theta$ of rounds of spin flips. It is clear that as soon as the
number $R_\mathrm{eff}$ of independent samples becomes too small, and hence the
population too strongly correlated, the blocking analysis becomes unreliable (we find
that $R_\mathrm{eff} \approx 600$ for $R=50\,000$ near criticality for $\theta=1$).
As in the analysis of MCMC simulations, it is hence quite easily possible to monitor
the self-consistency of the error analysis.

\begin{figure}[tb]
  \centering
  \includegraphics[clip=true,keepaspectratio=true,width=0.95\columnwidth]{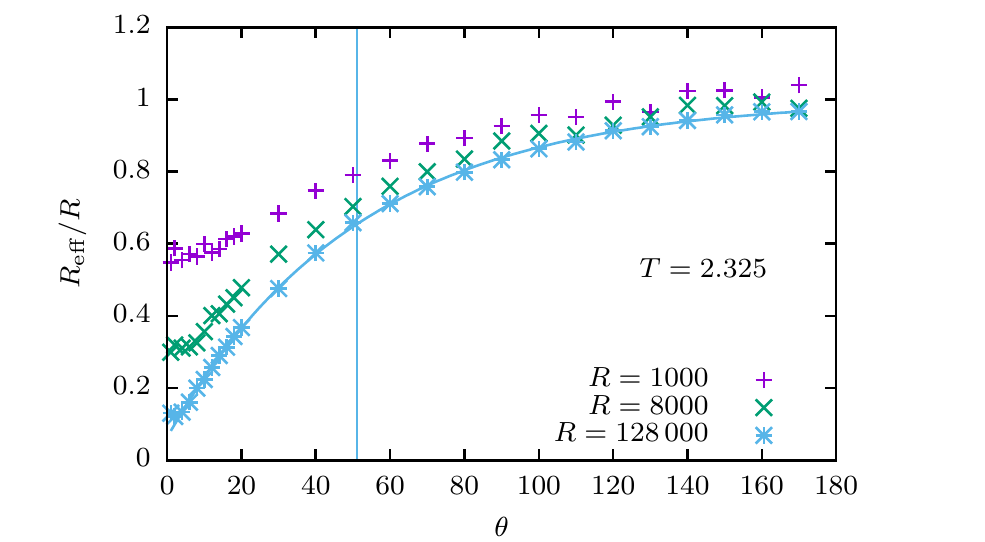}
  \includegraphics[clip=true,keepaspectratio=true,width=0.95\columnwidth]{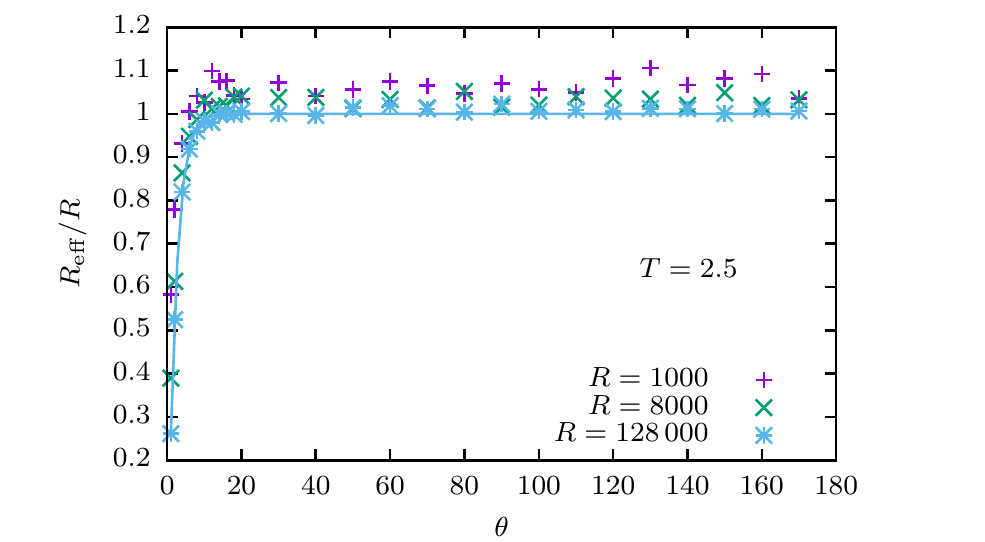}
  \caption
  {Effective relative number of independent members of the population,
    $R_\mathrm{eff}/R$, for PA simulations of the $L=64$ system with
    $\Delta\beta = 0.005$ and varying population sizes in dependence of the $\theta$
    values. Upper panel: runs at temperature $T=2.325$, close to the critical
    point. Lower panel: runs at $T=2.5$ in the disordered phase. The lines show fits
    of the functional form \eqref{eq:Reff} to the data with
    $\tau_\mathrm{eff} = 51.2$ (upper panel) and $\tau_\mathrm{eff} = 2.2$ (lower
    panel), respectively.
    \label{fig:Neff_theta_scaling}}
\end{figure}

It remains to discuss the dependence of statistical errors on the parameters $R$,
$\Delta\beta$ and $\theta$ of the PA simulation. At each resampling step,
correlations are introduced into the population through the creation of identical
copies of some members and the elimination of others, leading to a reduction of the
effective population size $R_\mathrm{eff}$. The subsequent application of spin flips
onto each replica, on the other hand, works towards removing such correlations
between population members with common ancestors. The quantitative effect of these
processes is discussed in more detail in Appendix
\ref{app:effective-population-size}. The corresponding correlation and decorrelation
of replicas depends on the model as well as on temperature, such that the overall effect
on $R_\mathrm{eff}$ after a number of temperature steps is hard to infer in closed
form. Overall, however, one clearly expects an exponential dependence of
$R_\mathrm{eff}$ on $\theta$, and we find the following relation to accurately
describe the data,
\begin{equation}
  R_\mathrm{eff} = R\left[1-c\exp(-\theta/\tau_\mathrm{eff})\right],
  \label{eq:Reff}
\end{equation}
where $c$ is some constant that might depend on further simulation parameters such as
$\Delta \beta$ (see below). To illustrate this, in Fig.~\ref{fig:Neff_theta_scaling}
we show the behavior of $R_\mathrm{eff}/R$ as a function of $\theta$ for temperatures
$T=2.325$ and $T=2.5$ and a range of different population sizes. As the population
size is increased, $R_\mathrm{eff}$ approaches the form \eqref{eq:Reff}, which is
very well observed for the largest population with $R=128\,000$ as is illustrated by
the fit of the form \eqref{eq:Reff} shown in the upper panel of
Fig.~\ref{fig:Neff_theta_scaling} that yields $\tau_\mathrm{eff} \approx 51.2$. Given
that we used $B=31$, $62$ and $250$ bins for $R=1000$, $8000$, and $128\,000$,
respectively, our self-consistency condition $R_\mathrm{eff} \gg n$, which in
practice we read as $R_\mathrm{eff} > 50 n$, implies that all estimates of
$R_\mathrm{eff}$ for $R=1000$ are unreliable, while this applies only to
$R_\mathrm{eff}/R \le 0.4$ for $R=8000$ and $R_\mathrm{eff}/R \le 0.09$ for
$R=128\,000$, which appears to be in line with the observed deviations in the upper
panel of Fig.~\ref{fig:Neff_theta_scaling} \footnote{Note that the ratio
  $R_\mathrm{eff}/R$ will always be overestimated if the blocks in the binning
  analysis are not sufficiently independent. This leads to an approach of the
  asymptotic form of Eq.~\eqref{eq:Reff} from above as is seen in
  Fig.~\ref{fig:Neff_theta_scaling}}. As is apparent from the lower panel, such
problems only occur for the smallest values of $\theta$ for the higher temperature
$T=2.5$. There, the functional form \eqref{eq:Reff} works well for all three
population sizes, and a fit yields $\tau_\mathrm{eff} \approx 2.2$.

\begin{figure}[tb]
  \centering
  \includegraphics[clip=true,keepaspectratio=true,width=0.95\columnwidth]{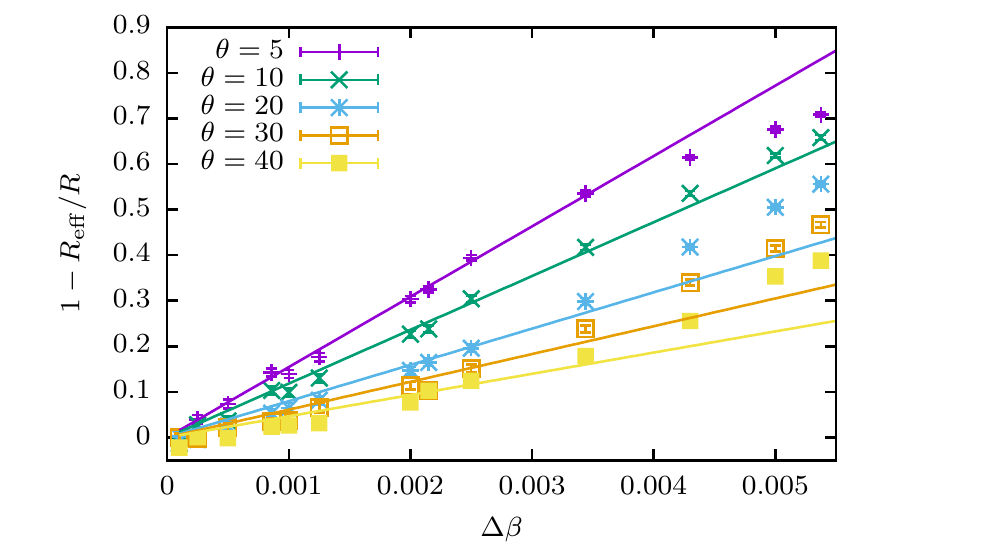}
  \caption
  {Relative deviation of the effective population size $R_\mathrm{eff}$ from the
    actual population size $R$ as a function of $\Delta\beta$ for an $L=64$ system
    close to criticality for a range of different values of $\theta$ ($\beta = 0.43$,
    $R=10\,000$). The data are averaged over 200 independent runs. The lines show
    one-parameter fits of the form $1-R_\mathrm{eff}/R = A \Delta\beta$.
    \label{fig:Neff_deltabeta_scaling}
  }
\end{figure}

Regarding the dependence of statistical errors on the (inverse) temperature step
$\Delta\beta$, it is clear that $R_\mathrm{eff}$ should approach $R$ as
$\Delta\beta\to 0$ since there is no correlating effect from resampling for
$\Delta\beta = 0$ and the number of Monte Carlo sweeps performed in a given
temperature interval increases inversely proportional to $\Delta\beta$. In
Fig.~\ref{fig:Neff_deltabeta_scaling} we show $1-R_\mathrm{eff}/R$ as a function of
$\Delta\beta$ for different choices of $\theta$, and it is clear that the behavior is
linear for small $\Delta\beta$. In fact, it is possible to derive this scaling for
the behavior in a single temperature step from the arguments laid out in
Appendix~\ref{app:effective-population-size}. We hence generalize relation
\eqref{eq:Reff} to include the effect of varying temperature steps,
\begin{equation}
  \label{eq:Reff-deltabeta}
  R_\mathrm{eff} = R\left[1-\frac{\Delta\beta}{\Delta\beta_0}
    \exp(-\theta/\tau_\mathrm{eff})\right],
\end{equation}
which is correct in the limit $\Delta\beta \to 0$. Here, $\Delta \beta_0$ is an
empirical constant, and from the fits shown in Fig.~\ref{fig:Neff_deltabeta_scaling}
we find $\Delta \beta_0 \approx 0.005$, which is in line with the results of
Fig.~~\ref{fig:Neff_theta_scaling}, where we found $c\approx 1$ for the step size
$\Delta \beta = 0.005$ used there.

\begin{figure}[tb]
  \centering
  \includegraphics[clip=true,keepaspectratio=true,width=0.95\columnwidth]{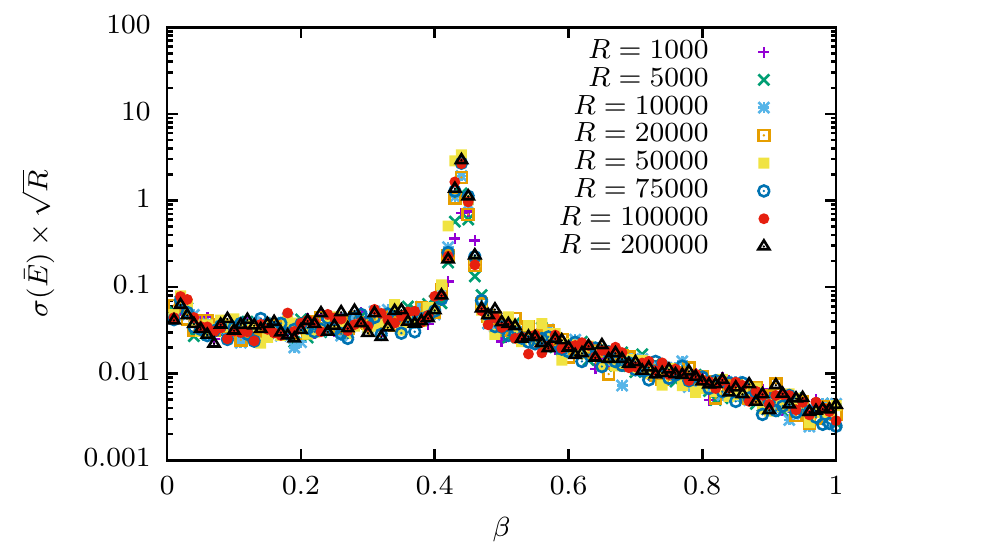}
  \includegraphics[clip=true,keepaspectratio=true,width=0.95\columnwidth]{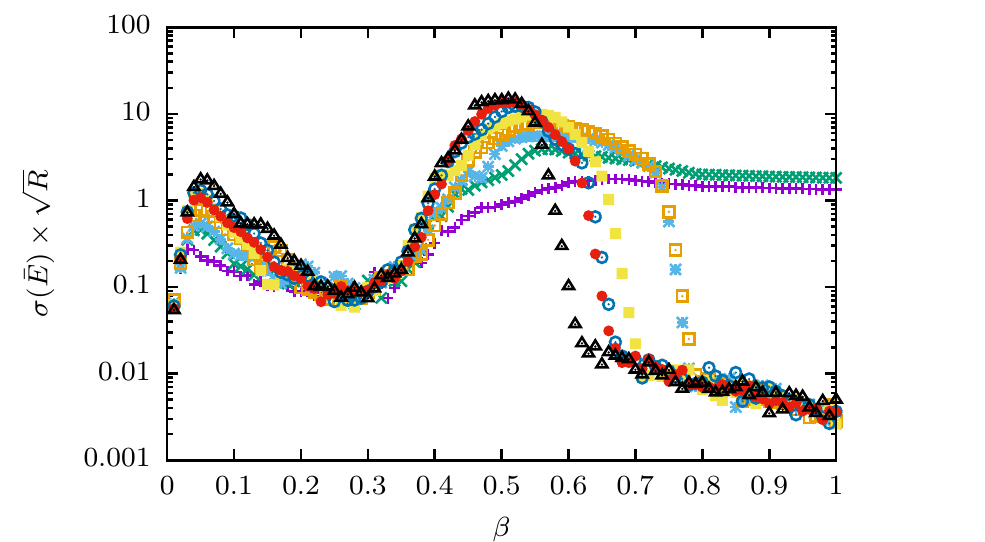}
  \caption
  {Estimated standard deviation of the mean  times $\sqrt{R}$ for the
    energy of the $L=64$ Ising model for PA simulations with a range of different
    population sizes $R$ for $\theta = 10$ (top) and $\theta = 1$ (bottom),
    respectively ($\Delta\beta = 0.01$).
    \label{fig:population_size}
  }
\end{figure}

The population size affects the statistical errors in the expected way. In well
equilibrated simulations statistical errors decay as $1/\sqrt{R}$ which is an
immediate consequence of the number of families growing linearly with $R$, such that
the number of independent samples must also grow linearly with the population
size. This fact is illustrated in the scaling plots for the error bars of the energy
in Fig.~\ref{fig:population_size}. For $\theta = 10$ the simulations are sufficiently
close to equilibrium everywhere and clear $1/\sqrt{R}$ scaling of statistical errors
is observed. For $\theta=1$, however, such behavior is almost nowhere observed: for
very high temperatures this is prevented by the non-ergodicity of the sequential
Metropolis update (cf.\ Sec.~\ref{sec:updates}), and in the critical region it is
prevented by critical slowing down. Only for very low temperatures a scaling collapse
is observed.

\section{Free energies}
\label{sec:free}

\subsection{Free-energy estimate}
\label{sec:free-energy}

As was already shown by Hukushima and Iba \cite{hukushima:03}, population annealing
naturally allows to estimate partition-function ratios or, equivalently, free-energy
differences. This can be motivated by the following telescopic product expansion of
the partition function,
\begin{equation}
\begin{split}
  -\beta_i F(\beta_i) &= \ln Z_{\beta_i} = \ln \left(\frac{Z_{\beta_i}}{Z_{\beta_{i-1}}}
  \frac{Z_{\beta_i-1}}{Z_{\beta_{i-2}}}\cdots  \frac{Z_{\beta_1}}{Z_{\beta_0}} Z_{\beta_0}\right)\\
  &= \ln Z_{\beta_0} + \sum_{j=0}^{i-1} \ln  \frac{Z_{\beta_{j+1}}}{Z_{\beta_j}}.
\end{split}
\end{equation}
The partition function ratios on the r.h.s.\ are equivalent to expectation values of
ratios of Boltzmann factors,
$\langle \exp[-(\beta_k-\beta_{k-1})E]\rangle_\beta = Z_{\beta_k}/Z_{\beta_{k-1}}$, which in turn
are estimated in PA by the normalization factors $Q(\beta_{k-1},\beta_k)$, cf.\
Eq.~\eqref{eq:Q}. As a consequence, a natural estimator of the free energy at inverse
temperature $\beta_i$, $i=0,\ldots,N_T$ (there are $N_T+1$ temperatures including
$\beta_0$) is given by \cite{machta:10a}
\begin{equation}
  \label{eq:free-emergy-estimator}
  -\beta_i \hat{F}_i = \ln Z_{\beta_0} + \sum_{k=1}^{i} \ln Q(\beta_{k-1},\beta_k).
\end{equation}
If $\beta_0 = 0$ is chosen, $ Z_{\beta_0}$ corresponds to the number of microstates,
which usually can be worked out exactly. For the present case of Ising systems, we
have $\ln Z_{\beta_0} = L^d\ln 2$.

While the form Eq.~\eqref{eq:free-emergy-estimator} might appear like an expression
that is specific to the PA method, it is in fact a slight generalization of what is
more traditionally known in the field of Monte Carlo simulations as {\em
  thermodynamic integration\/}. This is easily seen by noting that in the limit
$\Delta\beta\to 0$ of small (inverse) temperature steps we have
\begin{equation}
  \begin{split}
    -\beta_i F(\beta_i) &= \ln Z_0 + \ln \prod_{k=1}^i \langle
    e^{-(\beta_k-\beta_{k-1})E}\rangle_{\beta_{k-1}} \\
    \stackrel{\Delta\beta_k\to
      0}{\longrightarrow} & \sum_k (-\Delta\beta_k)\langle E\rangle_{\beta_{k-1}}+\ln Z_o \\
     \stackrel{\Delta\beta_k\to
      0}{\longrightarrow} & \int_{\beta_0}^{\beta_i} \langle -E(\beta')\rangle\,\d \beta' +
    \ln Z_0,
    \\
  \end{split}
  \label{eq:td-integration}
\end{equation}
where $\Delta \beta_k = \beta_k-\beta_{k-1}$. Equation~\eqref{eq:td-integration} is
the standard expression for calculating free energies via thermodynamic integration
\cite{janke:03,binder:book2,barash:17}. In fact, the above relation can be read in
the opposite direction also, telling us that a more accurate version of thermodynamic
integration that disposes of the requirement of taking {\em small\/} inverse
temperature steps is given by the first line of
Eq.~\eqref{eq:td-integration}. (Something which surely must have been observed
somewhere else before.)

\begin{figure}[tb]
  \centering
  \includegraphics[clip=true,keepaspectratio=true,width=0.95\columnwidth]{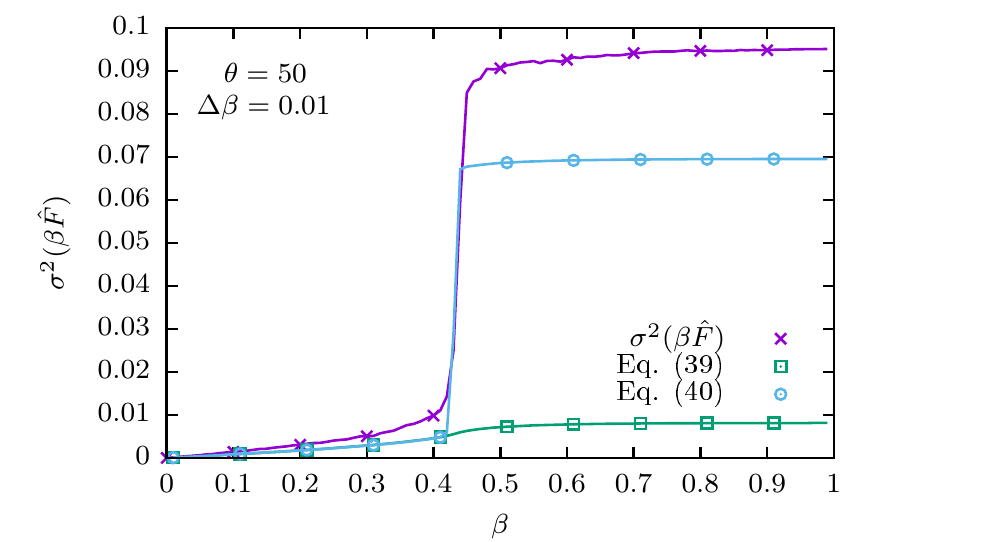}
  \includegraphics[clip=true,keepaspectratio=true,width=0.95\columnwidth]{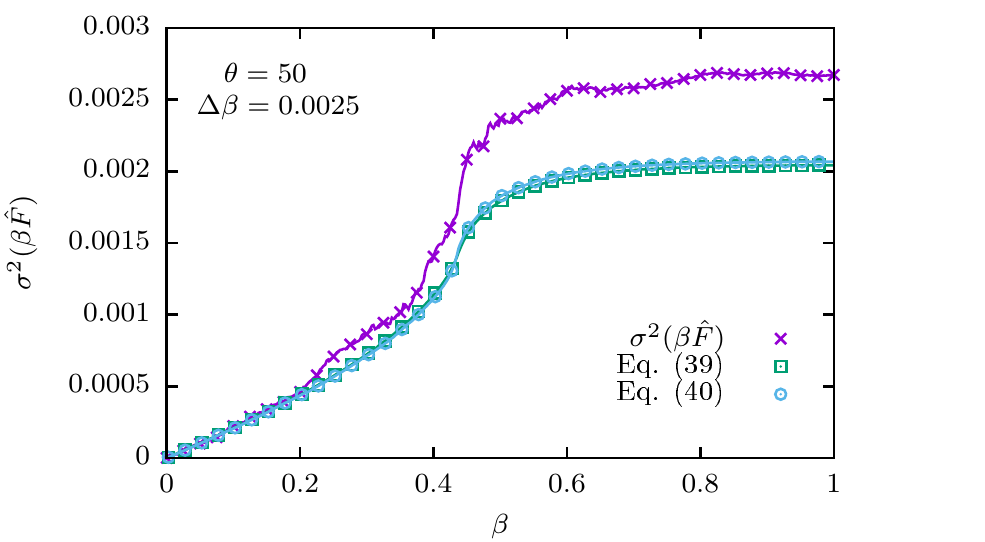}
  \includegraphics[clip=true,keepaspectratio=true,width=0.95\columnwidth]{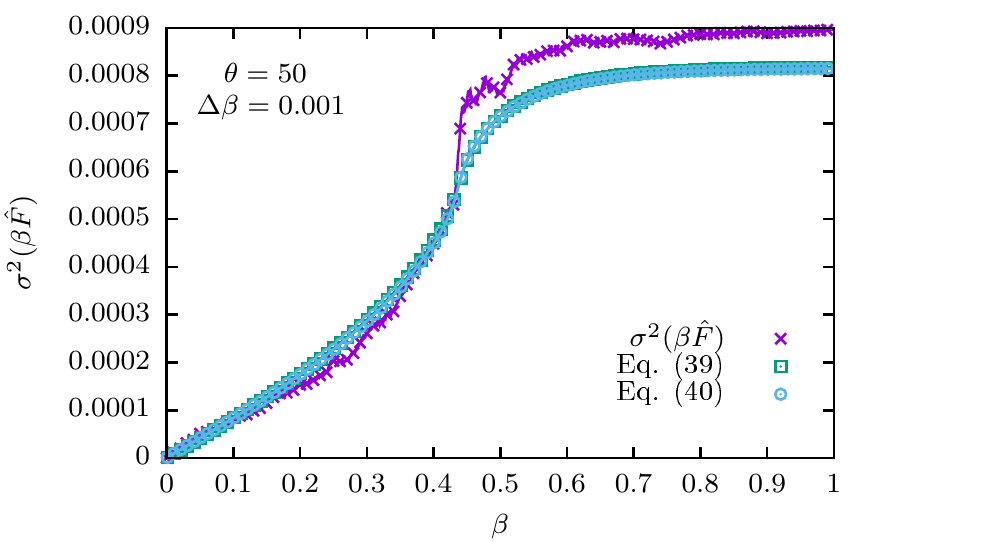}
  \caption
  {Variance of the free-energy estimator $\beta\hat{F}$ for different choices of the
    inverse temperature step $\Delta\beta$. The purple crosses show the variance of
    the quantity \eqref{eq:free-emergy-estimator} as estimated from 200 independent
    PA runs. The green squares and blue circles show the estimates derived via
    Eqs.~\eqref{eq:varbetaf2} and \eqref{eq:varbetaf3}, respectively, from the same
    simulations ($L=64$, $R=10\,000$, $\theta = 50$). For clarity of presentation,
    only every 10th temperature point is plotted with a symbol.
    \label{fig:td-integration}}
\end{figure}

Apart from being an interesting observation, relation \eqref{eq:td-integration}
provides a useful guideline allowing us to understand the behavior of PA in the limit
of small temperature steps. In the above limit $\Delta\beta\to 0$ of thermodynamic
integration, an alternative PA estimator of the free energy is given by
\begin{equation}
  -\beta\tilde{F}(\beta) = \ln Z_0 - \int_{\beta_0}^\beta \overline{E}(\beta')\,\d\beta',
  \label{eq:free-en-through-av-en}
\end{equation}
where $\overline{E}(\beta_i) = (1/R_i)\sum_j E_j$ is the population average of the
internal energy. As we shall see in Sec.~\ref{sec:weighted-averages} the variance of
the free-energy estimate is of some relevance for the reliability of PA. According to
Eq.~\eqref{eq:free-en-through-av-en} the variance of $\beta\tilde{F}$ is given by
\begin{equation}
  \sigma^2(\beta\tilde{F}) = \sigma^2\left[\int_{\beta_0}^\beta
    \overline{E}(\beta')\,\d\beta'\right].
\end{equation}
If the populations at successive temperatures are statistically independent of each
other, one can interchange the variance and the integral to find
\begin{equation}
  \sigma^2(\beta\tilde{F}) \approx \Delta\beta \int_{\beta_0}^\beta
  \sigma^2\left[\overline{E}(\beta')\right]\,\d\beta'.
  \label{eq:varbetaf}
\end{equation}
Hence, the variance of the free energy estimator corresponds to the integral (sum) of
the squared error bars of the energies along the trajectory in $\beta$. Clearly, the
variance of $\beta\tilde{F}$ is proportional to the temperature step $\Delta\beta$ in
this limit. As we shall see below in Sec.~\ref{sec:effect-of-resampling}, this
implies that the bias of PA is also linear in $\Delta\beta$. If also the members of
the population at a given temperature are uncorrelated to each other, one concludes
that
\begin{equation}
   \sigma^2(\beta\tilde{F}) \approx \frac{\Delta\beta}{R} \int_{\beta_0}^\beta
   \sigma^2\left[E(\beta')\right]\,\d\beta'.
   \label{eq:varbetaf1.5}
 \end{equation}
Finally, if the simulation is in equilibrium at all times, one can also write this as
\begin{equation}
  \sigma^2(\beta\tilde{F}) \approx  \frac{\Delta\beta}{R} \int_{\beta_0}^\beta
  \frac{C_V(\beta') L^d}{\beta'^2}\,\d\beta'.
  \label{eq:varbetaf2}
\end{equation}
The inversely linear dependence of the variance of the free-energy estimate on $R$ is
expected, and a more general argument in support of this relation is discussed below
in Sec.~\ref{sec:effect-of-resampling}. In the presence of correlations between
population members, Eq.~\eqref{eq:varbetaf1.5} becomes instead
\begin{equation}
  \sigma^2(\beta\tilde{F}) \approx \Delta\beta\int_{\beta_0}^\beta
  \frac{\sigma^2\left[E(\beta')\right]}{R_\mathrm{eff}(\beta')}\,\d\beta'.
  \label{eq:varbetaf3}
\end{equation}
This relation shows the intimate relation of $\sigma^2(\beta \tilde{F})$ and the
effective population size $R_\mathrm{eff}$. Effectively differentiating relation
\eqref{eq:free-en-through-av-en}, we find for the free-energy contribution of one
step in the limit $\beta\to 0$,
\begin{equation}
  \beta\Delta \tilde{F} \approx \Delta\beta \overline{E}(\beta),
  \label{eq:var-delta-F}
\end{equation}
and hence
\begin{equation}
  \begin{split}
    \sigma^2(\beta\Delta \tilde{F}) & \approx (\Delta\beta)^2
    \sigma^2[\overline{E}(\beta)]\\
    &= (\Delta\beta)^2\frac{\sigma^2(E)}{R_\mathrm{eff}} = (\Delta\beta)^2\frac{C_V L^d}{\beta^2
      R_\mathrm{eff}}.
  \end{split}
  \label{eq:varF-tdlimit}
\end{equation}
To illustrate the regime of validity of the thermodynamic integration approximation,
we show in Fig.~\ref{fig:td-integration} the variance of the free-energy estimator
\eqref{eq:free-emergy-estimator} as estimated from the statistics over 200
independent runs in comparison to the approximations of Eqs.~\eqref{eq:varbetaf2} and
\eqref{eq:varbetaf3}, respectively. For $\Delta\beta = 0.01$ the approximations track
the independent estimate quite well until reaching the critical regime, where
significant deviations start to appear. It is clear, however, that the expression
\eqref{eq:varbetaf3} involving $R_\mathrm{eff}$ is a more accurate description than
the estimator \eqref{eq:varbetaf2}. As the inverse temperature step is decreased to
$\Delta\beta = 0.0025$ and finally to $\Delta\beta = 0.001$ the agreement with the
independent estimate of $\sigma^2(\beta\hat{F})$ improves significantly.

\subsection{Weighted averages}
\label{sec:weighted-averages}

For technical reasons, it is not always possible to consider in a single PA run as
big a population size as would be desirable. In this case one may resort to
performing several independent runs with smaller populations and then averaging the
results. Instead of using a plain arithmetic average, it was proposed by Machta in
Ref.~\cite{machta:10a} to employ weighted averages of the independent runs to reduce
bias and statistical errors of the final answers. The necessity of such weighting
follows immediately from the configurational weights $W_i^j$ discussed in
Sec.~\ref{sec:algo}. For the version of the algorithm where resampling according to
$\tau_i(E_j)$ is performed at each temperature step, at inverse temperature $\beta_i$
the replicas carry a weight $\widetilde{W}_i^j$ according to
Eq.~\eqref{eq:resampling-weights},
\begin{equation}
  \label{eq:resampling-weights2}
  \widetilde{W}_i^j = \frac{1}{Z_{\beta_i}} \prod _{k=1}^i Q_k =
  \frac{1}{Z_0 Z_{\beta_i}} \exp(-\beta_i \hat{F}_i).
\end{equation}
While these weights are the same for all replicas of the same run, and so they do not
enter any of the thermal averages for one run, they should be taken into account when
combining data from different PA simulations. If we perform $M$ independent PA runs
with initial population sizes $R^m$, we hence should take a weighted average of
observables according to
\begin{equation}
\overline{O}(\beta_i) = \sum_{m=1}^M \omega_i^m \overline{O}_m(\beta_i),
\end{equation}
with
\begin{equation}
  \omega_i^m = \frac{R_i^m \exp(-\beta_i \hat{F}_i^m)}{\sum_m R_i^m \exp(-\beta_i \hat{F}_i^m)}.
  \label{eq:weighted-averages}
\end{equation}
Note that $\widetilde{W}_i^j$ refers to single replicas, and so the prefactors
$R_i^m$ in Eq.~\eqref{eq:weighted-averages} make sure that each replica gets the same
weight in the average over several runs. It is worthwhile to point out that these
weights are temperature dependent, in particular they are different at each
temperature step of the simulation. As resampling proportional to $\tau_i(E_j)$ is
only really reasonable for the case of a constant population size, in practice one
has $R_i^m = R^m$ in Eq.~\eqref{eq:weighted-averages}.

On the other hand, for resampling procedures with a fluctuating population size such
as the Poisson and nearest-integer schemes the considerations of
Ref.~\cite{machta:10a} need to be generalized. In this case, population members are
replicated proportional to $\hat{\tau}_i(E_j) = (R/R_{i-1})\tau_i(E_j)$, cf.\
Eq. \eqref{eq:taus}. As a consequence, in this case the weights $\widetilde{W}_i^j$
become
\begin{equation}
  \widetilde{W}_i^j = 
  \frac{1}{Z_0Z_{\beta_i}} \prod_{k=1}^i \left(\frac{R_{k-1}}{R}\right) \exp(-\beta_i \hat{F}_i),
  \label{eq:modified-weights}
\end{equation}
such that in this more general situation the weights of
Eq.~\eqref{eq:weighted-averages} turn into
\begin{equation}
  \omega_i^m = \frac{R_i^m \prod_{k=1}^i(R_{k-1}^m/R^m)\exp(-\beta_i \hat{F}_i^m)}
  {\sum_m R_i^m \prod_{k=1}^i(R_{k-1}^m/R^m)\exp(-\beta_i \hat{F}_i^m)},
  \label{eq:modified-weights2}
\end{equation}
and hence the standard choice \eqref{eq:weighted-averages} is formally not correct
for fluctuating population size. In practice the difference between the weights
\eqref{eq:weighted-averages} and \eqref{eq:modified-weights2} is rather small,
however \footnote{To see this, consider the variance of the product
  $\prod_{k=1}^i (R_{k-1}^m/R^m)$. If the population sizes $R_{k-1}^m$ are
  uncorrelated, we can approximate
  \[
    \sigma^2\left(\prod_{k=1}^i \frac{R_{k-1}^m}{R^m}\right) \approx
    \frac{i\sigma^2(R_i^m)}{(R^m)^2} \lesssim \frac{N_\beta}{R^m},    
  \]
  where we used the fact that $\unexpanded{\langle R_i^m \rangle} \approx R^m$ and
  $\sigma^2(R_i^m) \lesssim R^m$.  Hence the effect of the additional factors
  depending on $R_{k-1}^m$ is small whenever $N_\beta \ll R$, which should normally
  be the case.}. Note that the $R$ related factors in
Eq.~\eqref{eq:modified-weights2} incorporate the effect of two types of variations in
population size: (1) independent PA runs with different target population sizes
$R^m$ ({\em extrinsic\/} fluctuations), and (2) the fluctuations of actual
population size $R_i^m$ in a given simulation at inverse temperature $\beta_i$
induced by using a resampling method such as the Poisson or nearest-integer schemes
({\em intrinsic\/} fluctuations).

Regarding the behavior of the weights $\omega_i^m$, one sees from the small
$\Delta\beta$ expression \eqref{eq:free-en-through-av-en} that $\beta\hat{F}_i^m$
should follow a normal distribution for small $\Delta\beta$ and large $R$. We expect
this to be the case also for $\Delta\beta$ that are not very small.  Disregarding the
effect of the much more slowly fluctuating denominator in
Eq.~\eqref{eq:weighted-averages}, it is then clear that $\omega_i^m$ will follow a
log-normal distribution. (While this is for the case of constant population size,
similar conclusions would be reached when considering
Eq.~\eqref{eq:modified-weights2} representing the more general situation of
fluctuating $R_i^m$.) If $\hat{F}_i^m \sim {\cal N}(\mu,\sigma^2)$, we see that
\begin{equation}
  \omega_i^m =  \frac{R^m\exp(-\beta \hat{F}_i^m)}{\sum_m R^m\exp(-\beta \hat{F}_i^m)} =
   \frac{R^m\exp(-\beta \underline{\hat{F}}_i^m)}{\sum_m R^m\exp(-\beta \underline{\hat{F}}_i^m)},
\end{equation}
where $\underline{\hat{F}}_i^m \sim {\cal N}(0,\sigma^2)$. Checking the properties of
the log-normal distribution, we see that the mean of
$\exp(-\beta\underline{\hat{F}}_i^m)$ is
$\exp[\sigma^2(\beta\underline{\hat{F}}_i^m)/2]$ and the most likely value (mode) is
at $\exp[-\sigma^2(\beta\underline{\hat{F}}_i^m)]$. If
$\sigma^2(\beta\underline{\hat{F}}_i^m) \equiv \sigma^2(\beta\hat{F}_i^m)$ is at
least of order $1$, average and typical value are substantially different and hence
the weighted average will be dominated by the tails of the distribution. Numerical
estimates will then be unstable. Interestingly, there is no further scale in this
relation and it is indeed the comparison of $\sigma^2(\beta\hat{F}_i^m)$ and unity
that distinguishes the two limiting cases. Also note that it is the variance of the
total free energy and not the free energy per site that matters here, so there is an
important size dependence. For $\sigma^2(\beta\hat{F}_i^m) > 1$ weighted averages
will be poor, but it does {\em not\/} mean that bias and/or statistical error for any
other observable of a single run must be bad. A clear-cut case would be the Ising
model simulated with PA using cluster updates. In that case the dynamics are rapidly
mixing everywhere \cite{guo:17}, in particular also in the ordered phase where
single-spin flips are not able to connect configurations in the two pure phases in
polynomial time. Hence for a Swendsen-Wang update in the ordered phase there are no
biases if we simulate for long enough ($\theta$ sufficiently large), however,
depending on the other parameters it could well be that
$\sigma^2(\beta\hat{F}_i^m) > 1$. This clearly shows that the value of
$\sigma^2(\beta\hat{F}_i^m)$ is not suitable as a sole general measure of
equilibration. Numerically, we find that the range of situations where weighted
averaging is beneficial is rather limited as when $\sigma^2(\beta\hat{F}_i^m) \ll 1$
the weights are very nearly equal to each other, such that the weighted average
reduces to a plain average, whereas for $\sigma^2(\beta\hat{F}_i^m) > 1$ the
weighting scheme breaks down for the reasons outlined above.

It is also useful to revisit the estimates of $\sigma^2(\beta\tilde{F})$ found in the
previous Section. In the limit $\theta \to \infty$ where the population is perfectly
in equilibrium and perfectly uncorrelated at each step, Eq.~\eqref{eq:varbetaf2}
implies that there is still some variance of $\beta\tilde{F}$ which could well be
larger than one if the specific heat is large enough, although the population is
perfectly in equilibrium. Hence there is an intrinsic component of the variance of
the free energy that is independent of any correlations in the population, but which
might lead to biased estimates.

\section{Bias}
\label{sec:bias}

Bias in PA results from two sources, the finite population size affecting the
resampling step and the usual equilibration bias present in the MCMC subroutine. The
former is related to the reweighting bias well known from reweighting techniques in
MCMC \cite{ferrenberg}: on using the distribution at inverse temperature $\beta$ for
estimating that at $\beta' > \beta$, events in the relatively badly sampled wing of
the current distribution are amplified, whereas those in the peak are suppressed,
leading to bias from bad statistics in this wing, especially if
$\Delta \beta = \beta'-\beta$ is chosen (too) large. There is a second bias effect
connected to the resampling which is through the introduction of correlations in the
population effected by the resampling step, thus also deteriorating the quality of
the representation of the energy distribution $p_\beta(E)$ by the population of
replicas through a reduction of the effective population size (see the discussion in
Sec.~\ref{sec:correlations}).
 
\subsection{Behavior without resampling}

\begin{figure}[tb]
  \centering
  \includegraphics[clip=true,keepaspectratio=true,width=0.95\columnwidth]{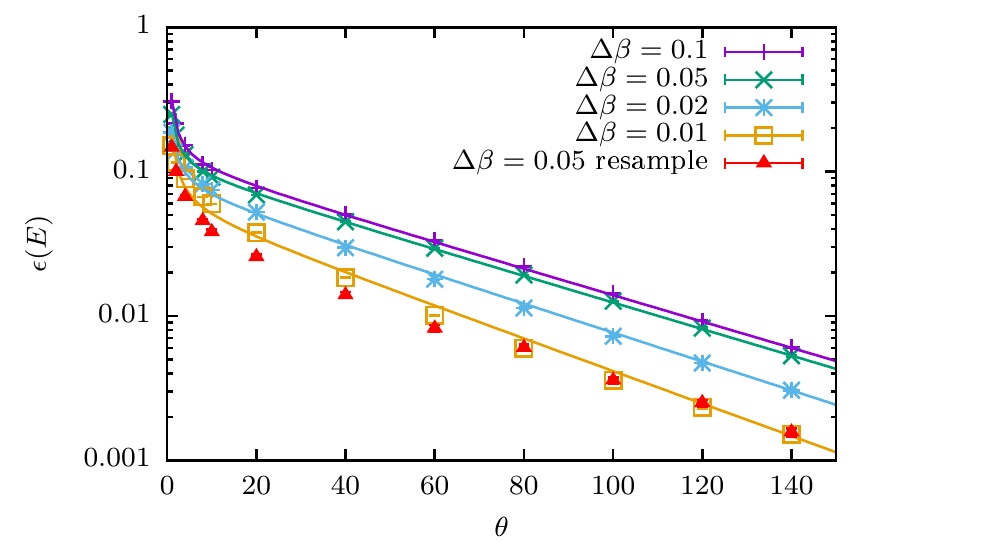}
  \includegraphics[clip=true,keepaspectratio=true,width=0.95\columnwidth]{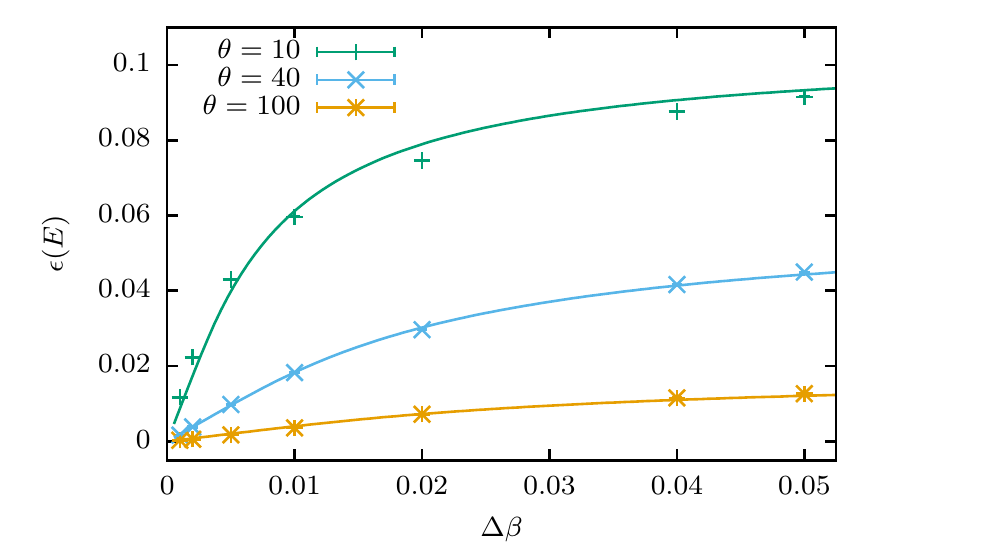}
  \caption
  {%
    Relative deviation $\epsilon(E) = (E-E_\mathrm{exact})/E_\mathrm{exact}$ of the
    internal energy at the critical inverse temperature
    $\beta_c = \frac{1}{2}\ln(1+\sqrt{2})$ for an $L=32$ system sampled in PA runs
    {\em without resampling\/} on a population of size $R=10\,000$. The results are
    averaged over $m=200$ independent runs to reduce statistical errors. Top:
    relative deviation as a function of $\theta$ together with fits of the functional
    form $\epsilon(E) = a e^{-\theta/\tau_\mathrm{rel}}(1+b/\theta)$ motivated by
    \eqref{eq:bias-simplified} to the data. Bottom: relative deviation as a function
    of $\Delta\beta$ with fits of the functional form
    $\epsilon(E) = a \Delta\beta \left(1-e^{-b/\Delta\beta}\right)$.
    \label{fig:theta_beta_scaling_noresampling}
  }%
\end{figure}
 
We first consider the case of the PA algorithm without resampling, where the only
source of bias is the relaxation process as the population is cooled in
steps. Assume that the population is in equilibrium at inverse temperature
$\beta$. If a temperature step $\Delta\beta$ is taken, the system needs to relax
towards the new equilibrium energy at $\beta+\Delta\beta$. Assuming a purely
exponential relaxation process \footnote{This is, in general, a simplification, but
  the scaling results derived below are expected to carry over to the case of a more
  general spectrum of superimposed exponential decays.}, the energy will decay as
\begin{equation}
E(t) = \langle E \rangle_{\beta+\Delta\beta} - [\langle E\rangle_{\beta+\Delta\beta} -
  \langle E\rangle_\beta]e^{-t/\tau_\mathrm{rel}},
\end{equation}
where $\tau_\mathrm{rel} = \tau_\mathrm{rel}^E(\beta+\Delta\beta)$ is the exponential
relaxation time of the internal energy at inverse temperature $\beta+\Delta\beta$
\cite{sokal:97}. For sufficiently small temperature steps, a first-order Taylor
expansion of the energy as a function of (inverse) temperature implies that
\cite{janke:08}
\begin{equation}
\langle E\rangle_{\beta+\Delta\beta} -
  \langle E\rangle_\beta \approx \frac{\partial \langle E\rangle}
    {\partial\beta}\Delta\beta =
    - \beta^{-2} L^d C_V \Delta\beta,
\end{equation}
and hence we find that the remaining bias $\Delta E$ after $\theta$ sweeps of
spin flips is
\begin{equation}
  \Delta E = E(\theta) -  \langle E \rangle_{\beta+\Delta\beta} \approx
  \beta^{-2} L^dC_V \Delta \beta e^{-\theta/\tau_\mathrm{rel}}.
  \label{eq:single-step}
\end{equation}
To simplify notation, in the following we use
$E' = \partial E/\partial \beta = - \beta^{-2} L^dC_V$.  For an annealing sweep starting
in equilibrium from temperature $\beta_0$ (for instance for $\beta_0 = 0$) that
arrives at temperature $\beta$, there are remaining biases from all previous
temperature steps,
\begin{equation}
\Delta E_i = -E'_i \Delta\beta e^{-\theta/\tau_{\mathrm{rel},i}} + \Delta E_{i-1} e^{-\theta/\tau_{\mathrm{rel},i}},
\end{equation}
where $E'_i=E'(\beta_i)$ refers to the slope of the energy curve at step $i$, and
$\tau_{\mathrm{rel},i} =\tau_\mathrm{rel}^E(\beta_i)$. Iterating one finds with
$\Delta E_0 = 0$ that
\begin{equation}
  \Delta E_i \approx -\sum_{j=0}^{i-1} E'_{i-j}\Delta \beta\,
  \exp\left[-\theta \sum_{k=0}^j 1/\tau_{\mathrm{rel},i-k}\right].
  \label{eq:bias-series}
\end{equation}
While this expression yields the expected bias at inverse temperature $\beta_i$, in
order to study the dependence on the inverse temperature step $\Delta\beta$, we need
to express the bias directly as a function of $\beta$,
\begin{equation}
  \Delta E(\beta) \approx -\sum_{j=1}^{n(\beta)} E'(\beta_0+j\Delta\beta)\Delta \beta\,
  \exp\left[-\theta \sum_{k=j}^{n(\beta)} 1/\tau_{\mathrm{rel},k}\right],
  \label{eq:bias-series2}
\end{equation}
where $n(\beta) = (\beta-\beta_0)/\Delta\beta$ is the number of temperature steps up
to inverse temperature $\beta$. Often one will use small inverse temperature steps
$\Delta \beta$ such that we can approximate the sums by integrals to find
\begin{equation}
\Delta E(\beta) \approx -\int_{\beta_0}^\beta E'(\tilde{\beta})\,
\exp\left[-\frac{1}{\kappa}\int_{\tilde{\beta}}^\beta
  \frac{{\mathrm d}\hat{\beta}}{\tau_\mathrm{rel}(\hat{\beta})}
  \right] {\mathrm d}\tilde{\beta},
\end{equation}
where $\kappa = \Delta\beta/\theta$ is the cooling rate. Inspecting
Eq.~\eqref{eq:bias-series2}, we immediately see that to leading order
$\Delta E_i \sim \Delta\beta$, but note that a change of inverse temperature step
changes the whole sequence of the $\beta_i$ and hence has additional effects on
$\Delta E_i$. Considering the dependence on $\theta$, we note that $\Delta E_i$
depends on a sequence of exponentials for all higher temperatures, that decay with
the harmonic mean of the corresponding relaxation times.

We cannot proceed any further in evaluating the bias of \eqref{eq:bias-series2}
without further simplifying assumptions. In the extreme case where all
$\tau_{\mathrm{rel},i} = \tau_\mathrm{rel}$ are equal and $E'$ is independent of
$\beta$, we have from Eq.~\eqref{eq:bias-series2}
\begin{equation}
  \begin{split}
    \Delta E(\beta) &\approx -E' \Delta\beta \sum_{j=1}^{n(\beta)} 
    \exp\left[-\frac{\theta}{\Delta\beta \tau_\mathrm{rel}}(\beta-[j-1]\Delta\beta)\right] \\
    & =  -E'\Delta\beta \frac{e^{-\theta/\tau_\mathrm{rel}}}{1-e^{-\theta/\tau_\mathrm{rel}}}
    \left[ 1-\exp\left(-\frac{\theta\beta}{\tau_\mathrm{rel}\Delta\beta}\right)\right] 
  \end{split}
  \label{eq:bias-simplified}
\end{equation}
where we have set $\beta_0 = 0$ for simplicity. We note that while these assumptions
are not accurate in general for the 2D Ising model studied here, they will be a good
approximation for small $\Delta\beta$ when the rates of change of $\tau_\mathrm{rel}$
and $E'$ are small. For small inverse temperature steps $\Delta\beta$ the term in
square brackets will be negligible. In this case, we expect the $\theta$ dependence
to be purely exponential $\Delta E = -E'\Delta\beta e^{-\theta/\tau_\mathrm{rel}}$
for $\theta/\tau_\mathrm{rel} \gg 1$, with a crossover to the inverse linear behavior
$\Delta E \approx -E'\Delta\beta\tau_\mathrm{rel}/\theta$ for
$\theta/\tau_\mathrm{rel} \ll 1$.  Note also that for this simple scenario the form
\eqref{eq:bias-simplified} ensures that $\Delta E(\beta_0 = 0) = 0$ (which will
always be the case by assumption) and bias increases away from $\beta_0$ in an
exponential fashion to its temperature independent limiting form.  Regarding the
dependence on temperature step, we see that for small $\Delta\beta$ this is linear,
with an exponential crossover to the constant
$\Delta E \approx -E' \beta(1-\theta/2\tau_\mathrm{rel})$ expected in the limit of
large steps $\Delta\beta$. In Fig.~\ref{fig:theta_beta_scaling_noresampling} we show
the relative deviation of internal energies at the critical coupling $\beta_c$ from
the exact result, $\epsilon(E) = (E-E_\mathrm{exact})/E_\mathrm{exact}$, calculated
from PA runs for $L=32$ with resampling turned off. The data have been averaged over
$m=200$ runs to reduce statistical errors, such that $\epsilon(E)$ is indeed
representative of the systematic error. As is seen from the fits shown together with
the data, which are of the form derived above but with independent amplitudes to take
account of the approximations involved, the simplified model fits the simulation data
very well. The effective relaxation time $\tau_\mathrm{rel} = 45.4(6)$ (extracted
from the fit for $\Delta\beta = 0.1$) is comparable, but somewhat smaller than the
integrated critical autocorrelation time of $\tau_\mathrm{int} = 72 \pm 14$ extracted
from a blocking analysis. Note that in general the bias is {\em not\/} a function of
the cooling rate $\kappa$ alone as might be naively assumed, although as the analysis
of the simplified form \eqref{eq:bias-simplified} above shows, this is the dependence
in certain limiting cases.

We note that while these calculations are for the energy bias $\Delta E$, similar
results hold for biases in other quantities with the energy derivative
$\partial \langle E\rangle/\partial \beta = -\beta^{-2} L^d C_V$ replaced by the
corresponding derivative of the observable considered and with using the
corresponding relaxation times.

\subsection{Effect of resampling}
\label{sec:effect-of-resampling}

\begin{figure}[tb]
  \centering
  \includegraphics[clip=true,keepaspectratio=true,width=0.95\columnwidth]{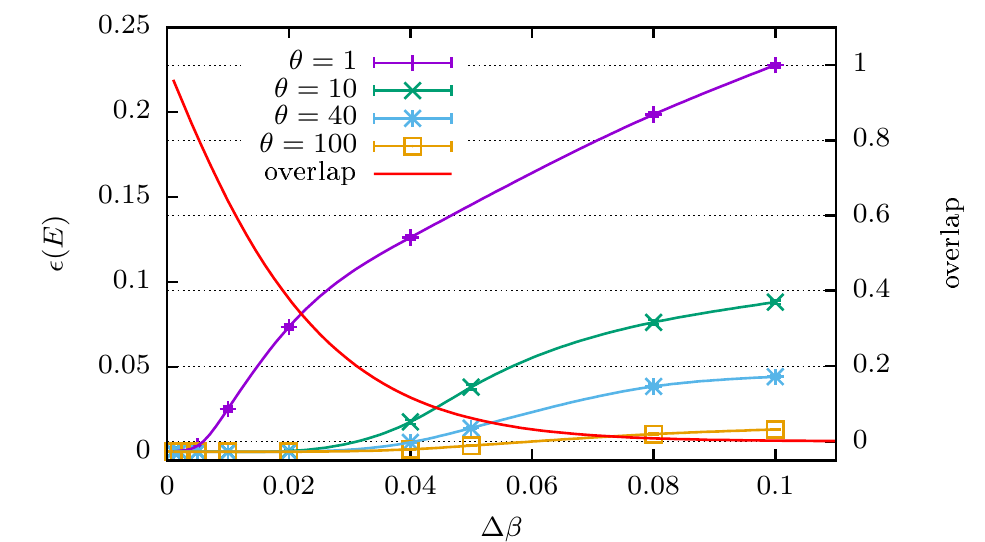}
  \includegraphics[clip=true,keepaspectratio=true,width=0.95\columnwidth]{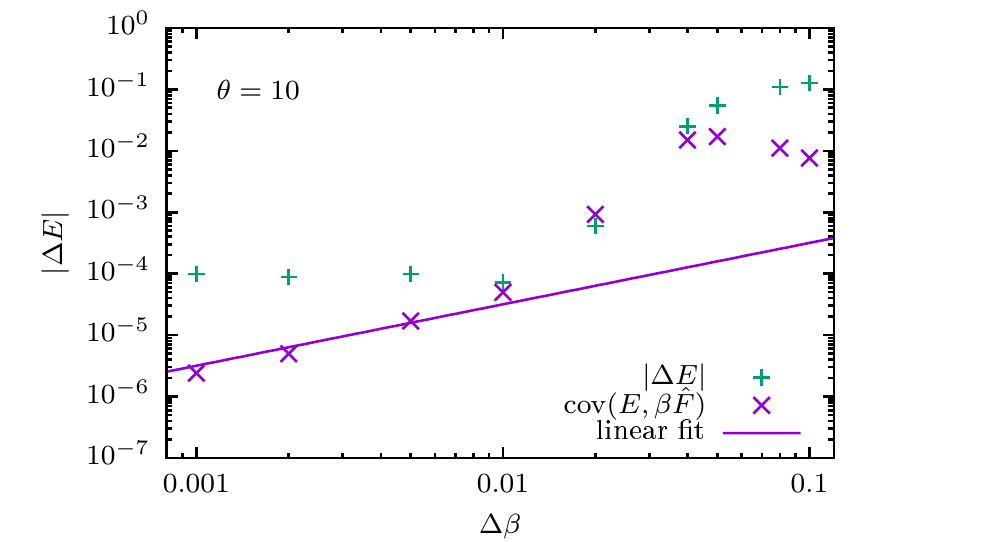}
  \caption
  {%
    Top: relative deviation $\epsilon(E) = (E-E_\mathrm{exact})/E_\mathrm{exact}$ of
    the internal energy at the critical inverse temperature $\beta_c$ for an $L=32$
    system sampled in PA runs with resampling on a population of size $R=10\,000$ as
    a function of inverse temperature step $\Delta\beta$. The interpolating lines are
    merely guides to the eye. The red line shows the exact histogram overlap $\alpha$
    at criticality as a function of step size (right scale). Bottom: bias
    $|\Delta E| = |E-E_\mathrm{exact}|$ for the run with $\theta=10$ as compared to
    the covariance $\operatorname{cov}(E,\beta\hat{F})$ of
    Eq.~\eqref{eq:bias_covariance} together with a linear fit for small values of
    $\Delta\beta$.
    \label{fig:theta_beta_scaling}
  }%
\end{figure}

We now turn to the situation of PA with the resampling step enabled. We find that
resampling leads to a reduction in bias that is almost independent of the number
$\theta$ of equilibration sweeps, such that in this respect it is similar to choosing
a reduced inverse temperature step. This is illustrated for $\Delta\beta = 0.05$ in
the additional data set in the upper panel of
Fig.~\ref{fig:theta_beta_scaling_noresampling}. The resampling procedure introduces
an additional dependence on the step size $\Delta\beta$ due to histogram overlap as
outlined above. To understand this effect, we extend the analysis proposed by Wang
{\em et al.} \cite{wang:15a}. It was shown there that in the limit of large
population sizes the bias
$\Delta {\cal O} = \langle \hat{\cal O} \rangle - \langle {\cal O} \rangle$ of an
observable ${\cal O}$, i.e., the difference of the expected value of the estimator
$\hat{\cal O}$ from PA runs with a given set of parameters and the thermal
expectation value of ${\cal O}$, is given by its covariance with the free-energy
estimate,
\begin{equation}
  \label{eq:bias_covariance}
  \Delta{\cal O} = \operatorname{cov}(\hat{\cal O}, \beta\hat{F}).
\end{equation}
In Ref.~\cite{wang:15a} it is argued that the size of this bias is essentially
determined by the variance $\sigma^2(\beta\hat{F})$, namely if one decomposes
\begin{equation}
  \Delta\mathcal{O} = \operatorname{cov}(\hat{\cal O}, \beta\hat{F}) =
  \sigma^2(\beta\hat{F}) \left[
    \frac{\operatorname{cov}(\hat{\cal O}, \beta\hat{F})}{\sigma^2(\beta\hat{F})}\right],
  \label{eq:bias_variance}
\end{equation}
the quantity in square brackets is claimed to be asymptotically independent of
$R$. If and when the estimator $\beta\hat{F}$ of Eq.~\eqref{eq:free-emergy-estimator}
[together with Eq.~\eqref{eq:Q}] is a sum of many uncorrelated contributions of
finite variance stemming from effectively uncorrelated sub-populations, the central
limit theorem implies that its variance $\sigma^2(\beta\hat{F}) \propto 1/R$. In these
cases, we can expect the bias in $R$ to decay as $1/R$. We will discuss the numerical
findings regarding this behavior below. For completeness, let us mention that it is
useful to define the quantity \cite{wang:15a}
\begin{equation}
  \label{eq:rho_f}
  \rho_f = R\,\sigma^2(\beta \hat{F}),
\end{equation}
which will attain a finite value in the limit $R\to\infty$ if the above scaling
holds. As was discussed above in Sec.~\ref{sec:weighted-averages}, weighted averages
are dominated by outliers for $\sigma^2(\beta\hat{F}) \gtrsim 1$, and it is hence
reasonable to demand that $R \gg \rho_f$ for reliable results, justifying the name
{\em equilibration population size\/} for $\rho_f$ \cite{wang:15a}.

Before turning to our numerical results for the $R$ dependence of bias, we study the
dependence of $\operatorname{cov}(\hat{\cal O}, \beta\hat{F})$ and
$\sigma^2(\beta\hat{F})$ on the inverse temperature step $\Delta\beta$. To
investigate $\sigma^2(\beta\hat{F})$, we note that the estimator
\eqref{eq:free-emergy-estimator} is a sum of $i+1$ terms. Neglecting correlations
between these terms, the variance of the sum will be the sum of the variances
\cite{feller:68}. While the constant $\ln Z_{\beta_0}$ does not contribute to the
variance, each of the other terms
$\delta \beta_k \hat{F}_k \equiv \ln Q(\beta_{k-1},\beta_k)$ yields
\begin{equation}
  \begin{split}
    \sigma^2(\delta\beta_k\hat{F}_k) &= \sigma^2[\ln Q(\beta_{k-1},\beta_k)] \\
    &\approx \sigma^2\left[\ln \sum_{j=1}^{R_{k-1}} e^{-\Delta\beta
        E_j}\right],\\
  \end{split}
\end{equation}
where we have used the fact that the prefactor $R_{k-1}$ in Eq.~\eqref{eq:Q} has only
tiny fluctuations or the algorithm could also be formulated for fixed
$R_{k-1} = R$. Error propagation implies that to leading order and neglecting
correlations between the random variables $E_j$, $j=1,\ldots,R_{k-1}$ we have
\begin{equation}
\sigma^2(\delta\beta_k\hat{F}_k) \approx (\Delta\beta)^2 \sum_j
\frac{e^{-2\Delta\beta E_j}}{\left[\sum_l e^{-\Delta\beta E_l}\right]^2} \sigma^2(E_j).
\end{equation}
Since the number of temperature steps up to a given, fixed inverse temperature
$\beta$ is inversely proportional to $\Delta\beta$, we find the leading behavior
\begin{equation}
  \sigma^2(\beta\hat{F}) \approx \sum_k \sigma^2(\delta\beta_k\hat{F}_k)
  \propto \Delta\beta.
\end{equation}
While in reality there will be correlations between the estimates of the different
free-energy differences as well as between the energies in the population, we do not
expect these to alter the leading scaling behavior. For the covariance
$\operatorname{cov}(\hat{\cal O}, \beta\hat{F})$ an analogous argument shows that
in one step
$\operatorname{cov}(\hat{\cal O}, \delta \beta_k\hat{F}_k) \propto \Delta\beta$.
However, in contrast to
$\sigma^2(\beta\hat{F}) = \operatorname{cov}(\beta\hat{F}, \beta\hat{F})$ where
${\cal O} = \beta\hat{F}$ depends on all previous temperature steps, for a
``regular'' observable ${\cal O}$ such as the energy or magnetization that is a
function only of the population at inverse temperature $\beta$, there are no
contributions of previous temperature steps to the covariance, and hence the
dependence of the total number of temperature steps on $\Delta\beta$ is not relevant
for the scaling of $\operatorname{cov}(\hat{\cal O}, \delta \beta_k\hat{F}_k)$ with
$\Delta\beta$ such that we have
\begin{equation}
  \operatorname{cov}(\hat{\cal O}, \beta\hat{F}) \propto
  \Delta\beta.
  \label{eq:deltabeta-scaling}
\end{equation}

\begin{figure}[tb]
  \centering
  \includegraphics[clip=true,keepaspectratio=true,width=0.95\columnwidth]{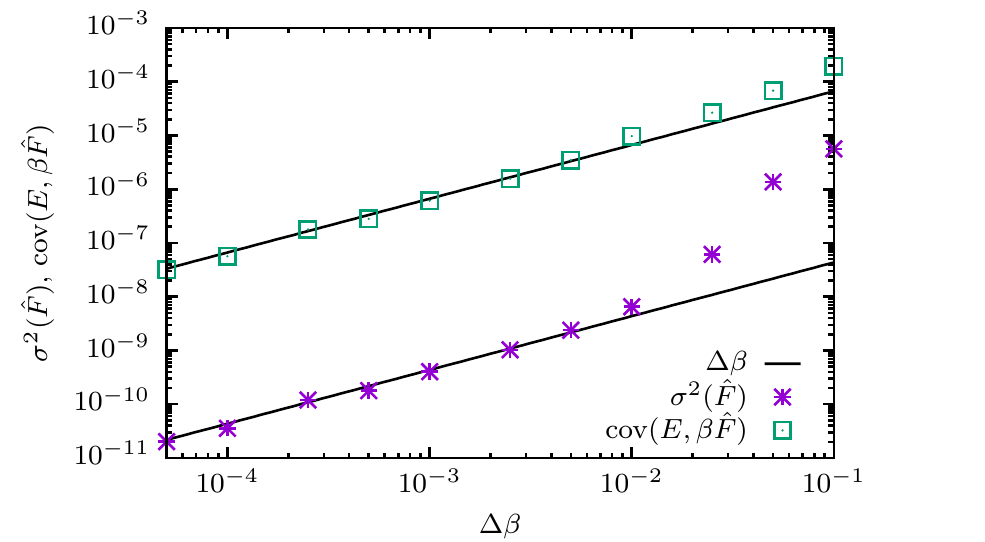}
  \caption
  {%
    Variance $\sigma^2(\beta\hat{F})$ and covariance
    $\operatorname{cov}(\beta\hat{F},E)$ averaged over the temperature range
    $0 \le \beta \le 1$ for $L=32$ as a function of the inverse temperature step
    $\Delta\beta$. Both scale proportional to $\Delta\beta$ in the limit of small
    steps as the linear fits illustrate.
    \label{fig:bias_deltabeta}
  }%
\end{figure}

To check these predictions of limiting behaviors we performed simulations for a wide
range of step sizes $0.5\times 10^{-5} \le \Delta\beta \le 10^{-1}$ and MCMC steps
$1\le \theta\le 200$. The results are summarized in the upper panel of
Fig.~\ref{fig:theta_beta_scaling}.  For larger values of $\Delta\beta$ we find a
moderate reduction of bias as compared to the algorithm without resampling (lower
panel of Fig.~\ref{fig:theta_beta_scaling_noresampling}), but substantially increased
fluctuations. (Note that the topmost data set in the lower panel of
Fig.~\ref{fig:theta_beta_scaling_noresampling} is for $\theta=10$ while that in the
top panel of Fig.~\ref{fig:theta_beta_scaling} is for $\theta=1$.) Such increased
fluctuations occur due to the loss of diversity in the population induced by the
resampling. For $\Delta\beta \ge 0.05$ we have an overlap of energy histograms at
inverse temperatures $\beta_c$ and $\beta_c+\Delta\beta$ of less than 10\% (right
scale of the top panel of Fig.~\ref{fig:theta_beta_scaling}), such that the amount of
statistically independent information in the population is reduced by more than a
factor of ten in each step --- an effect that is only partially made up by the
intermediate equilibration sweeps. Only for $\Delta\beta \lesssim 0.02$, is the
histogram overlap large enough to counterbalance this effect and lead to a
significantly reduced bias without an accompanying increase in statistical
fluctuation (cf.\ also the discussion of the balance of these effects in Appendix
\ref{app:effective-population-size}). We note that the histogram overlap decays
exponentially away from $\Delta\beta = 0$.  It is minimal around the critical point
and, as one reads off from Fig.~\ref{fig:theta_beta_scaling}, for the $L=32$ system
it is about $0.3$ for $\Delta\beta = 0.02$ at $\beta_c$, which could therefore be
considered a reasonable maximal inverse temperature step for this case.

To investigate the functional dependence of $\Delta E$ on $\Delta\beta$, consider the
lower panel of Fig.~\ref{fig:theta_beta_scaling}, where $\Delta E$ from the upper
panel is shown for the case of $\theta = 10$, but now in a doubly logarithmic plot,
displayed together with the expected bias $\operatorname{cov}(E, \beta\hat{F})$
according to Eq.~\eqref{eq:bias_covariance}. One can distinguish three regimes: for
$\Delta\beta \gtrsim 0.04$ the actual bias clearly exceeds
$\operatorname{cov}(E, \beta\hat{F})$, indicating that the assumptions made in the
derivation leading to the form \eqref{eq:bias_covariance} (in particular the Gaussian
nature of fluctuations) are not fulfilled there. For
$0.01\lesssim \Delta\beta \lesssim 0.04$, the measured bias agrees with the
prediction from the covariance. Finally, for $\Delta\beta \lesssim 0.01$ the bias in
the actual simulation drops below the noise level and hence its further reduction
cannot be observed. In this latter regime, the predicted bias
$\operatorname{cov}(E, \beta\hat{F})$ follows the linear decay
$\propto \Delta\beta$ expected from Eq.~\eqref{eq:deltabeta-scaling}. That the
variance $\sigma^2(\beta\hat{F})$ as well as the covariance
$\operatorname{cov}(E, \beta\hat{F})$ indeed decay proportional to $\Delta\beta$
for sufficiently small steps is more cleanly demonstrated by the data for the
temperature-averaged bias presented in Fig.~\ref{fig:bias_deltabeta} (the dependence
on $\Delta\beta$ is found to be uniform in $\beta$), showing the linear decay to hold
over several orders of magnitude for sufficiently small $\Delta\beta$ for both
quantities. It again turns out to be crucial to ensure sufficient histogram overlap
to observe this behavior, which is achieved for $\Delta\beta \lesssim 0.02$ for this
system size.

\subsection{Dependence on population size}

\begin{figure}[tb]
  \centering
  \includegraphics[clip=true,keepaspectratio=true,width=0.95\columnwidth]{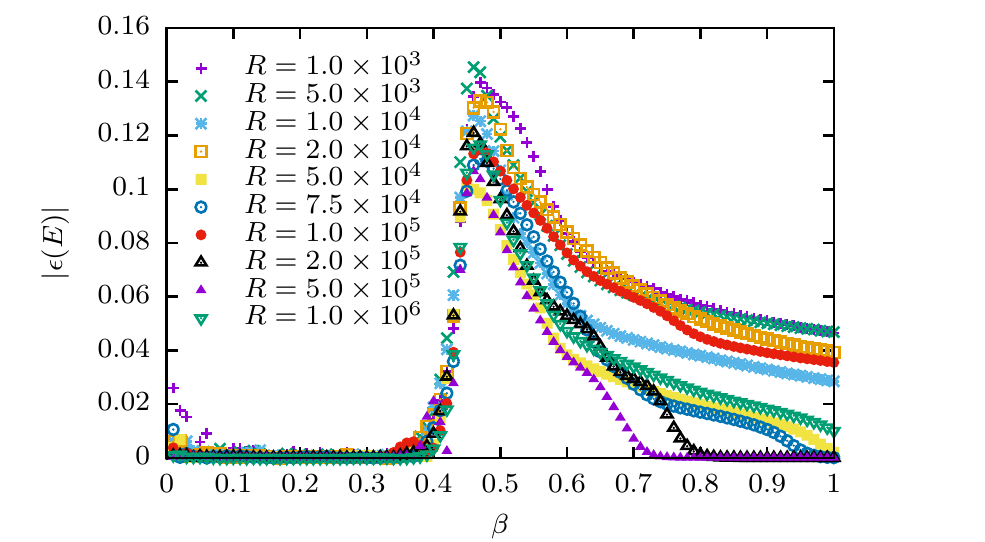}
  \includegraphics[clip=true,keepaspectratio=true,width=0.95\columnwidth]{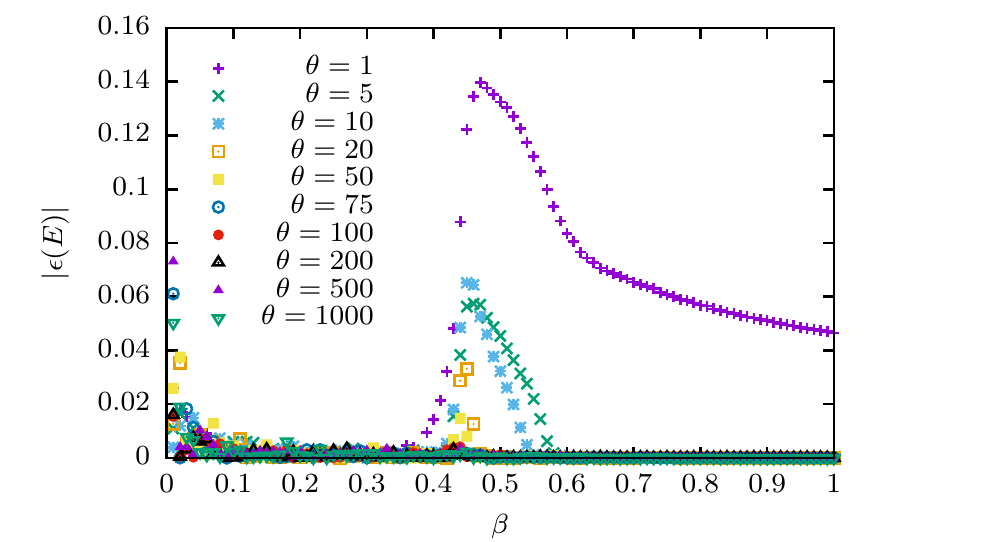}
  \includegraphics[clip=true,keepaspectratio=true,width=0.95\columnwidth]{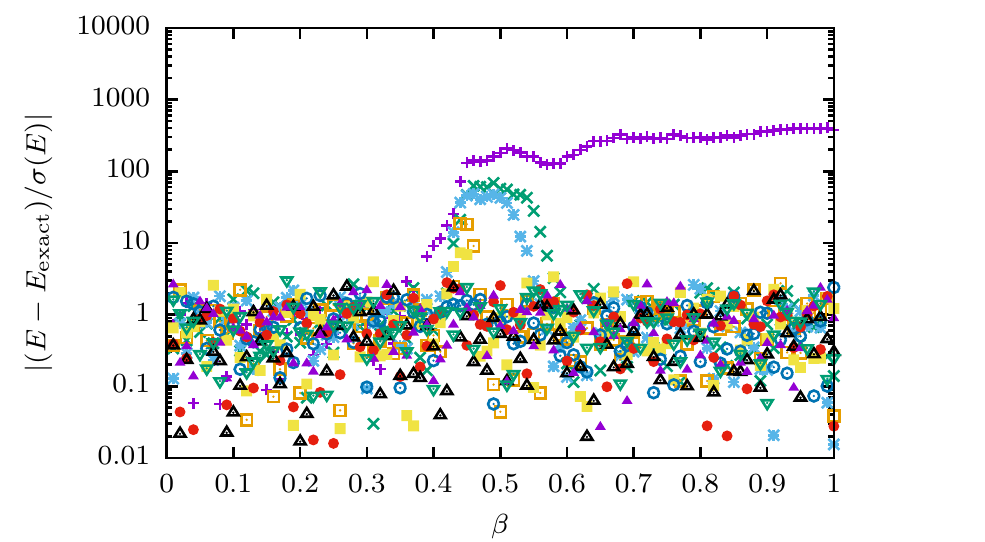}
  \caption
  {Top: Relative deviation $|\epsilon(E)| = |(E-E_\mathrm{exact})/E_\mathrm{exact}|$
    of the internal energy from PA simulations for $L=64$, $\Delta\beta = 0.01$, and
    $\theta=1$ (heatbath update) and population sizes ranging from $R=10^3$ to
    $R=10^6$. Middle: The same bias for a population $R=1000$ for different choices of
    $1 \le \theta \le 1000$. Bottom: Energy deviation relative to the statistical error
    in a logarithmic representation, illustrating that for $\theta \gtrsim 50$, the
    bias drops below the statistical fluctuations [symbols of data sets as in the
    middle panel]. Corresponding data sets in all panels amount to the same total
    computational effort.
    \label{fig:population2}
  }
\end{figure}

It remains to discuss the dependence of systematic errors on the population size. The
analysis in Ref.~\cite{machta:11} for a double-well model in the absence of any
autocorrelations as well as the arguments from Ref.~\cite{wang:15a} discussed in the
previous subsection suggesting that $\sigma^2(\hat{F}) \propto 1/R$ for large $R$
would indicate that bias decays inversely in $R$.  To scrutinize the behavior for the
PA simulations of the Ising model considered here, in the top panel of
Fig.~\ref{fig:population2} we show the relative deviation in the internal energy,
$\epsilon(E) = (E-E_\mathrm{exact})/E_\mathrm{exact}$, as a function of $R$. In the
critical region it decays much more slowly than $1/R$ and one sees hardly any
reduction in bias although $R$ is varied over three orders of magnitude. In contrast,
the middle panel shows the bias as a function of $\theta$, where corresponding data
sets in the two panels belong to calculations with the same computational effort (and
the scales on the axes are the same). Here, the decay is fast and consistent with an
asymptotically exponential drop as expected. As is illustrated in the bottom panel of
Fig.~\ref{fig:population2} showing the deviation relative to the statistical error,
for $\theta \gtrsim 50$ the bias drops below the level of the noise.

It is also instructive to examine the expression for the bias in the ``thermodynamic
integration limit'' $\Delta\beta\to 0$ discussed above. To the extent that the MC is
efficient and hence the populations at successive temperature points are not very
strongly correlated, one could replace the free-energy estimate $\beta\hat{F}$ in
Eq.~\eqref{eq:bias_covariance} by the last increment $\beta\Delta\hat{F}$,
\begin{equation}
  \Delta {\cal O} \approx \operatorname{cov}(\hat{\cal O},\beta\Delta\hat{F}).
\end{equation}
In the limit $\Delta\beta \ll 1$ one then finds from Eqs.~\eqref{eq:var-delta-F} and
recalling Eq.~\eqref{eq:variance-of-the-mean-correlated}
\begin{equation}
  \begin{split}
    \Delta {\cal O} &\approx \Delta \beta
    \operatorname{cov}[\hat{\cal O},\overline{E}(\beta)] = \Delta\beta \sigma^2(\overline{E})
    \frac{\operatorname{cov}(\hat{\cal O},\overline{E})}{\sigma^2(\overline{E})} \\
    & \approx \Delta\beta \frac{\sigma^2(E)}{R_\mathrm{eff}}
    \left[\frac{\operatorname{cov}(\hat{\cal
          O},\overline{E})}{\sigma^2(\overline{E})}\right].
  \end{split}
\end{equation}
Assuming that the term in square brackets has only a weak dependence on $R$
(analogous to the argument used above in Sec.~\ref{sec:effect-of-resampling}), one
would conclude that
\begin{equation}
  \Delta {\cal O} \propto \frac{1}{R_\mathrm{eff}}.
\end{equation}
The significance of this observation for the performance of the algorithm is
discussed in the following section.

\subsection{Pure resampling and effective population size}
\label{sec:bias2}

Above, we have considered the PA algorithm and its bias in the absence of
resampling. It is also possible and instructive to analyze the method in the opposite
limit of a pure resampling method, i.e., for $\theta \to 0$. In this case, the size
of the temperature step does not matter as one works only with the configurations of
the initial population. Since the resampling factors multiply over different
temperature steps,
\begin{equation}
  \begin{split}
    \exp(-\beta_k E_j) = & \exp[-(\beta_k-\beta_0) E_j] \\
    & = \prod_{i=1}^{k}
  \exp[-(\beta_i-\beta_{i-1})E_j],
  \end{split}
\end{equation}
the statistical weight of a configuration of the initial population ($\beta_0 = 0$)
at a given lower temperature $\beta_k$ is independent of the number (and spacing) of
temperature steps taken in between. Due to the normalization of resampling factors,
which suppresses fluctuations, the above identity is only approximately realized in
the actual PA method, but numerically we find that (for reasonable values of $R$) the
results for $\theta=0$ are almost perfectly independent of $\Delta\beta$.

\begin{figure}[tb]
  \centering
  \includegraphics[clip=true,keepaspectratio=true,width=0.95\columnwidth]{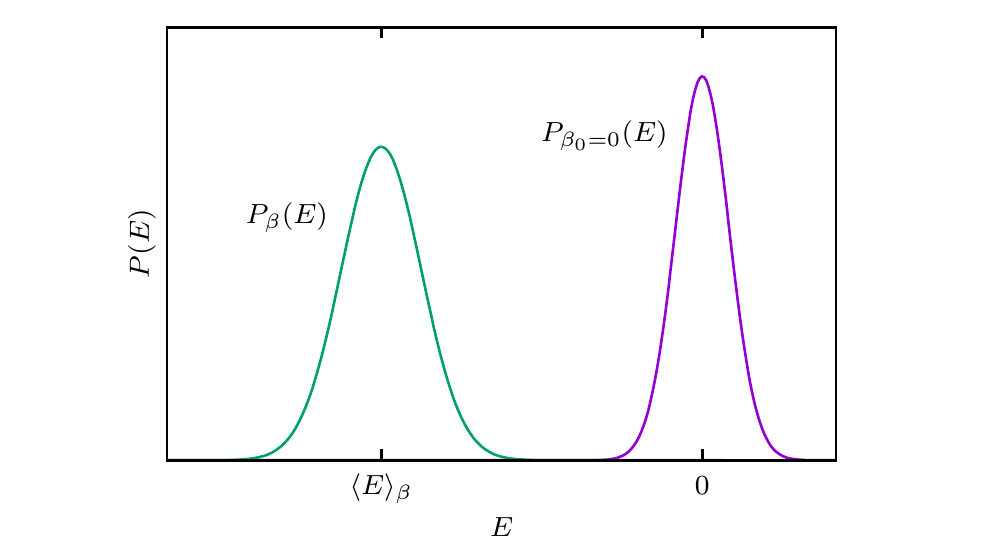}
  \includegraphics[clip=true,keepaspectratio=true,width=0.95\columnwidth]{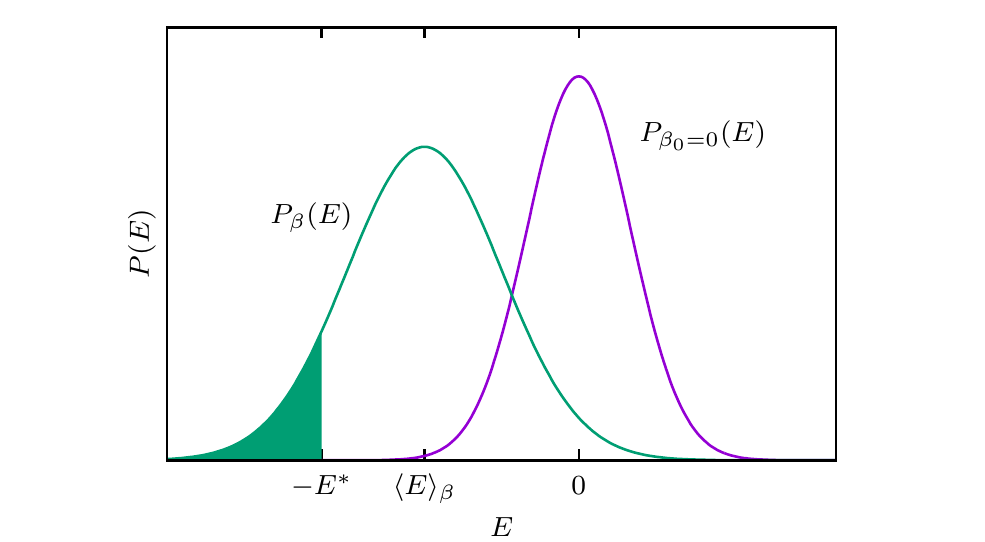}
  \caption
  { Schematic of energy distributions in $\theta=0$ population annealing. Top:
    $P_{\beta_0 = 0}$ corresponds to the initial population, $P_\beta$ to the
    distribution at a lower temperature. Bottom: in a finite population of size
    $R$ at $\beta_0 = 0$, there are typically no replicas with energies lower than
    $-E^\ast$ (and similarly none with energies higher than $E^\ast$), where
    $P_{\beta_0}(E^\ast) = 1/R$.
    \label{fig:distribution-sketch}
  }
\end{figure}

\begin{figure}[tb]
  \centering
  \includegraphics[clip=true,keepaspectratio=true,scale=0.95]{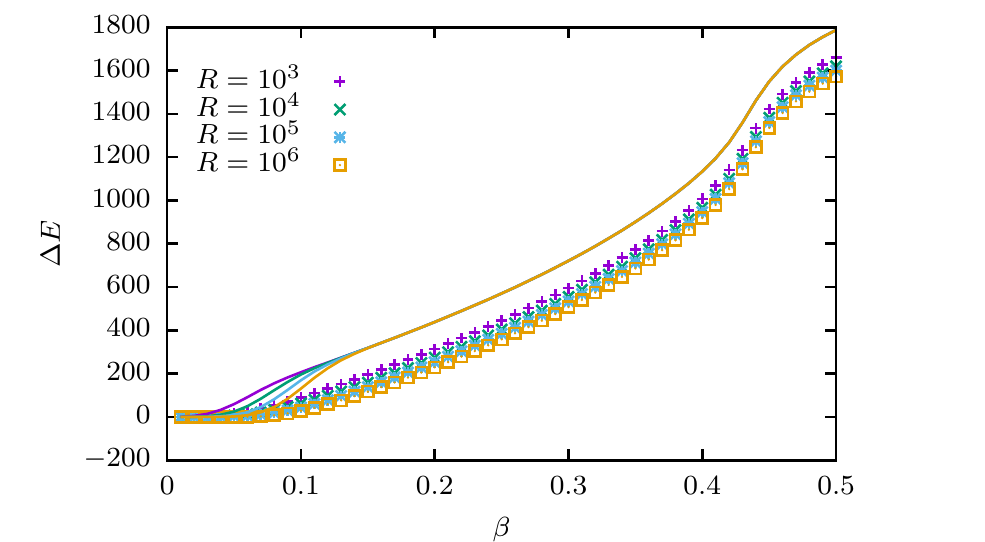}
  \includegraphics[clip=true,keepaspectratio=true,scale=0.95]{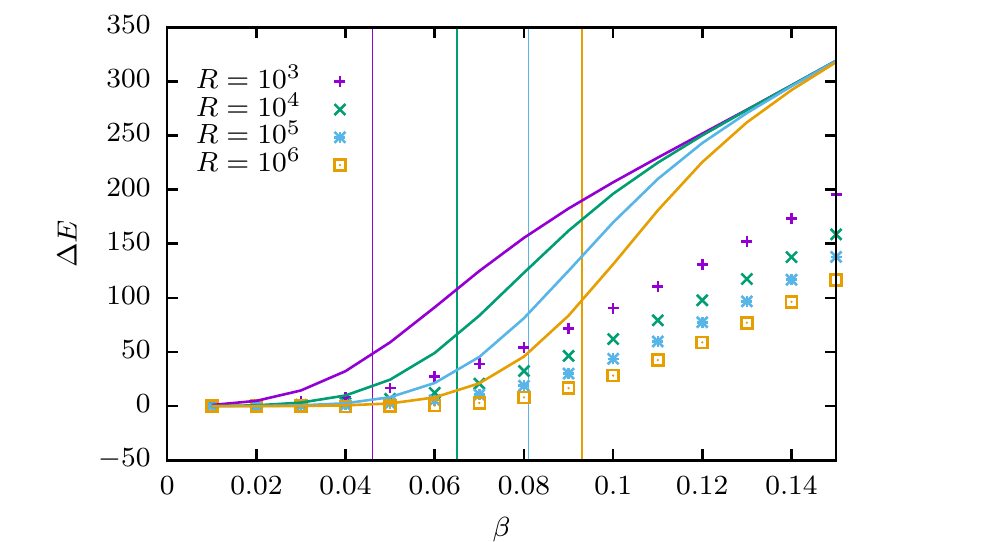}
  \includegraphics[clip=true,keepaspectratio=true,scale=0.95]{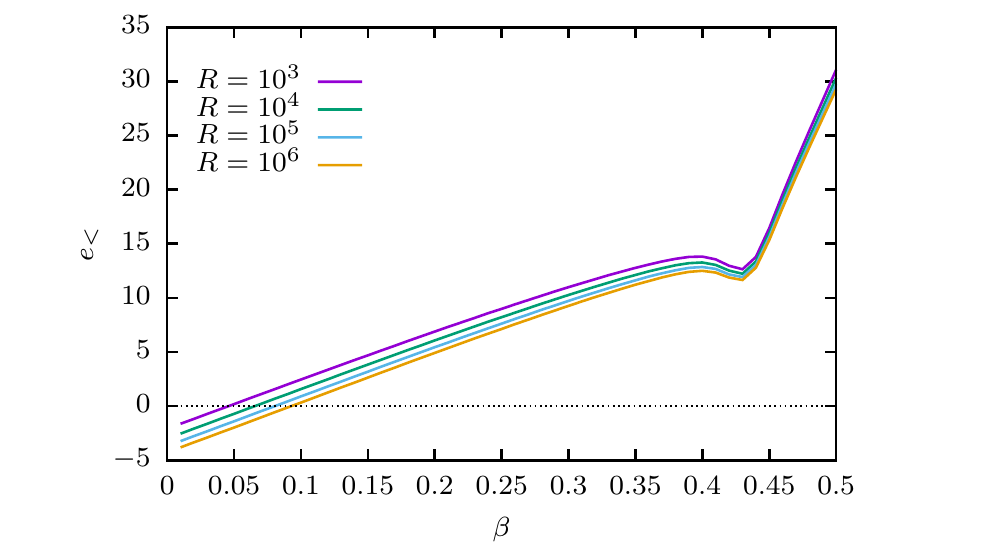}
  \caption
  { Top: Bias of the (total, not per site) internal energy estimate from PA
    simulations for $L=32$ and $\theta = 0$ and different population sizes. The lines
    show the conservative bias estimate resulting from
    Eq.~\eqref{eq:bias-estimate-analytical}. Middle: Detail of the upper panel. The
    vertical lines indicate the maximum inverse temperature, $\beta^\ast$ from which
    strong biases are expected according to
    Eq.~\eqref{eq:onset_of_strong_biases}. Bottom: Value of the normalized cut-off
    energy $e_< = (-E^\ast-\langle E\rangle_\beta)/\sigma_\beta$. 
    \label{fig:bias_noMC}
  }
\end{figure}

In PA with $\theta = 0$, estimating the energy distribution $P_\beta(E)$ (or any
derived quantity) amounts to reweighting from the distribution at $\beta_0 = 0$. This
is in fact just the situation encountered for simple-sampling Monte Carlo, and so the
following analysis applies to this problem as well. Because of the finite number $R$
of samples in the histogram $\hat{P}_{\beta_0}(E)$, there are no samples with
energies very far away from the peak of the distribution at $E=0$. We assume that the
width of these histograms is smaller than their distance, cf.\ the upper panel of
Fig.~\ref{fig:distribution-sketch}. If
$P_{\beta_0}(E=\langle E\rangle_\beta) \le 1/R$, there will essentially be no events
in $\hat{P}_{\beta_0}(E)$ that have substantial weight in the distribution
$P_\beta(E)$. In this case, the resampled histogram will be dominated by the few
replicas of smallest energies in $\hat{P}_{\beta_0}(E)$, and all replicas at inverse
temperature $\beta$ will be copies of these few replicas. Hence in this limit we have
$R_\mathrm{eff} \approx 1$. Since $P_{\beta_0}(E)$ is Gaussian (for not too small
system size), we can determine $\beta$ for the marginal case by requiring
\begin{equation}
  P_{\beta_0}(E=\langle E\rangle_\beta) = \frac{1}{\sqrt{2\pi\sigma_0^2}} e^{-\langle
    E\rangle_\beta^2/2\sigma_0^2} = \frac{1}{R_0},
\end{equation}
where $R_0$ is the population size starting from which one can expect to find a
reasonable result for $P_\beta(E)$ without MCMC steps, and $\sigma_0^2 = zL^d/2$ with
$z=4$ and $d=2$ for the square lattice. In other words, one expects substantial
biases to occur as soon as the point from which on there are no entries in the
population at $\beta_0$ reaches the peak of the distribution at $\beta$. The required
population size therefore grows as
\begin{equation}
  R_0 \sim e^{\langle E\rangle_\beta^2/2\sigma_0^2},
\end{equation}
i.e., exponentially in the total energy $\langle E\rangle_\beta \propto
L^d$. Conversely, for a given population size $R$ strong biases are expected for
inverse temperatures $\beta \ge \beta^\ast$ where
\begin{equation}
  \langle E\rangle_{\beta^\ast}^2 = 2\sigma_{0}^2\ln
  \frac{R}{\sqrt{2\pi\sigma_{0}^2}}.
  \label{eq:onset_of_strong_biases}
\end{equation}
This is illustrated in Fig.~\ref{fig:bias_noMC} showing results of PA simulations
with $\theta=0$. The vertical lines in the detailed view of the middle panel indicate
the values of $\beta^\ast$ corresponding to the chosen $R$.

To understand the behavior of the bias as a function of $\beta$ and $R$, we use a
simplified analysis based on the above argument for $R_0$ and a Gaussian shape of the
energy distribution. If we want to estimate $\langle E\rangle_\beta$ from the
population at $\beta_0$, the above arguments imply that there are no events in the
empirical (reweighted) histogram at $\beta$ for energies below $-E^\ast$ determined
by Eq.~\eqref{eq:onset_of_strong_biases} for
$E^\ast = \langle E\rangle_{\beta^\ast}$, see the bottom panel of
Fig.~\ref{fig:distribution-sketch}.
Hence the
PA estimate of the average energy will be systematically too large, namely
\begin{equation}
\hat{E}_\beta = \int_{-\infty}^\infty \hat{P}_\beta(E) E\,\d E \approx
\langle E\rangle_\beta - \int_{-\infty}^{-E^\ast} P_\beta(E) E\,\d E.
\end{equation}
If we assume that $P_\beta(E)$ is Gaussian, which is exact for $\beta = 0$, but
otherwise will be a good approximation for all temperatures apart from the critical
regime, the resulting bias is
\begin{equation}
  \begin{split}
    \Delta E &= \hat{E}_\beta - \langle E\rangle_\beta = 
    -\int_{-\infty}^{-E^\ast} P_\beta(E) E\,\d E \\
    &= -\frac{1}{\sqrt{2\pi \sigma_\beta^2}}\int_{-\infty}^{-E^\ast}
    e^{-\frac{(E-\langle E\rangle_\beta)^2}{2\sigma_\beta^2}} E\,\d E.
  \end{split}
\end{equation}
We note that $\sigma_\beta^2 = C_V(\beta)L^d/\beta^2$, where $C_V$ is the specific
heat. With the abbreviation
\begin{equation}
  e_<=(-E^\ast-\langle E\rangle_\beta)/\sigma_\beta
  \label{eq:eleft}
\end{equation}
we find
\begin{equation}
  \begin{split}
    \Delta E & = -\langle E\rangle_\beta\Phi(e_<)+\frac{\sigma_\beta}{\sqrt{2\pi}}
    \exp(-{e_<}^2/2) \\
    & \equiv \Delta E_1 + \Delta E_2.
  \end{split}
  \label{eq:bias-estimate-analytical}
\end{equation}
Here, $\Phi$ denotes the cumulative standard normal distribution function. Figure
\ref{fig:bias_noMC} shows the bias in energy observed from PA simulations with
$\theta=0$ performed for the $L=32$ model as well as the estimate from
Eq.~\eqref{eq:bias-estimate-analytical}. The latter follows the general behavior of
the actual bias, but systematically overestimates it. This is expected, however, as
in reality there can be occasional events with energies less than $-E^\ast$, just
with a probability less than one per energy bin. From the data of
Fig.~\ref{fig:bias_noMC} it seems clear that while for very small
$\beta \lesssim 0.1$ one can see a decay of bias towards zero, this is not the case
for significantly lower temperatures, where the bias is almost unchanged even on
increasing $R$ over three orders of magnitude. This is in quite strong contrast to
the general law of a $1/R$ decay of bias proposed in Ref.~\cite{wang:15a} (and
earlier in Ref.~\cite{machta:11}), but in line with the observations for the Ising
model shown above in Fig.~\ref{fig:population2}.

\begin{figure}[tb]
  \centering
  \includegraphics[clip=true,keepaspectratio=true,scale=0.95]{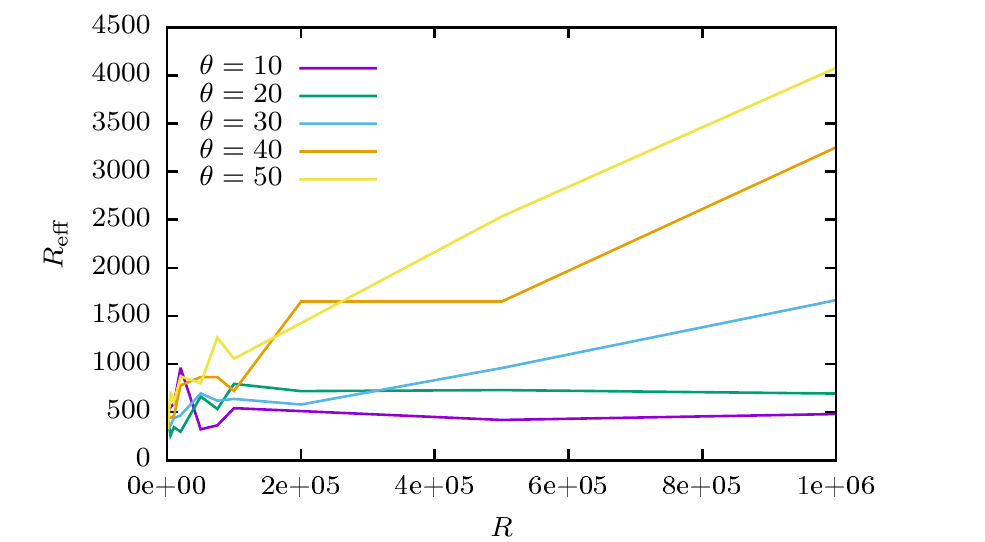}
  \caption
  { Effective population size $R_\mathrm{eff}$ at $\beta=0.44$ as estimated from the
    blocking analysis for PA runs for $L=64$ and the indicated values of $R$ and
    $\theta$. For $R < R_0(\theta)$, $R_\mathrm{eff}$ is essentially independent of
    $R$, while for $R \ge R_0(\theta)$, $R_\mathrm{eff} \propto R$.
    \label{fig:R0}
  }
\end{figure}

To understand the $R$ dependence of bias more systematically, we study the functional
form of Eq.~\eqref{eq:bias-estimate-analytical}. The behavior crucially depends on
the normalized cut-off energy $e_<$ of Eq.~\eqref{eq:eleft} which according to the
bottom panel of Fig.~\ref{fig:bias_noMC} changes sign on moving away from
$\beta=0$. This sign-change occurs when the cut-off energy $-E^\ast$ coincides with
the average energy $\langle E\rangle_\beta$, such that beyond that point there are
practically no relevant events in the histogram at $\beta_0 = 0$ that would allow to
estimate the energy at $\beta$. Hence, any reasonable reweighting can only occur in
the regime where $e_< < 0$. For very small $\beta$, we can use the asymptotic
expansion of $\Phi(z)$ \cite{abramowitz:book},
\begin{equation}
  \Phi(z) = -\frac{\exp(-z^2/2)}{\sqrt{2\pi}z}\left(1-\frac{1}{z^2}+\cdots\right),
\end{equation}
for $z\to-\infty$ to see that the leading-order $R$ dependence of both $\Delta E_1$
and $\Delta E_2$ is due to $\sim \exp(-e_<^2/2)$. On substituting $e_<$ from
Eq.~\eqref{eq:eleft} and using Eq.~\eqref{eq:onset_of_strong_biases} one finds
\begin{equation}
  e^{-e_<^2/2} = \left(\frac{\sqrt{2\pi\sigma_0^2}}{R}\right)^{\alpha(R,\beta)} e^{-\langle
    E\rangle_\beta^2/2\sigma_\beta^2},
\end{equation}
where the exponent $\alpha(R,\beta)$ is given by
\begin{equation}
  \alpha(R,\beta) =
  \frac{\sigma_0^2}{\sigma_\beta^2}-\frac{\sqrt{2\sigma_0^2}}{\sigma_\beta^2}|\langle
  E\rangle_\beta|\left(\ln \frac{R}{\sqrt{2\pi\sigma_0^2}}\right)^{-1/2}.
  \label{eq:bias-exponent}
\end{equation}
This is a rather interesting relation: asymptotically, it indicates power-law decay
in $1/R$ with an exponent that depends on the ratio $\sigma_0/\sigma_\beta$ of widths
of the corresponding energy distributions. In the limit of small $\beta$ considered
here, corresponding to small temperature steps in the $\theta > 0$ case,
$\sigma_0/\sigma_\beta \to 1$ and so the bias decays as $1/R$. However, due to the
second term in Eq.~\eqref{eq:bias-exponent} the crossover to the leading $1/R$
behavior is extremely (logarithmically) slow and, for example, for $R=10^6$ and
$\beta = 0.03$ one finds $ \alpha(R,\beta) \approx 0.36$ for $L=32$.

On the other hand, in the limit of large $\beta$, which for the $L=32$ Ising system
and population sizes between $10^3$ and $10^6$ already sets in for
$\beta \gtrsim 0.2$ where $e_< \gtrsim 5$ and hence $1-\Phi(e_<) \lesssim 10^{-7}$
(cf.\ the bottom panel of Fig.~\ref{fig:bias_noMC}), $\Phi(e_<) \approx 1$ to a high
accuracy, such that $\Delta E_1 \approx -\langle E\rangle_\beta$, while $\Delta E_2$
no longer decays (but note that it is strongly suppressed by a factor
$\exp[-\langle E\rangle_\beta^2/2\sigma_\beta^2]$). Hence the bias is essentially
independent of population size in this regime. Clearly, the onset of this regime
gradually shifts to larger $\beta$ for increasing $R$, but according to
Eq.~\eqref{eq:onset_of_strong_biases} this happens logarithmically slowly in $R$.

For a PA simulation with $\theta > 0$ the strength of bias effects will depend on the
efficiency of the Monte Carlo sweeps. In temperature regions where
$\theta \gg \tau_\mathrm{rel}$ we will always be in the weak-bias regime and the
$1/R$ decay can be observed at least asymptotically. In regions where
$\theta \ll \tau_\mathrm{rel}$, on the other hand, such as close to the critical
point in the Ising model for small values of $\theta$ or big systems, one is in the
strong-bias regime found above for $e_< > 0$, where there is essentially no
population-size dependence of the bias within practically achievable population
sizes. This is exactly the behavior found for simulations of the Ising model as
reported in Fig.~\ref{fig:population2}. Note that this effect only occurs for
problems where the energy is a relatively slow mode. For spin glasses this is not the
case, and so it is much easier to see the $1/R$ decay of the bias there (see also the
data presented in Ref.~\cite{wang:15a}).

Another effect of the tail domination of the resampling weights with a lack of
(efficient) MC moves is that for population sizes below $R_0$ the resampled
population is dominated by copies of one or a few replicas in the parent population
that happen to have the lowest energies. In these cases, the effective population
size following the discussion in Sec.~\ref{sec:correlation} is essentially
$R_\mathrm{eff} \approx 1$ \footnote{Note that for a determination of
  $R_\mathrm{eff}$ via the blocking method using $B$ blocks the estimate of
  $R_\mathrm{eff}$ is (up to fluctuations) bounded by $B$. In this case, the degree
  of correlation is actually too strong to be determined from the given population
  and number of blocks.}. Hence for $\theta=0$ the effective population size is
\begin{equation}
  R_\mathrm{eff} = \left\{
    \begin{array}{ll}
      \approx 1 & \mbox{for}\;R\le R_0 \\
      \mathrm{const.}\times R & \mbox{for}\;R > R_0
    \end{array}
    \right..
\end{equation}
For $\theta > 0$ we expect a similar behavior unless the MCMC alone is able to
equilibrate the replicas, i.e., we expect
$R_\mathrm{eff} = R[1-\exp(-\theta/\tau_\mathrm{eff})]$ to hold only for $R >
R_0$. (But note that $R_0$ should depend on $\theta$ too in this case.) This is
illustrated in Fig.~\ref{fig:R0} that shows the effective population size as
determined from the energy observable for $\beta = 0.44$, very close to the critical
point. We observe that $R_\mathrm{eff}$ is approximately constant and equal to a
minimal value (limited by the choice of the number of bins for the blocking analysis
which is $B=100$ here) for $R < R_0(\theta)$ and only proportional to $R$ for
$R \ge R_0$. For $\theta=30$, $R_0$ appears to be around $R=2\times 10^5$.

\section{Performance}
\label{sec:performance}

\begin{figure}[tb]
  \centering
  \includegraphics[clip=true,keepaspectratio=true,width=0.95\columnwidth]{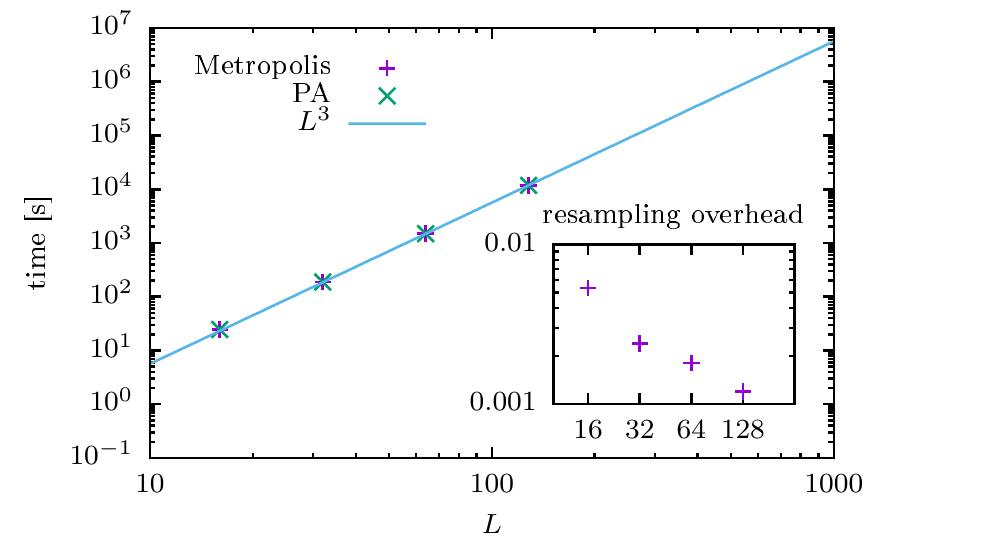}
  \caption
  {
    Run times of sample PA simulations for the 2D Ising model and system sizes
    $L=16$, $\ldots$, $128$ for a population size of $R=5000$, $\theta=10$,
    $\Delta\beta=16/75L$. The inset shows the relative excess time as compared to
    simulations with the resampling step turned off.
    \label{fig:runtimes}
  }
\end{figure}

Population annealing requires only relatively moderate modifications of standard
simulation codes that are typically based on MCMC, such as the single-spin flip
Metropolis or heatbath dynamics for the Ising model considered here. The main change
relates to the simulation of an ensemble of configurations rather than a single
copy. The resulting potential for the efficient utilization of highly parallel
architectures has been discussed elsewhere \cite{barash:16,christiansen:18}. The
computational overhead incurred by the resampling step results from the calculation
of the resampling weights $Q$ and $\hat{\tau}_i(E_j)$ of Eqs.~\eqref{eq:Q} and
\eqref{eq:taus}, drawing the numbers $r_i^j$ of copies from the chosen resampling
distribution, and the actual copy operations of configurations in memory or, for a
distributed implementation, over the network. In shared memory systems this overhead
is often rather moderate. For the serial CPU reference code used for the present
study, we show a comparison of a PA simulation and the pure single-spin flip code in
Fig.~\ref{fig:runtimes}. At the scale of the total simulation time, no difference is
visible for the chosen parameters. As the inset illustrates, the relative overhead of
performing the resampling step is below 1\% for $L=16$ and dropping to less than
1\textperthousand{} for $L>128$. For a discussion of the situation on GPU see
Ref.~\cite{barash:16}. As here the temperature step was chosen to scale with $L$ as
$\Delta\beta = 16/75 L$ to follow the expected scaling of the histogram overlap in
this model \cite{janke:08}, the overall runtime scales with $L^3$ as is illustrated
by the straight line in Fig.~\ref{fig:runtimes}.

\begin{figure}[tb]
  \centering
  \includegraphics[clip=true,keepaspectratio=true,width=0.95\columnwidth]{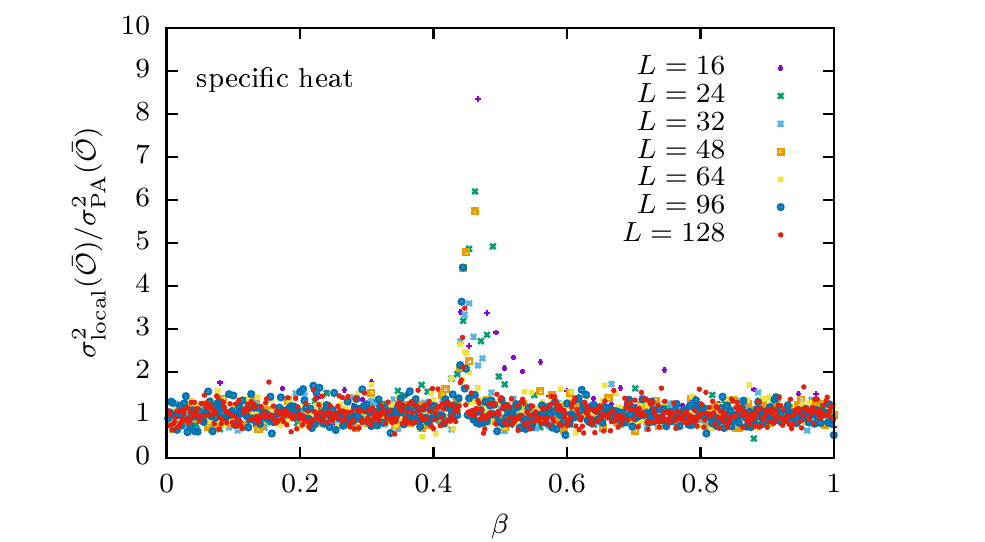}
  \includegraphics[clip=true,keepaspectratio=true,width=0.95\columnwidth]{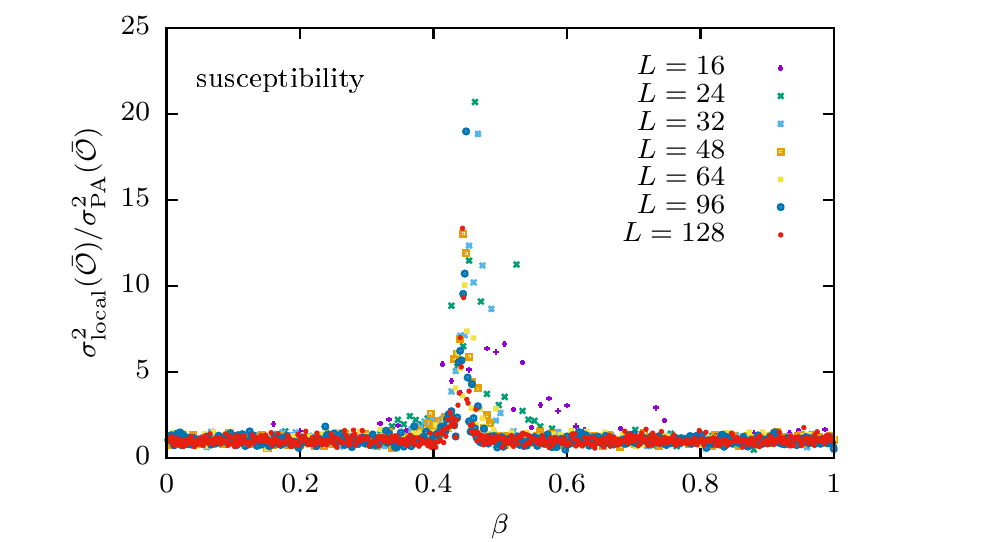}
  \includegraphics[clip=true,keepaspectratio=true,width=0.95\columnwidth]{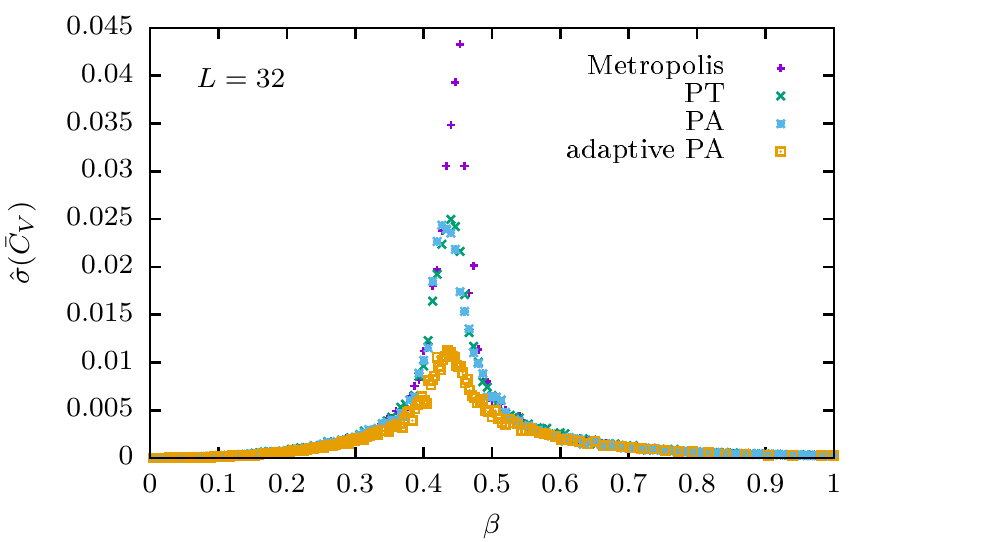}
  \caption
  {Relative squared error bars of observable estimates from Metropolis temperature
    sweeps against PA runs with the same number of spin flips. The resulting
    ``speedups'' relate to the specific heat (top panel) and the magnetic
    susceptibility (middle panel), respectively. PA updates are for $R=50\,000$,
    $\theta=(L/16)^2$, $\Delta\beta = 16/75L$ and Metropolis runs are for
    $R=1$, $\theta=50\,000 \times (L/16)^2$ and the same temperature
    sequence. Bottom panel: error bar of the specific heat estimate for $L=32$ and
    Metropolis and parallel tempering simulations as compared to standard and
    adaptive PA runs.
    \label{fig:speedup_plain}
  }
\end{figure}

The algorithmic performance of population annealing as a meta-algorithm is clearly
dependent on the model under consideration. For the reference case of the 2D Ising
model studied here, we do not expect massive improvements over the underlying MCMC
dynamics as the main difficulty in simulating the Ising model's continuous transition
lies in the critical slowing down near the transition, and not in a complex
free-energy landscape. The Ising model can in fact be very efficiently simulated with
the help of cluster algorithms \cite{swendsen-wang:87a,wolff:89a}, which can also be
combined with PA, but this is not the subject of the present study. Here, instead, we
focus on any possible reductions in bias and statistical errors that result from
implanting the MCMC into the PA framework. Figure~\ref{fig:speedup_plain} shows the
ratio of squared error bars for the specific heat (top) and susceptibility (middle)
for simulations using PA as compared to single-spin flip runs with the same
statistics. While for most temperatures where the MCMC is easily able to decorrelate
configurations the statistics are equivalent leading to a unit ratio of variances, in
the critical regime the consideration of an ensemble of configurations together with
resampling leads to decreased correlations as compared to the time series of a single
MCMC run and hence reduced error bars. The ``speedup'' displayed in
Fig.~\ref{fig:speedup_plain} corresponds to the number of such single-spin flip
simulations required to get the statistical errors to the same level as in a single
PA run. It is found to reach up to about 10 for the specific heat and up to about 20
for the susceptibility, but no particularly clear scaling behavior is found with
increasing the system size. Note that it is crucial for the relatively good
performance of the single-spin flip simulations that the final configuration at
$\beta_{i-1}$ is used as starting configuration at $\beta_i$ (hence these runs
correspond to what is called equilibrium simulated annealing in
Ref.~\cite{rose:19}). For the chosen parameter combinations, bias is far below the
threshold of statistical error, so we do not provide a detailed comparison of methods
in this respect, but overall for the Ising model we do expect the exponential decay
of bias with the number of MCMC sweeps as compared to the inverse decay with
population size as discussed in Sec.~\ref{sec:bias} to put the single-spin flip
simulations in a position of advantage as compared to PA runs in this respect.

It is further instructive to compare the effect of the PA meta-heuristic to that of
the more established parallel tempering method \cite{geyer:91,hukushima:96a}. To
allow for a relatively fair comparison, we employ the same temperature sequence and
the same total number of spin flips. As is clear from the presentation of the results
in the bottom panel of Fig.~\ref{fig:speedup_plain}, the reduction achieved in
statistical errors by applying parallel tempering on top of the Metropolis spin-flip
dynamics is practically identical to that seen for PA, at least for the parameters
used here. This is in line with previous observations indicating that the algorithmic
performance of PA in improving simulations and exploring the state space is quite
comparable to that effected by the more traditional parallel-tempering heuristic
\cite{wang:15,wang:15a,christiansen:18}. In this respect, the main advantage of the
PA method must be sought in the far superior parallel scaling in large simulations
\cite{christiansen:18}.

There is a great potential for further improvements to the PA method, however, and
while these haven been \cite{barash:16,barzegar:17,amey:18,barash:18} and will be
\cite{weigel:17a} discussed elsewhere, we show for comparison in
Fig.~\ref{fig:speedup_plain} the error bars achieved by combining an adaptive
temperature schedule \cite{barash:16} with overlap $\alpha = 0.9$ and an adaptive
flip schedule that dynamically modifies $\theta$ to ensure sufficient
decorrelation. In addition to the reduction in statistical errors, through the
dynamic flip schedule the adaptive simulation needs only about a fifth of the runtime
as compared to the other methods.

\section{Conclusions}
\label{sec:conclusions}

We have provided a detailed analysis of the properties of the population annealing
algorithm, using as a controlled and generic example the case of the two-dimensional
Ising model for which manifold exact and previous numerical results are
available. Our focus was on a systematic study of the dependence of systematic and
statistical errors on the parameters of the simulation, most notably the population
size $R$, the number of rounds of spin flips $\theta$, and the inverse temperature
step $\Delta\beta$.

At the core of population annealing is the resampling step that replicates
particularly well equilibrated replicas while eliminating those that are not
representative of the current temperature, cf.\ Fig.~\ref{fig:specific_heat}. While
selective replication helps to drive the simulation towards equilibrium, the
correlations between replicas built up in this process naturally increase statistical
errors. On the other hand, they also work towards increasing the fluctuations in the
distribution of configuration weights that are responsible for systematic error
(bias). The strength of such correlations is hence the central quantity for assessing
the quality of approximation. A particularly useful proxy of the actual correlations
taking the correlating effect of replication as well as the decorrelation of the
Monte Carlo moves into account is given by the effective population size
$R_\mathrm{eff}$ that can be readily estimated using a standard blocking
procedure. For the Ising model, $R_\mathrm{eff}(E)$ is dramatically reduced in the
critical regime --- indicative of the presence of critical slowing down --- but can
recover in the ordered phase, in contrast to correlation measures based entirely on
the analysis of the family tree.

Apart from providing $R_\mathrm{eff}$, the blocking and jackknifing procedure also
allows for estimates of statistical errors from within a single PA simulation. Such
error estimates are reliable as long as  $R_\mathrm{eff}$ is never less than a few
thousand replicas. It should hence be monitored in any PA simulation, preferably for
several relevant observables. We established an effective description of the
dependence of $R_\mathrm{eff}$ on the PA parameters, namely
\[
  R_\mathrm{eff} =
  R\left[1-\frac{\Delta\beta}{\Delta\beta_0}\exp(-\theta/\tau_\mathrm{eff})\right],
\]
that holds for small $\Delta\beta$ and $R_\mathrm{eff}$ satisfying the
self-consistency condition. Here, $\tau_\mathrm{eff}$ is an effective autocorrelation
time that is related to the relaxation time of the underlying MCMC algorithm. By
definition, statistical errors decay as $1/\sqrt{R_\mathrm{eff}}$.

For systematic errors $\Delta{\cal O}$ (bias), we have provided a description of PA
without resampling, where the behavior is determined by the spectrum of relaxation
times at all temperature points above the one considered. Including selective
replication does not affect the exponential functional dependence on $\theta$, but
leads to a much stronger sensitivity with respect to $\Delta\beta$ since alike to the
swap moves in parallel tempering it is only in the presence of sufficient overlap of
the energy histograms at neighboring temperatures that the resampling works
reliably. For small steps the bias is linear in $\Delta\beta$ --- we find this to
hold numerically and additionally derive it from the relation of bias to the
covariance of the considered observable with the free-energy estimator that was
previously suggested in Ref.~\cite{wang:15a}. Studying this estimator in the limit of
small (inverse) temperature steps reveals that it is in fact thermodynamic
integration in disguise, and it is possible to understand that in this limit the
variance of the free-energy estimator is proportional to
$\Delta\beta/R_\mathrm{eff} \propto \Delta\beta/R$. These findings can be summarized
as
\[
  \Delta{\cal O} \sim \frac{\Delta\beta}{R_\mathrm{eff}} e^{-\theta/\tau_\mathrm{rel}}.
\]
In practice, however, it can be quite difficult to reach the asymptotic regime where
$\Delta{\cal O} \propto 1/R$. In cases where the energy itself is slow to relax and
if $\theta$ is too small to keep the population in equilibrium at a given temperature
step, $R_\mathrm{eff}$ is effectively independent of population size up to large
values of $R$. Increasing the size of the population is an extremely inefficient way
of improving equilibration in such situations, and instead the only viable option is
to increase $\theta$ and/or choose a more efficient MCMC algorithm.

In view of the above, one might wonder how to best choose the simulation parameters
$R$, $\theta$ and $\Delta\beta$. Unfortunately, the above relations for statistical
error and bias are asymptotic, and hence rules derived from them might not yield the
best compromise for a given computational budget. Nevertheless, it is possible to
derive a number of guiding principles for the implementation of successful PA
simulations:
\begin{enumerate}
\item Choose $\Delta\beta$ to (just) ensure sufficient histogram overlap, for
  instance $\alpha > 0.7$. This might involve different inverse temperature steps at
  different temperatures. Ideally revert to using adaptive stepping
  \cite{barash:16,amey:18}.
\item In the regime where the MCMC is efficient, ensure that $\theta$ is chosen large
  enough to ascertain equilibration of the population at each temperature step. In a
  regime where the relaxation times become too large --- for example on entering a
  phase of broken ergodicity --- spin flips will likely become less relevant
  \cite{rose:19}. Potentially choose a more efficient MCMC algorithm if it is
  available.
\item With any remaining computational resources adapt the population size to bring
  down statistical errors (that will asymptotically dominate) to the desired level.
\item Monitor $R_\mathrm{eff}$ during the course of the simulation, ensuring that it
  is $\gtrsim 1000$--$5000$ at all times; potentially consider $R_\mathrm{eff}$ for
  different relevant observables, including the configurational overlap.
\item Make use of the potential for extensive parallelization of the algorithm,
  potentially allowing to arbitrarily reduce statistical and systematic errors of the
  estimates at the same overall wallclock time if and when sufficient parallel
  resources are available \cite{weigel:18}.
\item Evaluate statistical errors, including the variance and covariances involving
  the free-energy estimator using a jackknife block analysis of the tree-ordered
  population of replicas.
\end{enumerate}
When considering any such general guidelines for optimizing population annealing, it
is important to also keep in mind the different levels of relevance of bias and
statistical errors in different applications. While in studies of pure systems
systematic errors can typically be brought under control by the employment of
suitable MCMC algorithms and long runs, the sitution is different for simulations of
systems with quenched disorder, where the main source of error are sample-to-sample
fluctuations and hence runs for individual disorder configurations are relatively
short compared to the relevant relaxation times \cite{wang:15a}. Further, so far
unexplored applications might have yet different intricacies.

While the present study delivers a rather detailed picture of the performance of the
established population annealing algorithm following Ref.~\cite{machta:10a}, a range
of improvements have already been proposed
\cite{barash:16,barzegar:17,amey:18,barash:18}, and many more are conceivable,
including adaptive $\theta$ and $R$ schedules \cite{weigel:17a}, the study of further thermodynamic
ensembles \cite{rose:19}, or generalized resampling schemes \cite{gessert:prep}. This
flexibility, together with the essentially unlimited potential for parallelization
turn population annealing into one of the most versatile and promising
generalized-ensemble simulation schemes available.

\begin{acknowledgments}
  We thank Chris Amey, Helmut Katzgraber, Jonathan Machta, and Wenlong Wang for
  useful discussions. The work of L.B. and L.S. was supported by the grant
  14-21-00158 from Russian Science Foundation and finished within the framework of
  State Assignment of Russian Ministry of Science and Higher Education. The authors
  acknowledge support from the European Commission through the IRSES network DIONICOS
  under Contract No. PIRSES-GA-2013-612707. The simulations were performed on the HPC
  facilities of Coventry University and the Science Center in Chernogolovka.
\end{acknowledgments}

\appendix

\section{Effective population size}
\label{app:effective-population-size}

\begin{figure}[tb!]
  \centering
  \includegraphics[clip=true,keepaspectratio=true,width=0.95\columnwidth]{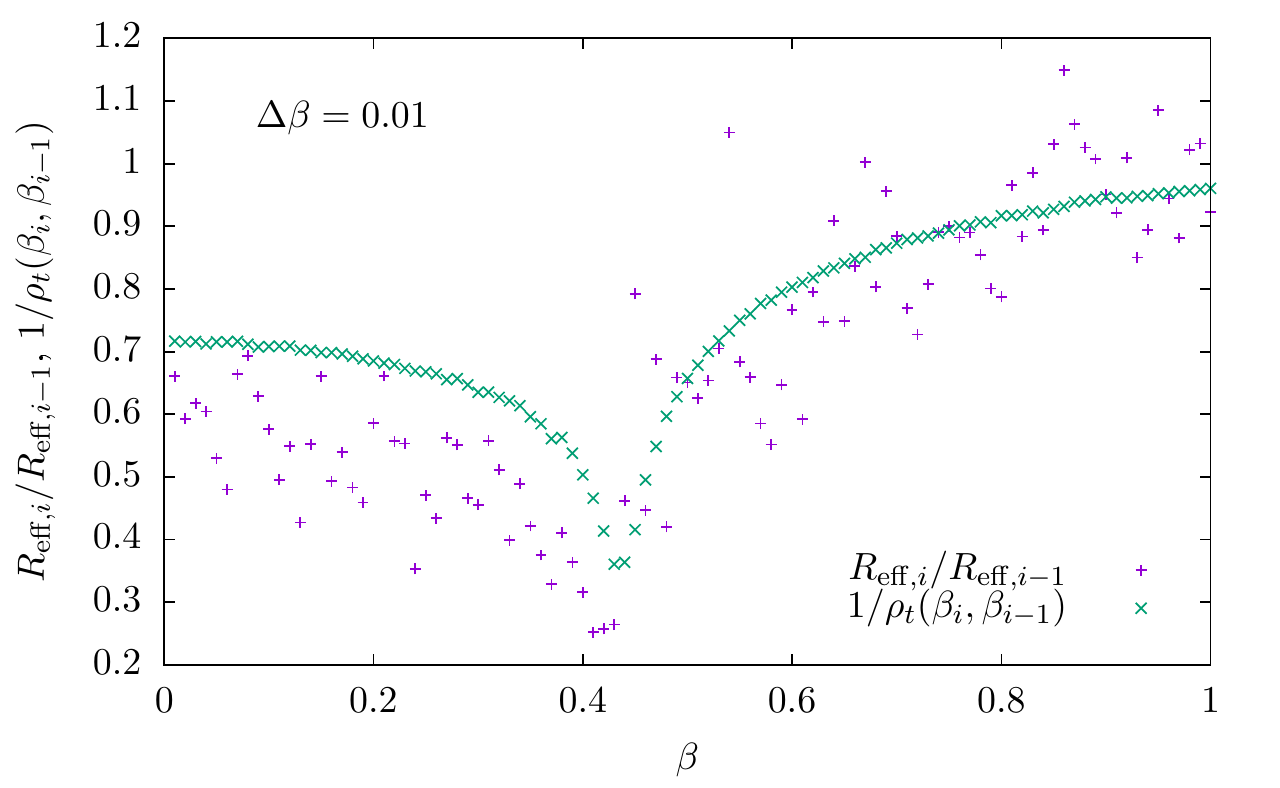}
  \includegraphics[clip=true,keepaspectratio=true,width=0.95\columnwidth]{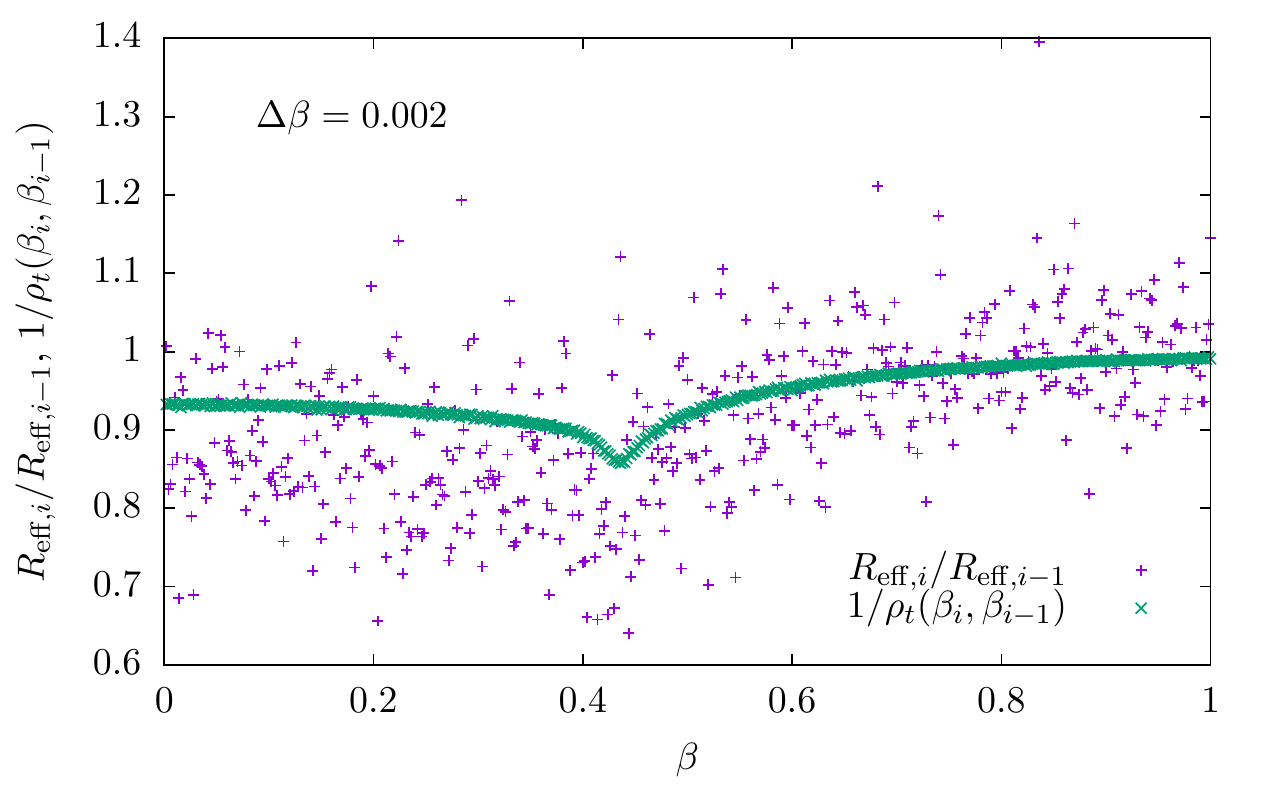}
  \caption
  {Ratio of effective population sizes before and after the resampling step as
    compared to the inverse of the one-step mean-square family size according to
    Eq.~\eqref{eq:one-step-Reff} for two sizes of the inverse temperature steps
    ($L=32$, $\theta=20$, $R=50\,000$).
    \label{fig:one-step-Reff}}
\end{figure}

In view of the observed relation Eq.~\eqref{eq:Reff} for $R_\mathrm{eff}$ it is
worthwhile to consider the effect of the individual elements of the algorithm on the
{\em effective\/} population size. The resampling step creates identical copies of
some replicas while eliminating other members of the population and hence must
normally reduce the effective population size. To understand this effect, we consider
the variance of the mean. In a population of perfectly uncorrelated members the
behavior follows Eq.~\eqref{eq:variance-of-the-mean} and hence $R_\mathrm{eff} = R$
initially \footnote{As in Sec.~\ref{sec:correlation} we ignore the effect of
  fluctuating population size here, but it is straightforward to adapt the resulting
  expressions to account for this factor.}. After resampling (but before applying the
spin flips) each copy $k$ of the original population at temperature step $i$ has been
replaced by $r^k_i$ identical copies in the descendant population (where $r^k_i = 0$
for replicas that have died out), cf.\ Eq.~\eqref{eq:resampling-factors}. If we
assume that $\sigma^2({\cal O})$ is the same for all replicas (i.e., the variance is
not correlated with the number of children), it is easy to see that the variance of
the mean becomes
\begin{equation}
  \sigma^2(\overline{\cal O}) = \sigma^2({\cal O}) \left(\sum_k \frac{{r^k_i}^2}{R^2}\right).
\end{equation}
It is then natural to define a generalization of the mean square family size $\rho_t$
as
\begin{equation}
  \rho_t(\beta_j, \beta_i) = R \sum_k {\mathfrak{n}_k(\beta_j, \beta_i)}^2,\;\;\; j < i,
  \label{eq:one-step-Reff}
\end{equation}
where $\mathfrak{n}_k(\beta_j, \beta_i)$ is the fraction of the population
at $\beta_i$ coming down from the $k$th member of the population at
$\beta_j$. Clearly, $\rho_t \equiv \rho_t(\beta_0, \beta_i)$. Since $r^k_i/R =
\mathfrak{n}_k(\beta_{i-1},\beta_i)$, we have
\[
  \sigma^2(\overline{\cal O}) = \frac{\sigma^2({\cal O})}{R/\rho_t(\beta_{i-1}, \beta_i)}.
\]
Repeating these arguments for a correlated population with $R_\mathrm{eff} < R$, one
concludes that after resampling
\begin{equation}
  R_{\mathrm{eff},i} = \frac{R_{\mathrm{eff},i-1}}{\rho_t(\beta_{i-1}, \beta_i)}.
  \label{eq:one-step-Reff2}
\end{equation}
As is shown in Fig.~\ref{fig:one-step-Reff} this relation indeed describes the
overall behavior correctly. The deviations, which are particularly visible for
$\beta < \beta_c$, remind us that the variance (and indeed the mean) of ${\cal O}$ is
{\em not\/} necessarily independent of the number of children. In particular, for
${\cal O} = E$ as considered here, we note that such correlations must exist since
replicas with lower energies will have more children than those with higher energies.

Note that there should be good reasons to assume that maximizing
$\rho_t(\beta_{i-1}, \beta_i)$ and hence minimizing the reduction of $R_\mathrm{eff}$
through the resampling step should be a desirable goal for choosing among different
possible resampling schemes (although the effect on bias is not completely clear at
present).

\begin{figure}[tb!]
  \centering
  \includegraphics[clip=true,keepaspectratio=true,width=0.95\columnwidth]{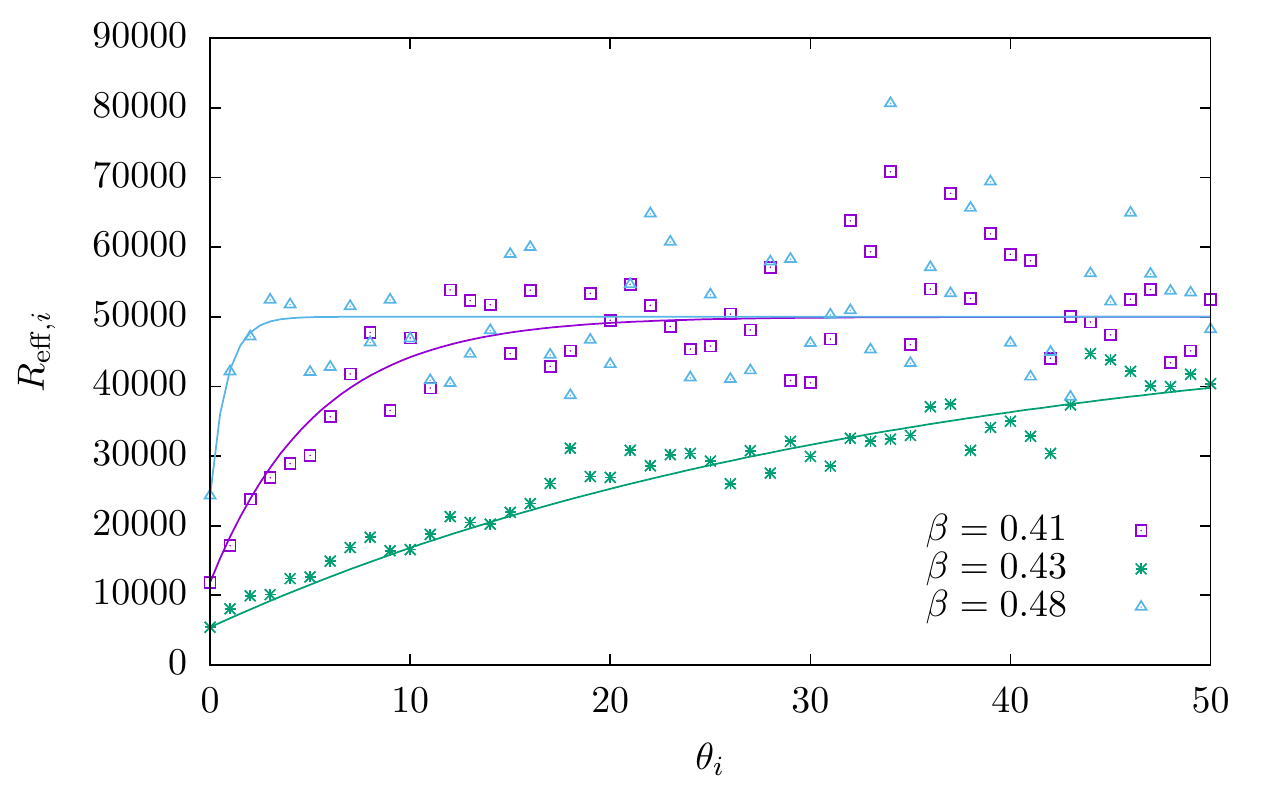}
  \caption
  {Effective population size $R_{\mathrm{eff},i}$ at inverse temperature
    $\beta_i = 0.41$, $0.43$ and $0.48$, respectively, after applying a number of
    $\theta_i$ MCMC steps. The effective population size relaxes to the total
    population size $R$ according to Eq.~\eqref{eq:one-step-Reff3} as $\theta_i$ is
    increased. The lines show fits of the form \eqref{eq:one-step-Reff3} with
    parameter $\tau_{\mathrm{rel},i}$ to the data, the best fit being achieved for
    $\tau_{\mathrm{rel},i} = 5.3$ ($\beta=0.41$), $33.9$ ($\beta = 0.43$), and $0.8$
    ($\beta=0.48$), respectively ($L=32$, $\theta=10$, $R=50\,000$,
    $\Delta\beta = 0.01$).
    \label{fig:one-step-Reff2}}
\end{figure}

After the resampling step that tends to reduce $R_\mathrm{eff}$, the spin flips serve
to remove some of the correlations in the population. As these are here
assumed to be implemented as an MCMC process, such decorrelation will have an
exponential time dependence. Focusing on the single leading exponential and taking
into account that $R_\mathrm{eff}\to R$ as $\theta\to\infty$, the expected behavior
is
\begin{equation}
  R_{\mathrm{eff},i} = R_{\mathrm{eff},i-1} +(R-R_{\mathrm{eff},i-1})(1-e^{-\theta_i/\tau_{\mathrm{rel},i}}).
  \label{eq:one-step-Reff3}
\end{equation}
Here, $\tau_{\mathrm{rel},i}$ denotes the relevant relaxation time at the inverse
temperature $\beta_i$. This functional form is indeed consistent with the observed
data, see the results presented in Fig.~\ref{fig:one-step-Reff2}, showing the
behavior of $R_{\mathrm{eff},i}$ as a function of the number $\theta_i$ of MCMC steps
applied. The lines show one-parameter fits of the functional form
\eqref{eq:one-step-Reff3} to the data, with $R=50\,000$ and $R_{\mathrm{eff},i-1}$
determined after resampling, but before the MCMC steps, via the blocking scheme of
Sec.~\ref{sec:correlation}.

From relations \eqref{eq:one-step-Reff2} and \eqref{eq:one-step-Reff3} one might
deduce conditions for adjusting the population size $R_i$ or the number of rounds of
spin flips $\theta_i$ at each temperature step such as to avoid a possible drop in
$R_\mathrm{eff}$, but such attempts are left for future work.

\begin{figure}[tb!]
  \centering
  \includegraphics[clip=true,keepaspectratio=true,width=0.95\columnwidth]{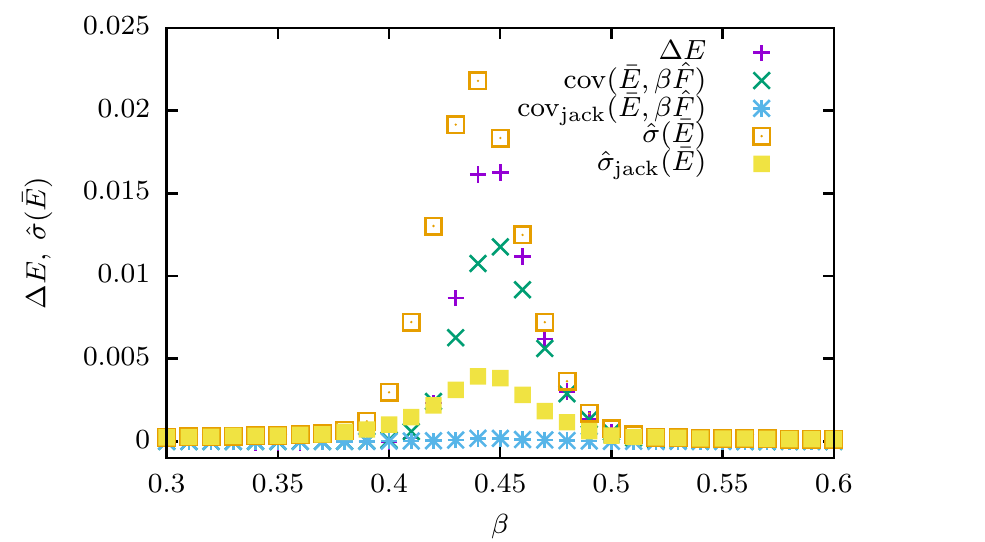}
  \includegraphics[clip=true,keepaspectratio=true,width=0.95\columnwidth]{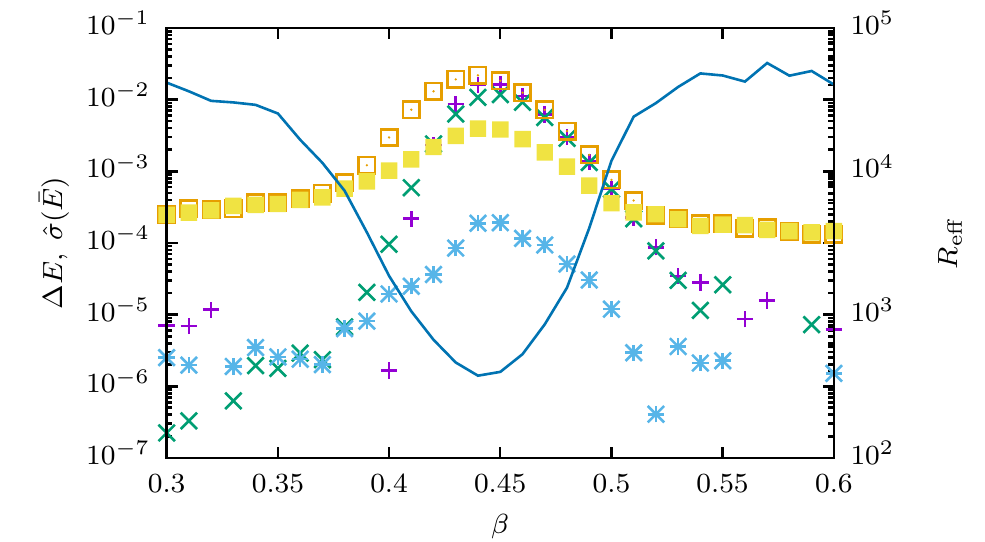}
  \caption
  { Top: Bias and statistical errors of the average internal energy $\overline{E}$
    for the $L=32$ square-lattice Ising model as estimated from independent runs as
    well as from the jackknife blocking method ($R=50\,000$, $\theta=1$,
    $\Delta\beta = 0.01$). Bottom: The same data in a logarithmic scale shown in
    comparison to the effective population size $R_\mathrm{eff}$ (solid line, right
    scale).
    \label{fig:bias-illustration}
  }
\end{figure}

\section{Bias --- further numerical observations}

Is the bias really represented fully by the covariance with the free-energy
estimator? How accurate are the estimates of covariance and variance from the
blocking method? These questions are addressed in the plots of
Fig.~\ref{fig:bias-illustration} showing the result of a PA run for $L=32$ with a
target population size $R=50\,000$ and fixed temperature steps $\Delta\beta = 0.01$
with $\theta = 1$. Here, we compare for the energy observable (1) the error bars
(standard deviation) as estimated from a jackknife block analysis as well as from
$m=200$ independent runs, and (2) the bias estimated via
$\operatorname{cov}(\overline{E}, \beta\hat{F})$, cf.\
Eq.~\eqref{eq:bias_covariance}, as determined either from the jackknife or from
independent runs with the actual bias as computed via comparison of the simulation
data to the exact result \cite{kaufman:49a,ferdinand:69a}. Relevant observations are:
\begin{itemize}
\item As expected, bias and statistical errors are small everywhere but in the
  critical region.
\item Bias is below statistical error {\em everywhere\/}, even in the critical
  region, such that (at least for the observable $E$ and for the choice of parameters
  considered here), a dynamical comparison of bias and statistical error in the sense
  of an adaptive algorithm would never have suggested to increase $\theta$.
\item The covariance with the free energy estimator (as estimated from independent
  runs) is consistent with the actual bias everywhere. (There might be some small
  deviations very close to the peak, though.)
\item The estimates of both the statistical error and the covariance from the
  blocking method are consistent with the results from independent runs apart from
  in the critical regime.
\item The deviations of the blocking estimates from the correct results set in
  exactly when $R_\mathrm{eff}$ gets small and hence estimates from blocking can
  naturally not be trusted.
\end{itemize}
We note that the data from ``independent runs'' was produced on GPU while the data
for the blocking method are from a CPU run. Since a checkerboard update is used on
GPU but sequential updates on CPU, this should lead to slightly different
decorrelation properties, which might have resulted in some differences between the
shown data sets.


%

\end{document}